\documentclass[aoas,preprint]{imsart}

\RequirePackage[OT1]{fontenc}
\RequirePackage{amsthm,amsmath}
\RequirePackage{natbib}
\RequirePackage[colorlinks,citecolor=blue,urlcolor=blue]{hyperref}

\usepackage{cancel}
\usepackage[pdftex]{graphicx}
\usepackage{tikz}
\usepackage{enumerate}
\usepackage{amsfonts}
\usepackage{multirow}
\usepackage{algorithmicx, algpseudocode}
\usepackage{algorithm}
\usetikzlibrary{calc,trees,positioning,arrows,chains,shapes.geometric,%
    decorations.pathreplacing,decorations.pathmorphing,shapes,%
    matrix,shapes.symbols}
\DeclareMathOperator*{\argmax}{arg\,max}
\arxiv{arXiv:1602.01462}

\startlocaldefs
\numberwithin{equation}{section}
\theoremstyle{plain}

\endlocaldefs

\begin{document}

\begin{frontmatter}
\title{Bayesian Estimates of Astronomical Time Delays between Gravitationally Lensed Stochastic Light Curves}
%\title{A Sample Document\thanksref{T1}}
\runtitle{Bayesian Time Delay Estimation}
%\thankstext{T1}{Footnote to the title with the ``thankstext'' command.}

\begin{aug}
\author{\fnms{Hyungsuk} \snm{Tak}\thanksref{t1}\ead[label=e1]{hyungsuk.tak@gmail.com}},
\author{\fnms{Kaisey} \snm{Mandel}\thanksref{t2}\ead[label=e2]{kmandel@cfa.harvard.edu}},
\author{\fnms{David A.} \snm{van Dyk}\thanksref{t3}\ead[label=e3]{dvandyk@imperial.ac.uk}},
\author{\fnms{Vinay L.} \snm{Kashyap}\thanksref{t2}\ead[label=e4]{vkashyap@cfa.harvard.edu}},
\author{\fnms{Xiao-Li} \snm{Meng}\thanksref{t4}\ead[label=e5]{meng@stat.harvard.edu}},
\and
\author{\fnms{Aneta} \snm{Siemiginowska}\thanksref{t2}\ead[label=e6]{asiemiginowska@cfa.harvard.edu}}

%\and
%\author{\fnms{Third} \snm{Author}\thanksref{t1,m2}
%\ead[label=e3]{third@somewhere.com}
%\ead[label=u1,url]{http://www.foo.com}}

\thankstext{t1}{Statistical and Applied Mathematical Sciences Institute, 19 T.W. Alexander Dr., Durham, NC, USA}
\thankstext{t2}{Harvard-Smithsonian Center for Astrophysics, 60 Garden St., Cambridge, MA, USA}
\thankstext{t3}{Statistics Section, Department of Mathematics, Imperial College London, London SW7 2AZ UK}
\thankstext{t4}{Department of Statistics, Harvard University, 1 Oxford St., Cambridge, MA, USA}
%\thankstext{t3}{Second supporter of the project}
\runauthor{Tak et al.}

\affiliation{Statistical and Applied Mathematical Sciences Institute, Harvard-Smithsonian Center for Astrophysics,  Imperial College London, and Harvard University}%\thanksmark{m1} and Another University\thanksmark{m2}}

\address{Hyungsuk Tak\\
SAMSI\\
19 T.W. Alexander Drive\\
Durham, NC, USA 27703\\
\printead{e1}\\
\phantom{E-mail:\ }}

\address{Kaisey Mandel\\
Harvard-Smithsonian Center for Astrophysics\\
Harvard University\\
60 Garden Street\\
Cambridge, MA, USA 02138\\
\printead{e2}\\
\phantom{E-mail:\ }}

\address{David A. van Dyk\\
Statistics Section\\
Department of Mathematics\\ 
Imperial College London\\
London SW7 2AZ UK\\
\printead{e3}\\
\phantom{E-mail:\ }}

\address{Vinay L. Kashyap\\
Harvard-Smithsonian Center for Astrophysics\\
Harvard University\\
60 Garden Street\\
Cambridge, MA, USA 02138\\
\printead{e4}\\
\phantom{E-mail:\ }}

\address{Xiao-Li Meng\\
Department of Statistics\\
Harvard University\\
1 Oxford street\\
Cambridge, MA, USA, 02138\\
\printead{e5}\\
\phantom{E-mail:\ }}

\address{Aneta Siemiginowska\\
Harvard-Smithsonian Center for Astrophysics\\
Harvard University\\
60 Garden Street\\
Cambridge, MA, USA 02138\\
\printead{e6}\\
\phantom{E-mail:\ }}

\end{aug}

\begin{abstract}
The gravitational field of a  galaxy can act as a lens and deflect the light emitted by a more distant object  such as a quasar.  Strong gravitational lensing causes multiple images of the same quasar to appear in the sky. Since the light in each gravitationally lensed image traverses a different path length from the quasar to the Earth, fluctuations in the source brightness are observed in the several images at different times.  The time delay between these fluctuations can be used to constrain cosmological parameters and can be inferred from the time series of brightness data or light curves of each image. To estimate the time delay, we construct a model based on a state-space representation for  irregularly observed time series generated by a latent continuous-time Ornstein-Uhlenbeck process. We account for microlensing, an additional source of independent long-term extrinsic variability, via a polynomial regression. Our Bayesian strategy adopts  a Metropolis-Hastings within Gibbs sampler. We improve the sampler by using an ancillarity-sufficiency interweaving strategy and adaptive Markov chain Monte Carlo. We introduce a profile likelihood of the time delay as an approximation of its marginal posterior distribution. The Bayesian and profile likelihood approaches  complement each other, producing almost identical results; the Bayesian method is more principled but the profile likelihood is simpler to implement.  We demonstrate our estimation strategy using simulated data of doubly- and quadruply-lensed quasars, and observed data from quasars \emph{Q0957+561} and \emph{J1029+2623}.
%When the quasar, intervening galaxy, and Earth are aligned, multiple images of the same quasar can appear in the sky
%scientifically motivated hyper-prior distributions and
%, including the current expansion rate of the Universe
%, and a tempered transition to efficiently sample the posterior distribution
%that mixes two approaches

\end{abstract}

%\begin{keyword}[class=MSC]
%\kwd[Primary ]{60K35}
%\kwd{60K35}
%\kwd[; secondary ]{60K35}
%\end{keyword}


\end{frontmatter}

%Quasars are a type of active galactic nuclei whose central black holes are accreting actively
\section{Introduction} 
\begin{figure}[b!]
\includegraphics[scale = 0.26]{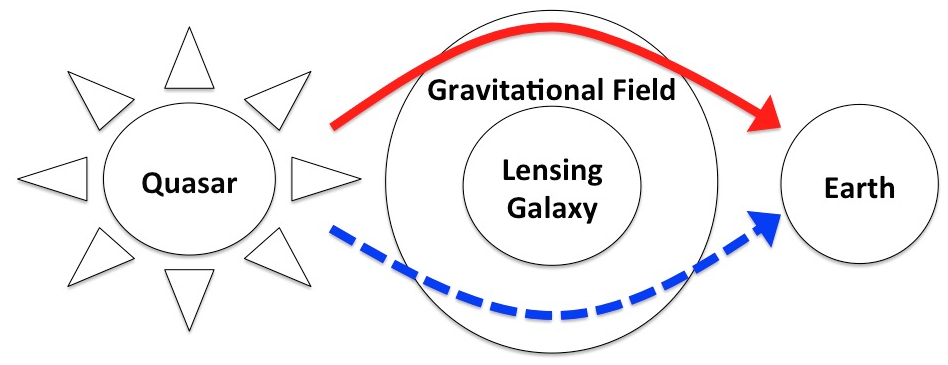}~~~~~~
\includegraphics[scale = 0.178]{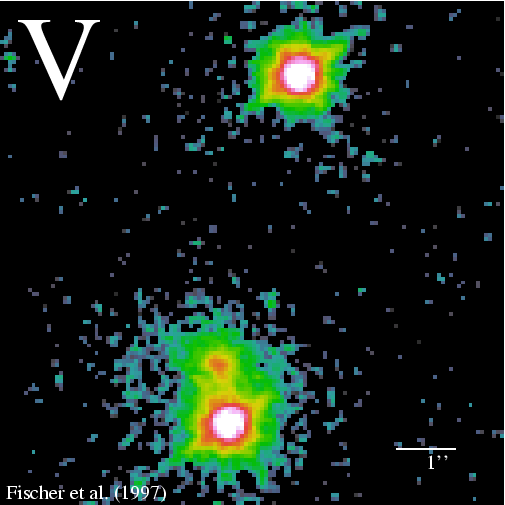}
\caption[]{The gravitational field of an intervening galaxy acts as a lens deflecting two light rays of a quasar image towards the Earth as shown in the left panel. The arrival times can differ owing to the different lengths of pathways and different gravitational potentials they pass through. An optical V-band image of the doubly-lensed quasar Q0957+561 obtained with the Canada France Hawaii telescope \citep{fischer1997, munoz1998castles} (https://www.cfa.harvard.edu/castles) appears in the right panel. The two bright sources at the top and bottom are the lensed images of the quasar, and the small red point towards the top-left of the lower quasar image is the lensing galaxy. }
\label{fig1}
\end{figure}

Quasars are the most luminous active galaxies in the Universe that host an accreting supermassive black hole at the center. The path that light takes from a quasar to Earth can be altered by the gravitational field of a massive intervening galaxy, acting as a lens and bending the trajectory of the emitted light; see the first panel of Figure~\ref{fig1}. When the quasar, lensing galaxy, and Earth are geometrically aligned, multiple images of the  quasar can appear in slightly different locations in the sky, from the perspective of an observer on Earth. This phenomenon is known as strong gravitational lensing \citep{schneider1992, schneider2006}.  In this case, there are typically two or more replicate  images, referred to as doubly- or multiply-lensed quasars. Since quasars are highly luminous, they can be seen at great distances, which both enhances the possibility of lensing by an intervening galaxy and makes them useful for cosmology.
%When the quasar (source), lensing galaxy, and Earth are aligned}, multiple images of the quasar can appear in slightly different locations in the sky, from the perspective of an observer on Earth, an effect known as strong gravitational lensing
%The path that light takes from a quasar to Earth can be altered by the gravitational field of a massive intervening galaxy which thus acts as a  lens, bending the trajectory of the emitted light

%The brightnesses of the two gravitationally lensed images at the top and bottom over time are measured with error. }

The light rays forming each of these gravitationally lensed quasar images take different routes from the quasar to Earth.   Since both the lengths of the pathways and the gravitational potentials they traverse differ, the resulting multiple images are subject to differing lensing magnifications and their light rays arrive at the observer at different times. Because of this, any fluctuations in the source brightness are observed in each image at different times. From a statistical perspective, we can construct a time series of the brightness of each image, known as  \emph{a light curve}. Features in these light curves  appear to be shifted in time and these shifts are called \emph{time delays}.

Obtaining accurate time delay estimates is important in cosmology because they can be used to address fundamental questions regarding the origin and evolution of the Universe. For instance, \citet{ref1964} suggested using time delay estimates to constrain the Hubble constant $H_0$, the current expansion rate of the Universe; given a model for the mass distribution and gravitational potential of the lensing galaxy, the time delay between multiple images of the lensed quasar is inversely proportional to $H_0$ \citep{blandford1992, suyu2013two, treu2016time}.  Also, \cite{Linder2011} showed that an accurate time delay estimate could substantially constrain  cosmological parameters and the equation of state of dark energy characterizing the accelerated expansion of the Universe.

%cosmological parameters in the equation of state of dark energy characterizing the accelerated expansion of the Universe.

The upcoming large-scale astronomical survey to be conducted with the Large Synoptic Survey Telescope \citep[LSST,][]{lsst2009lsst}  will monitor thousands of gravitationally lensed quasars beginning in 2022.  The LSST is the top-ranked ground-based telescope project in the 2010 Astrophysics Decadal Survey, and will produce extensive high-cadence time series observations of the full sky for ten years.  The LSST will produce multi-band light curves (observed via multiple optical filters centered at different wavelengths) that form a vector time series for each image.  In preparation for the era of the LSST, \cite{dobler2013} organized a blind competition called the Time Delay Challenge (TDC) which ran from October 2013 to July 2014 with the aim of improving time delay estimation methods for application to realistic observational data sets.  As a simplification for the first competition, the TDC organizers simulated thousands of single-band datasets, i.e., scalar time series for each image, that mimic real quasar data. We are among 13 teams who took part in the TDC, each of which analyzed the simulated data using their own methods to estimate the blinded time delays\footnote{In the last stage of the TDC (called \emph{rung4} in the TDC), an earlier version of our method achieved the smallest average coefficient of variation (\emph{precision}), the TDC target for the average error level (\emph{accuracy}) within one standard deviation, and acceptable average squared standardized residual ($\chi^2$) after analyzing the second highest number of data sets ($f$). See \cite{kai2014} for detailed results of the TDC.}.

\subsection{Data and challenges}\label{data_challenge} 
\begin{figure}[t!]
\includegraphics[scale = 0.3]{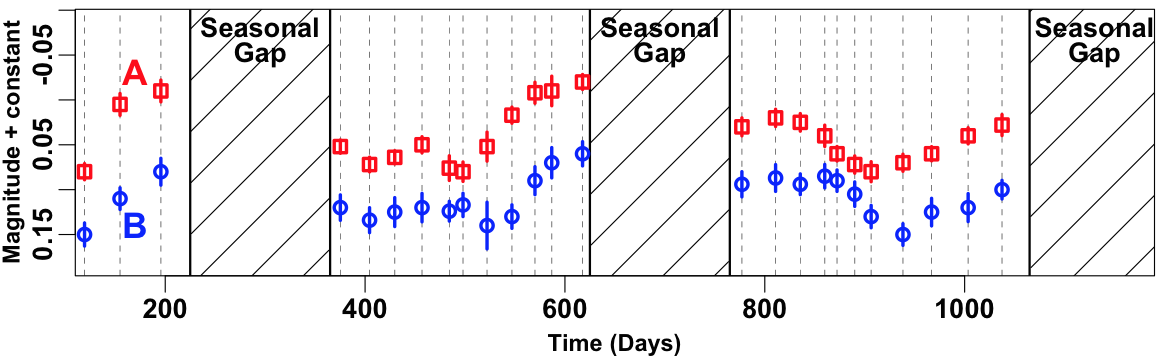}
\caption[]{The red squares and blue circles indicate the observed magnitudes  of the two simulated images at each observation time. The half lengths of vertical lines around the symbols represent the uncertainties (standard deviations) of the observed magnitudes. The convention in Astronomy is to plot the magnitude inversely so that  smaller magnitudes (brighter object) appear on the top and larger ones (fainter object) on the bottom.  The quasar magnitudes are vertically offset by an overall calibration constant, the value of which is unimportant for time delay estimation.}
\label{fig2}
\end{figure}

We plot a pair of  simulated light curves from a doubly-lensed quasar in Figure~\ref{fig2}; the light curves are labeled as $A$ and $B$.  Each observation time is   denoted by vertical dashed lines, at which the observer measures the brightness of each gravitationally lensed quasar image. In a real data analysis, these images would correspond to  the two bright sources in the second panel of Figure~\ref{fig1}. The brightness is reported on the magnitude scale, an astronomical logarithmic measure of brightness, in which smaller numbers correspond to brighter objects.  The magnitudes in Figure~\ref{fig2} are presented up to an overall additive calibration constant as was the case in the TDC. Since the time delay is estimated via relative comparison between fluctuations in the two light curves, our analysis is insensitive to this overall additive constant.
% only  to make both curves appear nicely in one plot

%Without loss of information, an arbitrary offset can be applied to the magnitude scale, and we have done so here. 

For a  doubly-lensed quasar, there are four variables recorded on an irregularly spaced sequence of observation times $\boldsymbol{t}=(t_1, t_2, \ldots, t_n)^\top$; the observed magnitudes $\boldsymbol{x}=(x_{1}, x_{2}, \ldots, x_{n})^\top$ for light curve $A$ and $\boldsymbol{y}=(y_{1}, y_{2}, \ldots, y_{n})^\top$ for light curve $B$ as well as standard deviations, $\boldsymbol{\delta}=(\delta_{1}, \delta_{2}, \ldots, \delta_{n})^\top$ and $\boldsymbol{\eta}=(\eta_{1}, \eta_{2}, \ldots, \eta_{n})^\top$, representing their uncertainties due to heteroskedastic measurement error. In Figure~\ref{fig2}, $\boldsymbol{x}$ and $\boldsymbol{y}$ are represented by red squares and blue circles, and their standard deviations by the half lengths of vertical lines around the symbols. Similarly, for a quadruply-lensed quasar, there are four light curves, each with their own measurement errors.

%\footnote{\textcolor{red}{Each standard error is the square root of the inverse of the observed information at the corresponding magnitude derived from  the likelihood function of the magnitude.}}

Since a quasar exhibits fluctuations in its brightness, it is possible to estimate time delays between different views of those fluctuations. In Figure~\ref{fig2}, for example, the bottom of the V-shaped valley of  light curve $A$ at around 900 days  precedes that of   light curve  $B$  by around 50 days. Other features in the light curves exhibit a similar time shift of about 50 days.

However, a number of aspects of the light curves in Figure~\ref{fig2} make  accurate time delay estimation statistically challenging. First, irregular observation times are inevitable because observations may be prevented in poor weather or during the day. Second, the motion of the Earth around the Sun causes seasonal gaps because the part of the sky containing the quasar is not visible at night from the location of a particular telescope during certain months. Third, since the light of each gravitationally lensed image traverses different paths through the gravitational potential, they are subject to differing degrees of lensing  magnification. Thus, the light curves often exhibit different average magnitudes. Finally,  observed magnitudes are measured with error, leading to relatively larger measurement errors for  fainter images.
%Second, the motion of the Earth around the Sun causes seasonal gaps because part of the sky is not visible at night during certain months. 

\begin{figure}[t!]
\includegraphics[scale = 0.3]{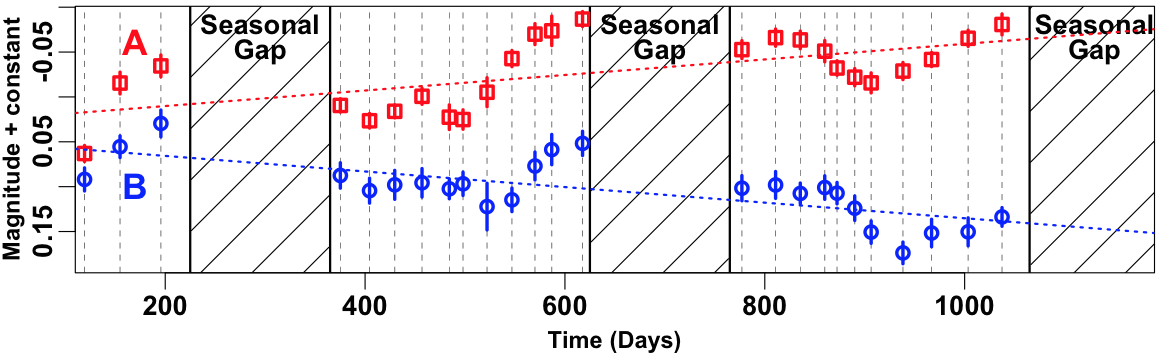}
\caption[]{The light curves of two lensed images can have different long-term trends caused by microlensing due to stars moving within the lensing galaxy.  This effect independently introduces a long-term magnification trend in each image. Here, we simulate the effect of two different long-term linear microlensing trends on the light curves in Figure \ref{fig2}. The dotted lines depict the linear microlensing trend for each image.}
\label{microlensing_intro}
\end{figure}

%
Moreover, some quasar images exhibit additional independent extrinsic variability, an effect called \emph{microlensing}\footnote{Microlensing is conceptually  similar to strong lensing except that the lens is a star moving within the intervening galaxy.  However, the lensed images produced by microlensing cannot be separately seen because their angular separation is too small for us to resolve with a telescope.  Instead,  astronomers observe only the combined magnification of both images, which changes with time due to the relative motions of the source and lensing star.}.  Significant microlensing occurs when a path of light passes unusually close to a star that is moving within the lensing galaxy.  Lensing by this star introduces independent brightness magnification variations into the corresponding image in addition to the overall magnifications caused by strong lensing of the galaxy \citep{chang1979flux, tewes2013a}. The timescale of the microlensing variability is typically much larger than that of the intrinsic quasar variability if the lens is on a galaxy scale \citep{kai2014}. Thus the individual light curves may exhibit different  long-term trends that are not related to the intrinsic variability of the source\footnote{\cite{macleod2010modeling} who analyzed about 9,000 quasars obtained from the Sloan Digital Sky Survey \citep{berk2004ensemble} show that the timescale of quasar intrinsic variability varies from days to years, and \cite{mosquera2011microlensing} indicate that the five shortest timescales of microlensing among 87 lensed quasars are between 8 and 12 years with respect to Einstein crossing timescales and  are between  1 and 8 weeks with respect to source crossing timescale. Since the microlensing timescale is not always longer than the quasar intrinsic variability timescale, it is not always the case that we see the extrinsic long-term trends in the presence of microlensing.}. In Figure~\ref{microlensing_intro}, as an illustration, we plot the same simulated light curves $A$ and $B$  plotted in Figure~\ref{fig2} but with different added linear trends to simulate the effect of microlensing.
%A component of the signal other than the intrinsic variability can be modeled via microlensing.  
%This occurs when gravitational lensing by stars moving inside the lensing galaxy independently introduces brightness magnification variations  into each path of of light,
%
%The fitted linear regression lines are denoted by dashed lines.
%that have different  long-term linear trends with almost identical short-term variabilities. 
% (see Figure~\ref{commingling_display}). 

%Grid-based searches for time delay estimates  are classic in this field. 
\subsection{Other time delay estimation methods}\label{previous_methods}   Conventional methods for time delay estimation have involved grid-based searches. One-dimensional grid methods estimate the time delay, $\Delta_{\textrm{AB}}$,\footnote{A positive value of $\Delta_{\textrm{AB}}$ indicates that features in light curve $A$ appear before they appear in light curve $B$.}   by minimizing the  $\chi^2$ distance  or by maximizing the cross-correlation between two light curves, $\boldsymbol{x}$ and $\boldsymbol{y}_{\Delta_{\textrm{AB}}}$,  on a grid of values of $\Delta_{\textrm{AB}}$  \citep{fassnacht1999}, where $\boldsymbol{y}_{\Delta_{\textrm{AB}}}$ denotes $\boldsymbol{y}$ shifted by $\Delta_{\textrm{AB}}$ days to the right. Both techniques require an interpolation scheme. The dispersion method \citep{pelt1994} combines two light curves by shifting one of them in  time and magnitude  by $\Delta_{\textrm{AB}}$ and $\beta_0$, respectively. This is called the \emph{curve-shifting} assumption. The method estimates $\Delta_{\textrm{AB}}$ and $\beta_0$ on a two dimensional grid by minimizing the sum of squared differences between consecutive pairs of magnitudes on the combined curve.  A bootstrapping method is used to produce standard errors of the time delay estimates. These methods account only for the intrinsic variability of a quasar. (When it is clear from the context, we suppress the subscript on $\Delta_{\textrm{AB}}$ and simply use $\Delta$.) %without considering the microlensing.
%the weighted sum of squared differences between consecutive pairs of magnitudes on the combined curve.
%The sampling distribution of $\Delta$ via t
%avoids interpolation by 

Model-based methods have also been proposed in past to avoid the computational burden of evaluating the fit on a fine grid. For example,  \cite{tewes2013a} model the  intrinsic and extrinsic variabilities of light curves using  high-order and low-order splines, respectively. They obtain the least square estimate of $\Delta$ by  iterating a two-step fitting routine in which splines are first fit given $\Delta$ and then $\Delta$ is optimized given the model fit. They also use parametric bootstrapping for the standard error of the time delay estimate. 
% (Here and elsewhere it is clear from context we suppress the subscript on $\Delta$.)
%However, a limitation of this model is that the number of parameters increases as the data size increases.
%\cite{hojjati2014next}  use the Gaussian process likelihood to model the light curves given $\Delta$. They obtain a maximum likelihood estimate of $\Delta$ by plugging in the best-fit estimates of the other model parameters. The inverse of the negative Hessian at the mode is used to estimate the standard error of the estimate. 
%; a fine grid may be a good approximation, but it increases the computational burden of the re-sampling procedure

\citet[hereafter H\&R]{harva2006} introduced the first fully Bayesian approach, though they do not account for  microlensing. They assume each observed light curve is generated by an unobserved underlying process. One of the latent processes is assumed to be a shifted and scaled version of the other, with the time and magnitude shifts and the magnitude scale  treated as unknown parameters. They use a collapsed Gibbs-type sampler for model fitting, with the latent process integrated out of the target posterior distribution. Unlike other existing methods this approach unifies parameter estimation and uncertainty quantification into a single coherent analysis based on the posterior distribution of $\Delta$.
%%a specific data generation process with a state-space model; 
% conditionally following a Gaussian distribution. 

%  Bayesian approach has several advantages. It  deepens scientific insights by modeling the data generation process with a few parameters.  Also, 
%because it treats the unknown time delay $\Delta$ as a random variable, it turns the separate optimization and re-sampling-optimization procedures for point and uncertainty estimates into one posterior sampling scheme for both estimates simultaneously from its posterior distribution.
% and explains how these parameters affect the time delay estimation

\subsection{Our Bayesian and profile likelihood approaches} 

The TDC motivated us to improve on H\&R's fully Bayesian model by taking advantage of  modeling and computational advances made since H\&R's 2006 proposal.  Specifically, we adopt an Ornstein-Uhlenbeck (O-U) process \citep{uhlenbeck1930theory} to model the latent light curve. The O-U process has been  empirically shown to describe the stochastic variability of quasar data well \citep{kelly2009variations, kozlowski2010quantifying, macleod2010modeling, zu2013quasar}. We address the effect of microlensing  by incorporating a polynomial regression on time into the model. We  specify scientifically motivated prior distributions and conduct a set of systematic sensitivity analyses; see Appendix~\ref{sec_sensitivity} for details of the sensitivity analyses. In contrast to H\&R's strategy of sampling from a marginal distribution with the latent process integrated out, we use a Metropolis-Hastings (M-H) within Gibbs sampler \citep{tierney1994markov} to sample the posterior in the full parameter space. We improve the convergence rate of our MCMC (Markov chain Monte Carlo) sampler by using an ancilarity-sufficiency interweaving strategy \citep{yu2011} and  adaptive MCMC  \citep{brooks2011handbook}.
%A Metropolis-Hastings (M-H) within Gibbs sampler \citep{tierney1994markov} is used to account for uncertainty in the latent process via posterior sampling in the full parameter space, rather than sampling from a marginal distribution with the latent process integrated out}.

To complement the Bayesian method, we introduce a simple profile likelihood approach that allows us to remove nuisance parameters and focus on $\Delta$ \citep[e.g., ][]{davison2003statistical}. We show that the profile likelihood function of $\Delta$ is  approximately proportional to the marginal posterior distribution of $\Delta$ when a Jeffreys' prior is used for the nuisance parameters \citep{berger1999integrated}, see Appendix \ref{proof_prof_likelihood}.  For the problems we investigate the profile likelihood is nearly identical to the marginal posterior distribution in most cases, validating the approximation. 
%approach produces nearly identical results as our Bayesian approach, providing nice confirmation to each other.
%, the shape of profile likelihood curve is almost identical to the marginal posterior distribution of $\Delta$, producing nearly the same estimates and uncertainties.

% without increasing the number of parameters according to the size of the data.
%, and optionally a tempered transition \citep{neal1996sampling} to handle  multimodality.
% available from an online supplement

Our time delay estimation strategy combines these two complementary approaches. We first obtain the  profile likelihood  of $\Delta$, which is simple to compute.  A more principled fully Bayesian analysis focuses on the dominant mode identified by the profile likelihood and provides joint inference for the time delay and other model parameters via the joint posterior distribution.
\\

%In the TDC, we removed the long-term trends from light curves via a regression, applying the Bayesian method to the residuals \citep{kai2014}.  This worked well because the intrinsic variability of quasar data remains even after the long-term polynomial trend is removed \citep{courbin2010}. However, this way does not take into account the additional uncertainties  involved in estimating the regression coefficients. Thus, we incorporate  the regression into the model.

The rest of this paper is organized as follows. We describe our Bayesian model in Section~\ref{sec4} and  the MCMC sampler that we use to fit it in Section~\ref{sec5}. In Section~\ref{sec_profile}, we introduce the profile likelihood approach. We then specify our estimation strategy and illustrate it via a set of numerical examples in Section~\ref{numerical}. An R package, \texttt{timedelay}, that implements the Bayesian and profile likelihood methods is publicly available at CRAN\footnote{https://cran.r-project.org/package=timedelay}.
%In Appendices, we specify conditional posterior distributions needed for the Gibbs sampler, the proof for the profile likelihood approximation to the marginal posterior distribution, and sensitivity analyses}. 
%, using simulated doubly- and quadruply-lensed  data sets and a observed data set of quasar \emph{Q0957+561}

\newpage
\section{A fully Bayesian model for time delay estimation}\label{sec4}

\subsection{Latent time series}\label{assumption} 

We assume that each time-delayed light curve is generated from a latent  curve  representing the true source magnitude in continuous time. We denote these latent curves by $\boldsymbol{X}=\{X(t),~ t\in \mathbf{R}\}$ and  $\boldsymbol{Y}=\{Y(t),~ t\in \mathbf{R}\}$, respectively, where $X(t)$ and $Y(t)$ are unobserved true magnitudes at time $t$.  We use the vector notation $\boldsymbol{X}(\boldsymbol{t})=(X(t_1), X(t_2), \ldots, X(t_n))^\top$ and $\boldsymbol{Y}(\boldsymbol{t})=(Y(t_1), Y(t_2), \ldots, Y(t_n))^\top$ to denote the $n$  magnitudes  of each latent light curve at the irregularly-spaced observation times $\boldsymbol{t}$. 
%For example, the solid red and dashed blue curves in Figure  \ref{fig3} connect the latent magnitudes at different times. 
% the latent light curves and are  denoted %We consider them as missing data that we would have observed without measurement errors. It is also convenient to distinguish them from model parameters.

%while $\boldsymbol{X}(\boldsymbol{t})$ and $\boldsymbol{Y}(\boldsymbol{t})$ are the latent magnitudes at observation times $\boldsymbol{t}$ that have generated the observed magnitude $\boldsymbol{x_t}$ and $\boldsymbol{y}$, respectively. 
\begin{figure}[b!]
\includegraphics[scale = 0.3]{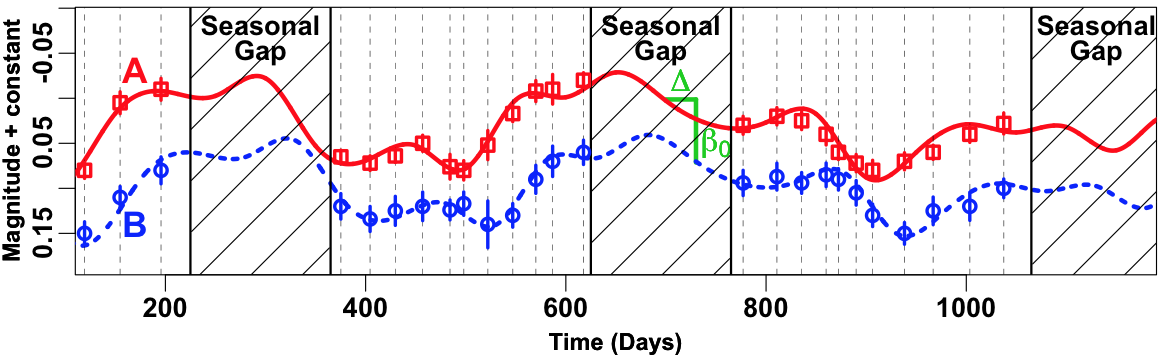}
\caption[]{The solid red  and dashed blue latent curves of images \emph{A} and \emph{B}, respectively, are generated under the model in \eqref{curve_shifted}.  These two curves are superimposed on Figure~\ref{fig2}. The curve-shifted model in  \eqref{curve_shifted}  specifies that the dashed blue  curve is a shifted version of the solid red curve by $\Delta$ (=70) days in time  and by $\beta_0$ (=0.07) in magnitude. For illustration purposes, $\boldsymbol{X}$ is depicted as a solid red smooth curve; a more realistic model is described in Section~\ref{priorsection1}.}
\label{fig3}
\end{figure}

A curve-shifted model \citep{pelt1994, kochanek2006time} assumes that one of the latent  light curves is a shifted version of the other, that is
\begin{equation}\label{curve_shifted}
Y(t) = X(t -\Delta) + \beta_0,
\end{equation}
where $\Delta$ is a shift in time and $\beta_0$ is a magnitude offset. For example, in Figure~\ref{fig3}, we  displayed  the solid red  and dashed blue latent curves of images \emph{A} and \emph{B}, respectively, generated under  the model in \eqref{curve_shifted}. Thus the two curves  exactly overlap if the solid red curve is shifted by $\Delta$ days and by $\beta_0$ magnitude units. (For illustration purposes, $\boldsymbol{X}$ is depicted as a solid red smooth curve; a more realistic model is described in Section~\ref{priorsection1}.) The key advantage of this model is that a single latent light curve, here $\boldsymbol{X}$, is sufficient to represent the true magnitude time series of the two (or more) lensed images.  This model is a special case of H\&R's scaled curve-shifted model, $Y(t) = s X(t - \Delta) + \beta_0$, where $s$ is a magnitude scale change, mentioned at the end of Section~\ref{previous_methods}. Setting  $s=1$ is  reasonable because  gravitational lensing only deflects the source light and magnifies it, i.e., multiplies the source flux. Because magnitude is on the $\log_{10}$ scale of  source flux, we expect an additive offset, i.e., $\beta_0$, rather than a scale change. The curve-shifted model captures the essential physical effects of strong gravitational lensing (at least in the absence of microlensing), and thus is an appropriate model for estimating the time delay.

%Since the curve-shifted model  reflects gravitational lensing well, it is appropriate for estimating $\Delta$, at least in the absence of microlensing.
%Adding a constant to either light curve does not affect the inference for $\Delta$ because $\beta_0$ accounts for such a magnitude offset. 

Microlensing causes additional long-term extrinsic variability unrelated to the intrinsic quasar variability driving the dynamics of $\boldsymbol{X}$. Thus, the curve-shifted model is not appropriate in the presence of microlensing. To account for microlensing, we assume that one of the latent  light curves is a time-shifted version of the other, but with an additional polynomial regression of order $m$ on $t-\Delta$, that is
\begin{equation}\label{fundamental}
Y(t) = X(t -\Delta) + \boldsymbol{w}_m^\top(t-\Delta) \boldsymbol{\beta},
\end{equation}
where $\boldsymbol{w}_m(t-\Delta)\equiv(1, t-\Delta, (t-\Delta)^2, \ldots, (t-\Delta)^m)^\top$ is a covariate vector of length $m+1$, and $\boldsymbol{\beta}\equiv\left(\beta_0, \beta_1, \beta_2, \ldots, \beta_m\right)^\top$ is a vector of regression coefficients\footnote{An orthonormal basis is more compatible with an independent prior on the regression coefficients and thus may be preferred if a higher degree polynomial regression is used.}. The polynomial regression term in \eqref{fundamental} accounts for the difference in the microlensing trends of the two light curves, i.e., the difference between the long-term trends of $Y(t)$ and $X(t -\Delta)$. The microlensing model in \eqref{fundamental} reduces to a curve-shifted model in \eqref{curve_shifted} if $\beta_1=\beta_2=\cdots=\beta_m=0$.   
% to reduce the correlation between regression coefficients
% when the timescale of microlensing is much larger than that of intrinsic quasar variability
%Thus,  accounts for polynomial long-term trends caused by microlensing, an additional variability that the curve-shifting model does not take into account. accounts for extrinsic variability  polynomial regression is to account for polynomial long-term trends, extrinsic variability additional to cu caused by microlensing that is  variability model is appropriate because micolensing causes long-term trends  model 

The best choice for the order of the polynomial regression  depends on the extent of microlensing, and this varies from quasar to quasar. We set $m=3$ as a default  because the third order polynomial regression has been successfully applied to model lensed quasars \citep{kochanek2006time, courbin2011, morgan2012further}.  If we find evidence via the profile likelihood  that a third order polynomial regression is not sufficient to reduce the effect of microlensing (see Section~\ref{case1} for details), we can impose a reasonable upper bound of $m$ by running preliminary regression on the observed light curves, and comparing the fits.

%\footnote{Magnitude is $ -2.5\log_{10}(\textrm{flux})$ up to an arbitrary offset.}
% in Figure~\ref{fig3}, i.e., $c$ in  \eqref{fundamental}. Thus, 

%However, if each light curve suffers from microlensing with a different long-term trend, the curve-shifted model in \eqref{fundamental} does not hold  because one latent curve is no longer a shifted version of the other. Considering that the microlensing effect is known to cause a linear or quadratic long-term trend, we incorporate a cubic regression into the model in \eqref{fundamental} as follows.
%\begin{equation}\label{fundamental2}
%Y(t) = X(t -\Delta) + \boldsymbol{t}(\Delta)^\top \boldsymbol{c},
%\end{equation}

%We treat both shifts as unknown parameters. . 

%idea shifting one of unobserved underlying light curves in time and magnitude axes, additionally considering a scale change $s$, \emph{i.e.}  $\boldsymbol{Y(t)} = s\cdot\boldsymbol{X(t - }\Delta\boldsymbol{)} + c$. 

\subsection{Distribution of the observed data}
Observing the gravitationally-lensed images with a telescope, an astronomer measures the magnitude in each image, $x_{j}$ and $y_{j}$, and reports standard deviations, $\delta_{j}$ and $\eta_{j}$, representing the uncertainties of the magnitudes due to measurement errors\footnote{The magnitude estimate and standard deviation typically summarize a Gaussian approximation to the likelihood of the latent magnitude for the flux data of an image.  The standard deviation does not necessarily represent a standard error of a repeated sampling measurement error distribution.} at time $t_j$, $j=1, 2, \ldots, n$.  We assume that these measurements have independent Gaussian errors  centered at the latent magnitudes $X(t_j)$ and $Y(t_j)$, i.e., 
\begin{align}
x_{j}\mid X(t_j) &\stackrel{\textrm{indep.}}{\sim} \textrm{N}[X(t_j),~  \delta^2_{j}],\label{lik1}\\
y_{j}\mid Y(t_j ) &\stackrel{\textrm{indep.}}{\sim} \textrm{N}[Y(t_j),~ \eta^2_{j}],\label{lik2}
\end{align}
where $\textrm{N}[M, V]$ is a Gaussian distribution with mean $M$ and variance $V$, and $\boldsymbol{x}$ and $\boldsymbol{y}$ are independent given their true magnitudes. Using the model in  \eqref{fundamental}, we can express  \eqref{lik2} as
\begin{equation}\label{lik3}
y_{j}\mid X(t_j - \Delta), \Delta, \boldsymbol{\beta} \stackrel{\textrm{indep.}}{\sim} \textrm{N}[X(t_j - \Delta) + \boldsymbol{w}_m^\top(t_j-\Delta)\boldsymbol{\beta},~ \eta^2_{j}].
\end{equation}

Given $\Delta$, we define $\boldsymbol{t}^\Delta=(t^\Delta_1, t^\Delta_2, \ldots, t^\Delta_{2n})^\top$ as  the sorted vector of  $2n$ times among the $n$ observation times, $\boldsymbol{t}$, and the $n$ time-delay-shifted observation times, $\boldsymbol{t}-\Delta$.  Also, $\boldsymbol{X}(\boldsymbol{t}^\Delta)=(X(t^\Delta_1), X(t^\Delta_2), \ldots, X(t^\Delta_{2n}))^\top$ is the vector of  $2n$ latent magnitudes at the times in $\boldsymbol{t}^\Delta$. The joint  density function  of the observed data given $\boldsymbol{X}(\boldsymbol{t}^\Delta)$, $\Delta$, and $\boldsymbol{\beta}$ is 
\begin{equation}\label{lik}
p(\boldsymbol{x}, \boldsymbol{y}\mid \boldsymbol{X}(\boldsymbol{t}^\Delta), \Delta, \boldsymbol{\beta})=\prod_{j=1}^{n} p\!\left(x_{j}\mid X(t_j)\right) \times p\!\left(y_{j}\mid X(t_j -\Delta), \Delta, \boldsymbol{\beta}\right)\!,
\end{equation}
where the two distributions in the product are given in  \eqref{lik1} and \eqref{lik3}.

%e.g., the solid red curve in Figure~\ref{fig3},}
\subsection{Prior distribution of the latent magnitudes}\label{priorsection1}
We assume  the latent continuous-time light curve,  $\boldsymbol{X}$,  is a realization of an O-U process \citep{uhlenbeck1930theory} as proposed in \cite{kelly2009variations}. The  stochastic differential equation, 
\begin{equation}\label{stochastic}
d X(t) =  - \frac{1}{\tau}\big(X(t) - \mu\big)d t +\sigma d B(t),
\end{equation}
defines the O-U process, where $\mu$ and $\sigma$ are on the magnitude scale and govern the overall mean and short-term variability of the underlying process, $\tau$ is a timescale (in days) for the process to revert to the long-term mean $\mu$,   $\{B(t),~t\ge0\}$ is a standard Brownian motion, and $d B(t)$ is an interval of the Brownian motion, whose distribution is Gaussian with mean zero and variance $dt$. We  denote the three O-U parameters by $\boldsymbol{\theta}=(\mu, \sigma^2, \tau)^\top$.

%100 MACHO (massive compact halo object)
%:
\cite{kelly2009variations} empirically demonstrated that the power spectrum of the O-U process is consistent with the mean power spectrum of 55 well-sampled quasar light curves  at a specific frequency range with timescales shorter than $\tau$. They also investigated  the associations between model parameters and the physical properties of quasars. For example, $\tau$  has a positive correlation with  black hole mass, which is consistent with previous astrophysical studies.  \cite{kozlowski2010quantifying} and \cite{macleod2010modeling} were concerned about a possible selection bias in  the sample of quasars used in \cite{kelly2009variations} and thus they analyzed thousands of light curves. \cite{kozlowski2010quantifying} found  further support for the O-U process in their analyses of about 2,700 quasars obtained from the Optical Gravitational Lensing Experiment \citep[OGLE,][]{kozlowski2009discovery}. They showed that the distribution of the goodness of fit statistic obtained by fitting the O-U process to their light curves was consistent with the expected distribution of the statistic under the assumption that the light curve variation was stochastic. \cite{macleod2010modeling} further verified  the argument about the correlations between model parameters and physical properties in \cite{kelly2009variations} by analyzing about 9,000 quasars obtained from the Sloan Digital Sky Survey \citep{berk2004ensemble}.   \cite{zu2013quasar} also supported the O-U process by comparing it to the Gaussian process with three different covariance functions in fitting about 200 OGLE light curves. Their numerical results based on the $F$-test and Bayesian information criterion supported the O-U process.  These studies popularized the O-U process among astrophysicists to the extent that the TDC   simulated its quasar light curves under an O-U process\footnote{The TDC organizers generated  500 10-year-long light curves by using the O-U process ($\mu=0$, $\log(\tau)\in[1.5, 3.0]$, and $\log(\sigma)\in[-1.1, -0.3]$), and  re-used these to make about 5,000 doubly- or quadruply-lensed light curves with different starting points, different seasonal gaps, etc. Microlensing is  simulated via a catalog convergence of \cite{oguri2010gravitationally}, shear, and surface density. The measurement errors were heteroskedastic Gaussian. The organizers intentionally contaminated the data to make the time delay estimation difficult; the reported standard deviations may be under-estimated, measurement errors may be correlated due to time-dependent calibration error, and magnitudes may be temporarily offset due to time-dependent systematic effects in the telescope optics.} \citep{dobler2013}.  The earlier approach of H\&R (2006) preceded these more recent advances in astrophysical and statistical modeling of quasars.
%H\&R  could not take advantage of these advances in astrophysical modeling because their work preceded the relevant astrophysical probes.
%Modeling the variability They also looked into the associations between model parameters and physical properties of quasars; $\tau$ ($\sigma$) had positive (negative) correlations with luminosity and mass of the black hole. 

%at irregularly-spaced observation times
The solution of the stochastic differential equation in  \eqref{stochastic} provides the prior distribution for the time-sorted latent magnitudes $\boldsymbol{X}(\boldsymbol{t}^\Delta)$  via its Markovian property. Specifically,
\begin{align} 
\begin{aligned}
X(t_1^\Delta)\mid \Delta, \boldsymbol{\theta} &~\sim~ \textrm{N}\!\left[\mu,~ \frac{\tau\sigma^2}{2}\right]\!, ~\textrm{and for}~ j=2, 3, \ldots, 2n,\\
~~X(t_j^\Delta)\mid X(t_{j-1}^\Delta), \Delta, \boldsymbol{\theta} &~\sim~ \textrm{N}\!\left[\mu + a_j\big(X(t_{j-1}^\Delta) - \mu\big) ,~\frac{\tau \sigma^2}{2}(1-a^2_j)\right]\!,
\end{aligned}\label{conditionalOU1}
\end{align}
where $a_j\equiv \exp(-(t_j^\Delta - t_{j-1}^\Delta)/\tau)$ is a shrinkage factor that depends on the observational  cadence and $\tau$. If two adjacent latent  magnitudes are close in time, i.e., $t_j^\Delta-t_{j-1}^\Delta$ is small, $a_j$ is close to unity and under this prior $X(t_j^\Delta)$  borrows more information or shrinks  more towards the previous latent magnitude, $X(t_{j-1}^\Delta)$, and exhibits less uncertainty. On the other hand, if neighboring latent magnitudes are distant in time, e.g., due to a seasonal gap, $a_j$ is close to zero, and under this prior $X(t_j^\Delta)$  borrows little information from the distant value $X(t_{j-1}^\Delta)$ and instead  approaches the overall mean $\mu$ with more uncertainty. This is known as \emph{the mean reversion} property of the O-U process.

The joint prior density function of the $2n$ latent  magnitudes  is
\begin{equation}\label{jointmissing}
p(\boldsymbol{X}(\boldsymbol{t}^\Delta)\mid \Delta, \boldsymbol{\theta}) = p(X(t^\Delta_1)\mid \Delta, \boldsymbol{\theta})\times\prod_{j=2}^{2n} p(X(t^\Delta_j)\mid X(t^\Delta_{j-1}), \Delta, \boldsymbol{\theta}),
\end{equation}
where the distributions on the right-hand side are given in \eqref{conditionalOU1}.
%They modeled the dependence between latent true magnitudes as 
%\begin{equation}
%X(t_j^\Delta)\mid X(t_{j-1}^\Delta), \Delta, w \sim \textrm{N}\big[X(t_{j-1}^\Delta) ,~(t_j^\Delta-t_{j-1}^\Delta)^2 e^w\big],
%\end{equation}
%with a Gaussian prior distribution on $w$, but their model did not account for the mean-reversion property of quasar light curves.
% without (astro) physical interpretation on $w$.

%\cite{harva2006} modeled the dependence between the neighboring missing data differently as $X(t_j^\Delta)\mid X(t_{j-1}^\Delta), \Delta, w \sim \textrm{N}\big[X(t_{j-1}^\Delta) ,~(t_j^\Delta-t_{j-1}^\Delta)^2 e^w\big]$ with a Normal hyper-prior distribution on $w$. Although they modeled the time dependence making its variance proportional to the squared time difference, they did not justify why it was a good way to model quasar data.  Instead, we adopted the O-U process known to describe the stochastic variability of quasars well \citep{kelly2009variations} to model the dependence in a more principled way as described in Equation (\ref{conditionalOU1}).

%Note that this underlying model reduces to the autoregressive model of order 1, called AR(1), when the data are regularly-observed time series ($t_j^\Delta-t_{j-1}^\Delta=1$, for all $j$). Hence, this model is also called continuous autoregressive model of order 1, CAR(1).

\subsection{Prior distributions for the time delay and the magnitude offset}\label{priorsection2} 
We adopt  independent proper prior distributions for $\Delta$ and $\boldsymbol{\beta}$,
\begin{equation}\label{priordeltac}
p(\Delta, \boldsymbol{\beta})= p(\Delta)p(\boldsymbol{\beta})\propto I_{\{u_1\le\Delta\le u_2\}} \times \textrm{N}_{m+1}(\boldsymbol{\beta}\mid \boldsymbol{0}, 10^5 \times I_{m+1}),
\end{equation}
where $I_{\{D\}}$ is the indicator function of $D$, $\textrm{N}_{m+1}(\boldsymbol{\beta}\mid \boldsymbol{0}, 10^5 \times I_{m+1})$ is an $m+1$ dimensional Gaussian density evaluated  at $\boldsymbol{\beta}$ whose mean is $\boldsymbol{0}$, a vector of zeros with length $m+1$, and variance-covariance matrix is $10^5 \times I_{m+1}$, with an $m+1$ dimensional  identity matrix $I_{m+1}$.  We put a diffuse Gaussian prior on $\boldsymbol{\beta}$ to minimize impact on the posterior inference and to ensure posterior propriety.

The range of the uniform prior distribution on $\Delta$, $[u_1, u_2]$, reflects the range of interest. One choice is the \emph{entire feasible range} (or feasible range) of $\Delta$, $[t_1-t_n, t_n-t_1]$;  only values of $\Delta$ in this range can  correspond to adjusted light curves that overlap by at least one data point.  (H\&R uses a diffuse Gaussian prior distribution on $\Delta$ that is defined even outside this range.)

In some cases, information about the likely range of $\Delta$ is available from  previous analyses or possibly from astrophysical probes. For example, we can find the likely range of $\Delta$ using a physical model for the mass and gravitational potential of the lens,  as well as the redshifts (an astronomical measure of distance) and relative spatial locations of a quasar and lens.

In reality, the time delay and lensing magnification may be correlated a priori. We assume a priori independence, however, because it is difficult to construct an informative joint prior distribution without more information about the lens system, i.e., image positions, distances, and a lens model.
%they are independent a priori 
%we cannot figure out their relationship unless we know the mass distribution of the lens and geometrical image separation. Thus, 

% measured the Einstein radius, i.e., the angular separation of the two images. we were not given any information about the lens model, nor the geometrical image separation, so there is not much you can say a priori about the correlation between magnification and time delay
%, then you could calculate the expected range of magnifications and time delays and their relationship.  to let the data speak more. No evidence about the strong correlation between posterior samples of $\Delta$ and $\boldsymbol{\beta}$ has been found throughout our numerical illustrations, e.g., see the first row of Figure~\ref{fig_corr}. 

\subsection{Prior distributions for the parameters in the O-U process}\label{hyper_prior_description}
Considering both scientific knowledge and the dynamics of the O-U process, we put a uniform  distribution on the $\textrm{O-U}$ mean $\mu$, an independent inverse-Gamma  (IG) distribution, IG(1, $b_{\sigma}$), on its short-term variance $\sigma^2$, and an independent IG(1, $b_{\tau}$) distribution on its timescale $\tau$, i.e., %considering both scientific knowledge and the dynamic of the O-U process. 
\begin{align}
\begin{aligned}
p(\mu, \sigma^2, \tau)=p(\mu)p(\sigma^2)p(\tau)&~\propto~\frac{\exp(-b_{\sigma}/\sigma^2)}{(\sigma^{2})^{2}} \times \frac{\exp(-b_{\tau}/\tau)}{\tau^2}\\
& ~~~~~\times I_{\{-30\le\mu\le30\}}\times I_{\{\sigma^2>0\}}\times I_{\{\tau>0\}}.
\end{aligned}\label{hyp}
\end{align} 
The units of $b_\sigma$ are magnitude squared per day, hereafter mag$^2/$day, and the  scale parameter of the IG distribution on $\tau$ is fixed at one day, i.e., $b_{\tau}=1$ day.
%(1~\textrm{day}})

Here the uniform distribution on $\mu$ encompasses a magnitude range from that of the Sun (magnitude $=-26.74$) to that of the faintest object visible with the Hubble Space Telescope (magnitude $=30$).  The IG distributions on  $\tau$  and $\sigma^2$  set  soft lower bounds\footnote{Because the density function of IG($a, b$)  decreases exponentially from its mode, $b/(a + 1)$, toward zero and geometrically decreases with a power of $a + 1$ towards infinity, it is relatively unlikely for the random variable to take on values much smaller than its mode.} to focus on practical solutions in which $\Delta$ can be constrained. For example, in the limits when $\tau$ is much less than the observation cadence or when $\sigma^2$ is much smaller than the measurement variance divided by the cadence, the discrete observations of the continuous latent light curve  appear as  serially uncorrelated white noise sequence. In these limiting cases it is impossible to estimate $\Delta$ by matching serially correlated fluctuation patterns. The soft lower bounds for $\tau$ and $\sigma^2$ discount these limiting cases, and allow us to focus on the relevant parameter space in which we  expect time delay estimation to be feasible. 
%The soft lower bounds for $\tau$ and $\sigma^2$  prevent these limiting cases, enabling estimation of $\Delta$.

%The relationship between the IG and scaled inverse-$\chi^2$ distributions allows us to interpret the shape parameter of the IG as half the number of directly observed pseudo realizations of the O-U process that would carry equivalent information as the prior distribution. (See, e.g., \cite{bernardo1994bayesian} or \cite{gelman2013bayesian} for a discussion of the pseudo observation interpretation of prior distributions.) Thus,
%this corresponds to two pseudo observations and can be interpreted as an indication that the prior distribution is relatively weak. 

We set  the shape parameter of the IG prior distribution on $\tau$  to unity and the scale parameter $b_\tau$ to one day to obtain a weakly informative prior. The resulting soft lower bound on $\tau$ is 0.5 day and is smaller than all  of the estimates of $\tau$ in \cite{macleod2010modeling}, who analyzed 9,275 quasars.

For the IG prior distribution of $\sigma^2$, we set the shape parameter  to unity and the scale parameter  to  (Mean measurement standard deviation)$^2$ / (Median cadence), i.e.,
\begin{equation}\label{scale_sigma}
b_\sigma=\frac{[\{\sum_{j=1}^n \delta_{j} + \sum_{j=1}^n \eta_{j}\}/2n]^2 }{\textrm{Median}(t_2-t_1, t_3-t_2, \ldots, t_n-t_{n-1})}.
\end{equation}
This scale parameter  enables us to search for solutions for which we can  constrain $\Delta$ by avoiding the above limiting case. Another viable choice for the scale parameter is $b_\sigma=2\times 10^{-7}$ because all  estimates of $\sigma^2$ in \cite{macleod2010modeling} are larger than this value.  Sensitivity analyses for the choice of prior distributions of $\tau$  and $\sigma^2$ appear in Appendix \ref{sec_sensitivity}.

\section{Metropolis-Hastings within Gibbs sampler}\label{sec5} Our overall hierarchical model is specified via the observation model in \eqref{lik1} and \eqref{lik3}, the O-U process for the latent light curve  in \eqref{conditionalOU1}, and the prior distributions given in \eqref{priordeltac} and \eqref{hyp}. Our first approach to model fitting uses a Gibbs-type sampler to explore the resulting full posterior distribution. It is possible to integrate out the latent magnitudes analytically and use a collapsed sampler based on the marginalized joint posterior distribution specified in Appendix~\ref{appcond} as H\&R did. However,  we treat $\boldsymbol{X}(\boldsymbol{t}^\Delta)$ as latent variables, alternatively updating $\boldsymbol{X}(\boldsymbol{t}^\Delta)$ and the other model parameters. (We could formulate our approach as data augmentation with $\boldsymbol{X}(\boldsymbol{t}^\Delta)$ as the missing data,  see \citet{van2001art}.)

%&~~~~~~~~~~~~~=p(\boldsymbol{X}(\boldsymbol{t}^\Delta)\mid \Delta, c, \mu, \sigma^2, \tau,  D_{\textrm{obs}})\cdot p(\Delta \mid c, \mu, \sigma^2, \tau, D_{\textrm{obs}})\nonumber\\
Specifically, we use a Metropolis-Hastings within Gibbs (MHwG)  sampler \citep{tierney1994markov} that iteratively samples five complete conditional distributions of the full joint posterior density, $p(\boldsymbol{X}(\boldsymbol{t}^\Delta), \Delta, \boldsymbol{\beta}, \boldsymbol{\theta} \mid\boldsymbol{x}, \boldsymbol{y})$, proportional to the product of densities of observed data in \eqref{lik}  and prior densities in \eqref{jointmissing}, \eqref{priordeltac} and \eqref{hyp}. Iteration $l$ of our sampler is composed of five steps. 
\begin{align}
&\textrm{Step 1: Sample } (\boldsymbol{X}^{(l)}(\boldsymbol{t}^{\Delta^{(l)}}), \Delta^{(l)})\sim  p(\boldsymbol{X}(\boldsymbol{t}^\Delta), \Delta \mid \boldsymbol{\beta}^{(l-1)}, \boldsymbol{\theta}^{(l-1)})\label{eq:step1}\\
&~~~~~~~~=p(\boldsymbol{X}(\boldsymbol{t}^\Delta)\mid \Delta, \boldsymbol{\beta}^{(l-1)}, \boldsymbol{\theta}^{(l-1)})\times p(\Delta \mid \boldsymbol{\beta}^{(l-1)}, \boldsymbol{\theta}^{(l-1)})\textrm{ by M-H}\nonumber\\
&\textrm{Step 2: Sample } \boldsymbol{\beta}^{(l)}\sim p(\boldsymbol{\beta} \mid \boldsymbol{\theta}^{(l-1)}, \boldsymbol{X}^{(l)}(\boldsymbol{t}^{\Delta^{(l)}}), \Delta^{(l)})\label{eq:step2}\\
&\textrm{Step 3: Sample } \mu^{(l)} \sim p(\mu \mid (\sigma^{2})^{(l-1)}, \tau^{(l-1)}, \boldsymbol{X}^{(l)}(\boldsymbol{t}^{\Delta^{(l)}}), \Delta^{(l)}, \boldsymbol{\beta}^{(l)})\label{eq:step3}\\
&\textrm{Step 4: Sample } (\sigma^{2})^{(l)}\sim p(\sigma^2 \mid \tau^{(l-1)}, \boldsymbol{X}^{(l)}(\boldsymbol{t}^{\Delta^{(l)}}), \Delta^{(l)}, \boldsymbol{\beta}^{(l)}, \mu^{(l)})\label{eq:step4}\\
&\textrm{Step 5: Sample }  \tau^{(l)}\!\sim p(\tau \mid {\small\boldsymbol{X}^{(l)}(\boldsymbol{t}^{\Delta^{(l)}})}, \Delta^{(l)}, \boldsymbol{\beta}^{(l)}, \mu^{(l)}, (\sigma^{2})^{(l)}) \!\textrm{ by \!M-H}\label{eq:step5},
\end{align} 
where we suppress conditioning on $\boldsymbol{x}$ and $\boldsymbol{y}$ in all five steps. The conditional distributions  in \eqref{eq:step2}, \eqref{eq:step3}, and \eqref{eq:step4}, are standard families that can be sampled directly, whereas those in \eqref{eq:step1} and \eqref{eq:step5}  require M-H updates.  We use the factorization in \eqref{eq:step1} to construct a joint proposal, ($\tilde{\boldsymbol{X}}(\boldsymbol{t}^{\tilde{\Delta}}), \tilde{\Delta}$), for  ($\boldsymbol{X}(\boldsymbol{t}^{\Delta}), \Delta$) and  calculate its acceptance probability. 
First, $\tilde{\Delta}$ is proposed from  N$(\Delta^{(l-1)}, \psi^2)$, where $\psi$ is a proposal scale and is set to produce a reasonable acceptance rate. Given $\tilde{\Delta}$, we  propose $\tilde{\boldsymbol{X}}(\boldsymbol{t}^{\tilde{\Delta}}) \sim p(\boldsymbol{X}(\boldsymbol{t}^{\tilde{\Delta}})\mid\tilde{\Delta},  \boldsymbol{\beta}^{(l-1)}, \boldsymbol{\theta}^{(l-1)}, \boldsymbol{x}, \boldsymbol{y})$; this is a Gaussian distribution and is  specified in Appendix \ref{app1}. Because the  proposal for $\Delta$ and that for $\boldsymbol{X}(\boldsymbol{t}^{\Delta})$ given $\Delta$ are symmetric, ($\tilde{\boldsymbol{X}}(\boldsymbol{t}^{\tilde{\Delta}}), \tilde{\Delta}$) is accepted with a probability $\textrm{min}(1,~ r)$, where
\begin{equation}\label{MH}
r=\frac{p(\tilde{\Delta}\mid \boldsymbol{\beta}^{(l-1)}, \boldsymbol{\theta}^{(l-1)}, \boldsymbol{x}, \boldsymbol{y})}{p(\Delta^{(l-1)}\mid \boldsymbol{\beta}^{(l-1)}, \boldsymbol{\theta}^{(l-1)}, \boldsymbol{x}, \boldsymbol{y})}.
\end{equation}
Details of the marginalized density $p(\Delta\mid \boldsymbol{\beta}, \boldsymbol{\theta}, \boldsymbol{x}, \boldsymbol{y})$ in \eqref{MH} appear in Appendix \ref{appcond} and details of Steps 2--5 appear in Appendix \ref{detail_metro}.
%Because the joint update for $\boldsymbol{X}(\boldsymbol{t}^\Delta)$ and  $\Delta$ is controlled solely by  the ratio in \eqref{MH}, Step 1 updates $\boldsymbol{X}(\boldsymbol{t}^\Delta)$  only when it accepts $\tilde{\Delta}$. 
%due to the symmetry of the Gaussian distribution for $\boldsymbol{X}(\boldsymbol{t}^{\Delta})$ given all the model parameters. 

%It is easy and fast to implement the MHwG sampler  for the DA model in \eqref{full} compared to the MHwG for H\&R's non-DA (or collapsed) approach based on \eqref{marginal_data_augmentation}. 

\begin{figure}[b!]
\includegraphics[scale = 0.238]{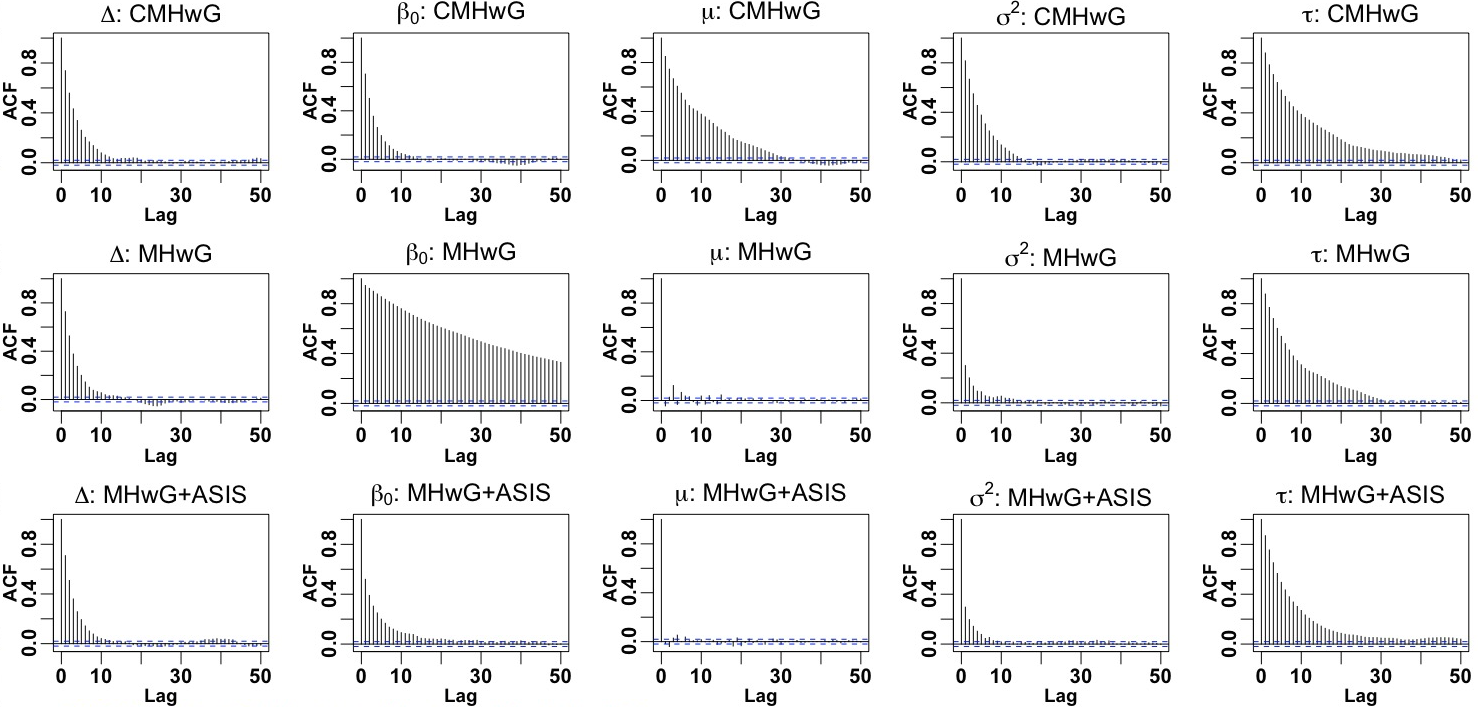}
\caption[]{The autocorrelation functions for $\Delta, \beta_0, \mu, \sigma^2$, and $\tau$ (columns from  left to right) based on 10,000 posterior samples after a burn-in of 10,000. Results are  obtained using three different posterior samplers  (CMHwG, MHwG, and MHwG + ASIS, rows from top to bottom). We use the curve-shifted model for simplicity and  the data from quasar Q0957+671 analyzed in Section~\ref{sec7}.}
\label{acf_comparison}
\end{figure}

The direct updates for $\boldsymbol{\beta}$, $\mu$, and $\sigma^2$ are based on  standard families that are not available using H\&R's  collapsed approach. Thus the collapsed approach must update each of the model parameters via a Metropolis or M-H update, which can slow down convergence. (Collapsing Gibbs-type samplers, however, is known to improve their rate of convergence \citep{liu2008monte} if the complete conditionals can be sampled directly.) Also, the collapsed MHwG (CMHwG) sampler requires about three times more CPU time per iteration than the (non-collapsed) MHwG sampler that we propose.  In Figure~\ref{acf_comparison}, we compare the autocorrelation functions (ACFs) of $\Delta$, $\beta_0$, $\mu$, $\sigma^2$, and $\tau$ obtained by the CMHwG sampler (first row) and those obtained by our MHwG sampler (second row).  The sampler in the third row is discussed in Section~\ref{sec_asis}. All algorithms are run using the curve-shifted model in \eqref{curve_shifted}  fit to data for quasar \emph{Q0957+561} \citep{hainline2012new}. Except for that of $\beta_0$, the ACFs generated with CMHwG (first row),  decay more slowly than those obtained with MHwG (second row).  The effective sample sizes per second (ESS/sec) tend to improve with MHwG over CMHwG. For example, for $\Delta$ the ESS/sec is 5.23 with CMHwG and 21.09 with MHwG. The exception is $\beta_0$, for which ESS/sec is 6.33 with CMHwG, but only 1.74 with MHwG. The mixing for $\boldsymbol{\beta}$ in our microlensing model is slow in general. In the following section, we discuss a way to improve the convergence rate of $\beta_0$ (or $\boldsymbol{\beta}$  in general) for the  MHwG sampler, while retaining  its fast running time. 
%However, the convergence rate  of Markov chains generated by the collapsed MHwG sampler is known to be faster  than that of Markov chains generated by the DA MHwG sampler that groups $\boldsymbol{X}(\boldsymbol{t}^\Delta)$ and $\Delta$ for their joint update 

\subsection{Ancillarity-sufficiency interweaving strategy}\label{sec_asis} To improve the convergence rate of $\boldsymbol{\beta}$, we adopt the ancillarity-sufficiency interweaving strategy \citep[ASIS,][]{yu2011}.  In a general hierarchical modeling setting, ASIS interweaves trajectories of the Markov chains   obtained by two discordant  parameterizations of the unknown quantities, which reduces dependence between the adjoining iterates. A different parameterization for the location parameters, e.g., $\boldsymbol{\beta}$  in our case, can be derived by shifting, and that for scale parameters by rescaling.   The two parameterizations are designed so that the original and transformed parameters can be viewed as ancillary and sufficient statistics for $\boldsymbol{\beta}$, respectively.  ASIS is always  faster to converge than the slower of the data augmentation samplers based on either of the two parameterizations and is geometrically convergent even when neither of the two data augmentation samplers is.

%One can benefit from ASIS by coming up with AA and SA of the unknown quantities.
%Yu and Meng (2011) have conducted intense theoretical investigation on  ASIS; ASIS is at least better than the worse of the AA and SA; ASIS becomes the optimal algorithm among a large class of data augmentation under a certain condition; MCMC of the ASIS converges geometrically even when MCMCs of the AA and SA do not.  See Yu and Meng (2011) for more theoretical probes. 

%Specifically, we use the ASIS  because our DA MHwG sampler sometimes leads to its slow convergence as shown in the $(2, 2)$ panel of Figure~\ref{acf_comparison}. %gain of convergence is large when one parameterization converges faster than the other.

% 
In the parameterization used up until now, $\boldsymbol{X}(\boldsymbol{t}^\Delta)$ is an \emph{ancillary augmentation} (AA) for $\boldsymbol{\beta}$ in that it is an ancillary statistic for $\boldsymbol{\beta}$. That is, the distribution of $\boldsymbol{X}(\boldsymbol{t}^\Delta)$ in \eqref{conditionalOU1} does not depend on $\boldsymbol{\beta}$. On the other hand, a \emph{sufficiency augmentation} (SA) for $\boldsymbol{\beta}$ is based on a transformation of $\boldsymbol{X}(\boldsymbol{t}^\Delta)$ that have sufficient information to estimate $\boldsymbol{\beta}$, that is, a sufficient statistic for $\boldsymbol{\beta}$. To derive an SA for $\boldsymbol{\beta}$, we introduce the parameterization,
\begin{equation}\label{sa}
K(t^\Delta_j)\equiv X(t^\Delta_j)+\boldsymbol{w}_m^\top(t^\Delta_j)\boldsymbol{\beta}\times I_{\boldsymbol{t}-\Delta}(t^\Delta_j), \textrm{ for } j=1, 2, \ldots, 2n,
\end{equation}
where 
\begin{equation}\label{time_indicator}
I_{\boldsymbol{t}-\Delta}(t^\Delta_j) = \left\{ \begin{array}{ll}
1, & \textrm{if $t^\Delta_j\in \boldsymbol{t}-\Delta,$}\\
0, & \textrm{if $t^\Delta_j\in \boldsymbol{t}.$}
\end{array} \right.
\end{equation}
This indicator  is one if $t^\Delta_j$ is an element of $ \boldsymbol{t}-\Delta=\{t_1-\Delta, t_2-\Delta, \ldots, t_n-\Delta\}$ and zero otherwise. Thus, $\boldsymbol{K}(\boldsymbol{t}^{\Delta})$  represents the time-sorted latent magnitudes of $\boldsymbol{X}(\boldsymbol{t})$ and of microlensing-adjusted $\boldsymbol{Y}(\boldsymbol{t})$, i.e., $\boldsymbol{X}(\boldsymbol{t}-\Delta) + \boldsymbol{w}_m^{\top}(\boldsymbol{t}-\Delta) \boldsymbol{\beta}$. Using \eqref{sa}, we express the observation model in  \eqref{lik1} and \eqref{lik3} as
\begin{align}
x_{j}\mid K(t_j) &\stackrel{\textrm{indep.}}{\sim} \textrm{N}[K(t_j),~ \delta^2_{j}].\label{lik5}\\
y_{j}\mid K(t_j-\Delta), \Delta &\stackrel{\textrm{indep.}}{\sim} \textrm{N}[K(t_j-\Delta),~ \eta^2_{j}].\label{lik4}
\end{align}
The  distributions for the latent light curve in \eqref{conditionalOU1} are replaced by
\begin{align}
\begin{aligned}
K(t_1^\Delta)\mid \Delta, \boldsymbol{\beta}, \boldsymbol{\theta} &~\sim~ \textrm{N}\!\left[\mu+\boldsymbol{w}_m^\top(t^\Delta_1)\boldsymbol{\beta} \times I_{\{\boldsymbol{t}-\Delta\}}(t_1^\Delta),~ \frac{\tau\sigma^2}{2}\right]\!,\\
K(t_j^\Delta)\mid K(t_{j-1}^\Delta), \Delta, \boldsymbol{\beta}, &\boldsymbol{\theta} ~\sim~ \textrm{N}\bigg[\mu+\boldsymbol{w}_m^\top(t^\Delta_j)\boldsymbol{\beta} \times I_{\{\boldsymbol{t}-\Delta\}}(t_j^\Delta)\\
+ a_j\big(K(t_{j-1}^\Delta) -\mu- &\boldsymbol{w}_m^\top(t^\Delta_{j-1})\boldsymbol{\beta} \times I_{\{\boldsymbol{t}-\Delta\}}(t_{j-1}^\Delta)\big), ~ \frac{\tau \sigma^2}{2}(1-a^2_j)\bigg].
\end{aligned}\label{newparak}
\end{align}

Under this reparameterization of the  model in terms of $\boldsymbol{K}(\boldsymbol{t}^\Delta)$, $\boldsymbol{\beta}$ appears only in \eqref{newparak}, which means that $\boldsymbol{K}(\boldsymbol{t}^\Delta)$ contains sufficient information to estimate $\boldsymbol{\beta}$ and thus $\boldsymbol{K}(\boldsymbol{t}^\Delta)$ is an SA for $\boldsymbol{\beta}$. In contrast, $\boldsymbol{\beta}$  appears only in the distribution of observed magnitudes in \eqref{lik}, not in that of latent magnitudes in \eqref{conditionalOU1}, and thus $\boldsymbol{X(t^{\Delta})}$ is an AA for $\boldsymbol{\beta}$. Because the parameterization does not affect the prior distributions of the model parameters in \eqref{priordeltac} and \eqref{hyp},  the full joint posterior density in terms of $\boldsymbol{K}(\boldsymbol{t}^\Delta)$, i.e., $p(\boldsymbol{K}(\boldsymbol{t}^\Delta), \Delta, \boldsymbol{\beta}, \boldsymbol{\theta} \mid\boldsymbol{x}, \boldsymbol{y})$, is proportional to the product of densities of observed data given in \eqref{lik5} and \eqref{lik4} and prior densities in \eqref{newparak}, \eqref{priordeltac} and \eqref{hyp}. Consequently, the marginal posterior distribution of the model parameters, $\{\Delta, \boldsymbol{\beta}, \boldsymbol{\theta}\}$, is unchanged.

ASIS interweaves the trajectory of $\boldsymbol{\beta}$ from a sample constructed under AA and that constructed under SA. This can be accomplished by replacing Step 2 in \eqref{eq:step2} with the following four steps:
\begin{align}
{\small\textrm{Step 2$_a$}}&:  {\small\textrm{Sample }} \boldsymbol{\beta}^{(l)}_{\textrm{AA}}\sim p(\boldsymbol{\beta} \mid \boldsymbol{\theta}^{(l-1)}, \boldsymbol{X}^{(l)}(\boldsymbol{t}^{\Delta^{(l)}}), \Delta^{(l)})\label{eq:asis_step1}\\
{\small\textrm{Step 2$_b$}}&:  {\small \textrm{Set }} K^{(l)}(t^{\Delta^{(l)}}_j){\small  = }~ X^{(l)}(t^{\Delta^{(l)}}_j)+\boldsymbol{w}_m^\top(t^{\Delta^{(l)}}_j)\boldsymbol{\beta}^{(l)}_{\textrm{AA}} I_{\boldsymbol{t}-\Delta^{(l)}}(t^{\Delta^{(l)}}_j)\label{eq:asis_step2}\\
{\small\textrm{Step 2$_c$}}&:   {\small \textrm{Sample }} \boldsymbol{\beta}^{(l)}_{\textrm{SA}}\sim p(\boldsymbol{\beta} \mid \boldsymbol{\theta}^{(l-1)}, \boldsymbol{K}^{(l)}(\boldsymbol{t}^{\Delta^{(l)}}), \Delta^{(l)})\label{eq:asis_step3}\\
{\small \textrm{Step 2$_d$}}&:  \textrm{Set }X^{(l)}(t^{\Delta^{(l)}}_j){\small=}~ K(t^{\Delta^{(l)}}_j) - \boldsymbol{w}_m^\top(t^{\Delta^{(l)}}_j)\boldsymbol{\beta}^{(l)}_{\textrm{SA}}  I_{\boldsymbol{t}-\Delta^{(l)}}(t^{\Delta^{(l)}}_j)\label{eq:asis_step4}
\end{align}
Again, we suppress the conditioning on $\boldsymbol{x}$ and $\boldsymbol{y}$. In Step 2$_c$, we set $\boldsymbol{\beta}^{(l)}$ to $\boldsymbol{\beta}^{(l)}_{\textrm{SA}}$ sampled from its conditional posterior distribution specified in \eqref{suff_c}. In Step 2$_d$, ASIS updates $\boldsymbol{X}^{(l)}(\boldsymbol{t}^{\Delta^{(l)}})$ to adjust for the inconsistency between the updates sampled in \eqref{eq:step3}--\eqref{eq:step5} that are based on  $\boldsymbol{X}^{(l)}(\boldsymbol{t}^{\Delta^{(l)}})$ and the update $\boldsymbol{\beta}^{(l)}$ that  is based on $\boldsymbol{K}^{(l)}(\boldsymbol{t}^{\Delta^{(l)}})$. Updating $\boldsymbol{X}^{(l)}(\boldsymbol{t}^{\Delta^{(l)}})$ in \eqref{eq:asis_step4}  synchronizes this inconsistency and preserves the stationary distribution \citep{yu2011}.  The additional computational cost of  ASIS  is negligible because the conditional updates in \eqref{eq:asis_step1} and \eqref{eq:asis_step3} include quick multivariate Gaussian sampling; see  \eqref{ancil_c} and \eqref{suff_c} for details.
% to the update of $\boldsymbol{\beta}$ at iteration $l$, i.e., 

% without the ASIS
ACFs of the model parameters  obtained by MHwG equipped with ASIS, denoted by MHwG$+$ASIS,  appear on the third row of  Figure~\ref{acf_comparison}; the ACF of $\beta_0$ in the second column shows a noticeable improvement compared to that obtained by MHwG sampler. The ESS/sec for $\beta_0$ is 20.95 with MHwG$+$ASIS and 1.74 with MHwG. In general, ASIS improves   the mixing of all the regression coefficients  in our microlensing model. Although it improves the ACF for the components of $\boldsymbol{\beta}$, ASIS has little effect on the ACF of $\Delta$. The ESS/sec for $\Delta$ is 21.35 with MHwG$+$ASIS and 21.09 with MHwG. This small improvement implies that the dependence between $\Delta$ and  $\boldsymbol{\beta}$ may  be weak a posteriori; this is confirmed by our data analyses in Section~\ref{case1}. Nevertheless, ASIS improves overall convergence of the chain which we expect to improve the reliability of all inferences based on the chain. 
%
% the convergence rate overall.
%Naively, it would seem there isn't much advantage in
%the ASIS modification if the focus of inference is the posterior for
%$\Delta$.  More carefully, one should be cautious about using a sampler
%that did not mix well for all parameters, unless one could somehow show
%there is not significant joint dependence on poorly mixed nuisance
%parameters.  So the attention to ASIS is still well-motivated.

% This makes intuitive sense because the ASIS interweaves seemingly unrelated trajectories of Markov chains for $\beta_0$ obtained under two discordant parameterizations, AA and SA. 
%Figure~\ref{acf_comparison} also indicates that the enhanced convergence rate of $c$ slightly improves the convergence rates of $\Delta$ and $\tau$, which reflects on that they are in the same Gibbs cycle and the convergence of one component dramatically improves.

\subsection{Adaptive MCMC}\label{adaptivemcmc} Our  MHwG sampler (either with or without  ASIS) requires  a proposal distribution in each of its two Metropolis steps, that is, N$\left[\Delta^{(l-1)}, \psi^2\right]$ used to update $\Delta^{(l)}$ in \eqref{eq:step1} and N$\left[\log(\tau^{(l-1)}), \phi^2\right]$ used to update $\log(\tau^{(l)})$ in \eqref{eq:step5}, where $\psi$ and $\phi$ are the proposal scales. To avoid burdensome off-line tuning of the proposal scales, we implement an adaptive MCMC sampler \citep{brooks2011handbook} that allows  automatic adjustment during the run. The steps of the adaptive MHwG+ASIS sampler are specified in Algorithm~\ref{algorithm_sampler1}. Specifically, we implement an algorithm that updates the two proposal scales every 100 iterations, based on the most recent 100 proposals as outlined in Step 6 of Algorithm~\ref{algorithm_sampler1}.  The Markov chains equipped with the adaptive MCMC converge to the stationary distribution because the adjustment factors, $\exp(\pm\min(0.01,~ 1/\sqrt{i}))$, in Step 6 of of Algorithm~\ref{algorithm_sampler1}   approach unity as $i$ goes to infinity. This condition is called diminishing adaptation condition \citep{roberts2007coupling}.  We set the lower and upper bounds of the acceptance rate to 0.23 and 0.44, respectively \citep{gelman2013bayesian}.

%The adjusted proposal scale $\psi$ for $\Delta$ is applied to the next one hundred iterations.  

All of the numerical results presented in Figure~\ref{acf_comparison} were obtained using algorithms that similarly adapted their M-H updates, i.e., the M-H updates of all the parameters in CMHwG and of  $\Delta$ and $\tau$ in both MHwG and MHwG+ASIS.
\begin{algorithm}[t!]
\begin{algorithmic}
\State Set $\boldsymbol{X^{(0)}(t^{\Delta^{(0)}})}$, $\Delta^{(0)}$, $\boldsymbol{\beta}^{(0)}$, $\mu^{(0)}$, $(\sigma^{2})^{(0)}$, $\tau^{(0)}$, $\psi^{(0)}$, $\phi^{(0)}$.
\State For $l=1, 2, \ldots $
\State Step 1: Sample $\Delta^{(l)}$ using a Metropolis step with proposal rule N$[\Delta^{(l-1)}, (\psi^{(l-1)})^2$].  
\\
~~~~~~~~~~If a new proposal for $\Delta^{(l)}$ is accepted, then sample $\boldsymbol{X}^{(l)}(\boldsymbol{t}^{\Delta^{(l)}})$,\\
~~~~~~~~~~or otherwise set $\boldsymbol{X}^{(l)}(\boldsymbol{t}^{\Delta^{(l)}})$ to $\boldsymbol{X}^{(l-1)}(\boldsymbol{t}^{\Delta^{(l-1)}})$.
\State Step 2: (ASIS) Update $\boldsymbol{\beta}^{(l)}$ and $\boldsymbol{X}^{(l)}(\boldsymbol{t}^{\Delta^{(l)}})$ via \eqref{eq:asis_step1}--\eqref{eq:asis_step4}.
\State Step 3: Sample  $\mu^{(l)}$ via  \eqref{eq:step3}.
\State Step 4: Sample $(\sigma^{2})^{(l)}$ via  \eqref{eq:step4}.
\State Step 5: Sample $\tau^{(l)}$ using an M-H step with proposal rule N$[\log(\tau^{(l-1)}), (\phi^{(l-1)})^2$].
\State Step 6: (Adaptation) If $l$ mod 100 = 0
\begin{algorithmic}
\If{the acceptance rate of $\Delta$ in iterations $l-99, l-98, \ldots, l$  $>$ 0.44} \State $\psi^{(l)}\gets \psi^{(l-1)}\times\exp(\min(0.01,~ 1/\sqrt{(l/100)}))$ \ElsIf{the acceptance rate of $\Delta$ in iterations $l-99, l-98, \ldots, l$  $<$ 0.23} \State $\psi^{(l)}\gets \psi^{(l-1)}\times\exp(-\min(0.01,~ 1/\sqrt{(l/100)}))$
\EndIf
\end{algorithmic}
\begin{algorithmic}
\If{the acceptance rate of $\tau$ in iterations $l-99, l-98, \ldots, l$  $>$ 0.44} \State $\phi^{(l)}\gets \phi^{(l-1)}\times\exp(\min(0.01,~ 1/\sqrt{(l/100)}))$ \ElsIf{the acceptance rate of $\tau$ in iterations $l-99, l-98, \ldots, l$  $<$ 0.23} \State $\phi^{(l)}\gets \phi^{(l-1)}\times\exp(-\min(0.01,~ 1/\sqrt{(l/100)}))$
\EndIf
\end{algorithmic}
Otherwise $\psi^{(l)}=\psi^{(l-1)}$ and $\phi^{(l)}=\phi^{(l-1)}$.
\end{algorithmic}
\caption{\!\!. Steps of the adaptive MHwG+ASIS sampler.}
\label{algorithm_sampler1}
\end{algorithm}

\section{Profile likelihood of the time delay}\label{sec_profile} We use the profile likelihood of $\Delta$  \citep[e.g.,][]{davison2003statistical} to obtain  a simple approximation to its marginal posterior distribution, $p(\Delta\mid \boldsymbol{x}, \boldsymbol{y})$. This profile likelihood is 
\begin{equation}\label{profile}
L_{\textrm{prof}}(\Delta)\equiv\max_{\boldsymbol{\beta}, \boldsymbol{\theta}}L(\Delta, \boldsymbol{\beta}, \boldsymbol{\theta})=L(\Delta, \hat{\boldsymbol{\beta}}_\Delta, \hat{\boldsymbol{\theta}}_\Delta),
\end{equation}
where $L(\Delta, \boldsymbol{\beta}, \boldsymbol{\theta})$ is the marginal likelihood function of the model parameters with the latent light curve integrated out, i.e.,
\begin{align}\label{marginal_data_augmentation}
L(\Delta, \boldsymbol{\beta}, \boldsymbol{\theta})&=p(\boldsymbol{x}, \boldsymbol{y}\mid \Delta, \boldsymbol{\beta}, \boldsymbol{\theta})\\
&=\int  p(\boldsymbol{x}, \boldsymbol{y}\mid \boldsymbol{X}(\boldsymbol{t}^\Delta), \Delta, \boldsymbol{\beta}) \times p(\boldsymbol{X}(\boldsymbol{t}^\Delta)\mid \Delta, \boldsymbol{\theta})~d\boldsymbol{X}(\boldsymbol{t}^\Delta),\nonumber
\end{align}
and ($\hat{\boldsymbol{\beta}}_{\Delta}$, $\hat{\boldsymbol{\theta}}_{\Delta}$) are the values of ($\boldsymbol{\beta}$, $\boldsymbol{\theta}$) that maximize $L(\Delta, \boldsymbol{\beta}, \boldsymbol{\theta})$ for each $\Delta$. 
%=\int  p(\boldsymbol{x}, \boldsymbol{y}, \boldsymbol{X}(\boldsymbol{t}^\Delta)\mid \boldsymbol{\theta})~d\boldsymbol{X}(\boldsymbol{t}^\Delta)
 
The profile likelihood of a parameter, say $\varphi$, may asymptotically approximate its marginal posterior distribution with a uniform prior on $\varphi$. This happens, for example, if the log likelihood of the model parameters is approximately quadratic given $\varphi$ under standard asymptotic arguments. The prior distribution on the parameters other than $\varphi$ is chosen in such a way as to approximately cancel the determinant of Hessian matrix of the log likelihood, e.g., as happens asymptotically with the Jeffreys' prior, see Appendix~\ref{proof_prof_likelihood} for details. 
%(We can also use the Laplace approximation by putting a flat prior on $\varphi$, but it additionally requires calculating a Hessian matrix.)} % ) 
 
%We show that $L_{\textrm{prof}}(\Delta)$ is an approximation to the Jeffreys-integrated likelihood   that integrates out $\boldsymbol{\beta}$ and  $\boldsymbol{\theta}$ from $L(\Delta, \boldsymbol{\beta}, \boldsymbol{\theta})$ with Jeffreys' prior on $\boldsymbol{\beta}$ and  $\boldsymbol{\theta}$ \citep{berger1999integrated}. We use the  Laplace approximation for the integration after replacing Fisher information in the integrand with observed information. Because the Jeffreys-integrated likelihood is proportional to $p(\Delta\mid \boldsymbol{x_t}, \boldsymbol{y})$ with a uniform prior on $\Delta$, $L_{\textrm{prof}}(\Delta)$ is approximately proportional to $p(\Delta\mid \boldsymbol{x_t}, \boldsymbol{y})$. See Appendix \ref{proof_prof_likelihood} for its proof.  

Treating $L_{\textrm{prof}}(\Delta)$ as an approximation to $p(\Delta\mid \boldsymbol{x}, \boldsymbol{y})$, we evaluate $L_{\textrm{prof}}(\Delta)$ on a fine grid of values over the interesting range of $\Delta$. We set $w$ values from $\Delta_1$ to $\Delta_w$, i.e.,  $\{\Delta_1, \Delta_2, \ldots, \Delta_w\}$, where e.g., $\Delta_j-\Delta_{j-1}= 0.1$ ($j=2, 3, \ldots, w$) for a high-resolution mapping. Unfortunately this can be computationally burdensome due to the large number of values on the grid.  For example, if the feasible range for $\Delta$ is $[-1500, 1500]$, the grid consists of 30,001 values.  At one second per evaluation  this requires about 8 hours and 20 minutes. Though computationally expensive, the high-resolution mapping of $L_{\textrm{prof}}(\Delta)$ is  useful because it clearly identifies the likely (modal) values of $\Delta$. In practice,  we use multiple cores in parallel to reduce the computation time and optimization is implemented using a general-purpose quasi-Newton method, \texttt{optim}, in R  \citep{r2014}.  Initial values for numerical optimization at the first grid point are set just as with the Bayesian method described in Section~\ref{numerical} and the initial values for subsequent grid point are set to the values that maximize the profile likelihood at the previous grid point.

% burden; each core computes a partition of the grid of values of $\Delta$. 
%, to draw the  profile likelihood curve over the interesting range of $\Delta$
%to maximize the profile likelihood with respect to $\mu, \sigma^2, \tau$, and $\boldsymbol{\beta}_{(4\times1)}$ given each  $\Delta_j$, then
%For example, if two cores are available, then we can divide the entire feasible range into $[-1,500, 0]$ and $[0, 1,500]$, letting each core compute the profile likelihood on one of the ranges.   %This  curve can be used to check  the multimodal behavior of $p(\Delta\mid D_{\textrm{obs}})$ in a fast way. 
%
%. 

The profile likelihood evaluated on the grid can be used to approximate the posterior mean $E(\Delta\mid  \boldsymbol{x}, \boldsymbol{y})$,
\begin{equation}\label{mean_approx}
\hat{\Delta}_\textrm{mean}\equiv \frac{\sum_{j=1}^w\Delta_j\times L_{\textrm{prof}}(\Delta_j)}{\sum_{j=1}^w L_{\textrm{prof}}(\Delta_j)},
\end{equation}
and the posterior variance $\textrm{Var}(\Delta\mid  \boldsymbol{x}, \boldsymbol{y})$,
\begin{equation}\label{var_approx}
\hat{V}\equiv\frac{\sum_{j=1}^w\Delta_j^2\times L_{\textrm{prof}}(\Delta_j)}{\sum_{j=1}^w L_{\textrm{prof}}(\Delta_j)}-\left[\frac{\sum_{j=1}^w \Delta_j\times L_{\textrm{prof}}(\Delta_j)}{\sum_{j=1}^w L_{\textrm{prof}}(\Delta_j)}\right]^2.
\end{equation}
Moreover, the posterior mode of $\Delta$ can be approximated by a value of $\Delta$ in the grid that maximizes the profile likelihood, which is a discrete approximation to the maximum  likelihood estimator, $\hat{\Delta}_{\textrm{MLE}}\equiv\argmax_{\Delta}~ L_{\textrm{prof}}(\Delta)$.  If the profile likelihood exhibits multiple modes, however, the (approximate) posterior mean, mode, and variance can be misleading. Instead each mode requires separate investigation based on their (approximate) relative size.
%The profile likelihood can give us .
%these are appropriate only when there is a dominant mode;
%it is better to investigate each mode when there are multiple modes.
%A closed-form for the second derivative of $\log(L_{\textrm{prof}}(\Delta))$ is not analytically available.  We find that estimating the asymptotic uncertainty from a numerical approximation to second derivative at the $\hat{\Delta}_{\textrm{MLE}}$ is numerically unstable.

%Using the asymptotic uncertainty based on the second derivative of $\log(L_{\textrm{prof}}(\Delta))$ at $\hat{\Delta}_{\textrm{MLE}}$ can be unstable because the closed-form second derivative is not available and its numerical approximation  can be unstable.

%the (approximate) posterior mean  and variance can be misleading because 

\section{Time delay estimation strategy and numerical illustrations}\label{numerical} 
The first step of our analysis is to plot $L_{\textrm{prof}}(\Delta)$ over the  range of $\Delta$ to check for multi-modality that may indicate multiple modes \citep[e.g.,][]{brooks1997finite} in the marginal posterior distribution of $\Delta$.  For some quasars, the interesting range of $\Delta$ can be narrowed using the results of past analyses or information from other astrophysical probes as discussed in Section \ref{priorsection2}.  If prior information for $\Delta$ is unavailable, we explore the feasible range.
%, which we show via  examples in Section~\ref{numerical}

In our numerical studies, we find that when $L_{\textrm{prof}}(\Delta)$ is dominated by one mode, the moment estimates of $\Delta$ based on $L_{\textrm{prof}}(\Delta)$, i.e., $\hat{\Delta}_{\textrm{mean}}$ in \eqref{mean_approx} and $\hat{V}$ in \eqref{var_approx}, are almost identical to the posterior mean and variance obtained via MCMC. On the other hand, modes near the margins of the range of $\Delta$ may indicate microlensing; see Section \ref{case1}. In this case, the order of polynomial regression must be increased.  If there are multiple modes that are not near the margins of the feasible range, each mode merits investigation; evaluating $L_{\textrm{prof}}(\Delta)$ divided by the square root of the observed Fisher information  at each mode  provides an approximation of the relative size of each mode. If the modes are so close that the MCMC chain readily jumps between them, it is easy to estimate their relative size; see Section~\ref{sec7}.
%if $L_{\textrm{prof}}(\Delta)$ reveals multiple modes near margins of the entire feasible range, it can be an indication of 

%the MCMC estimates for the relative sizes of the modes are reliable
%If multiple modes are close enough for Markov chains to jump between modes frequently, we run MCMC chains near the highest mode, not near every mode.}
As a cross-check,  in all of our numerical examples we run three MCMC chains near each of the major mode(s) identified by $L_{\textrm{prof}}(\Delta)$;  The three starting values for each mode are  \{mode, mode $\pm~20$ days\}.  Each chain is run for 510,000 iterations and the first 10,000 iterations are discarded as burn-in; the Gelman-Rubin diagnostic statistics \citep{gelman1992inference} of all of the model parameters computed from the post burn-in chains in  all of our numerical examples  are smaller than 1.001, which justifies our burn-in size. Because the smallest effective sample size of the parameters computed from the post burn-in chains across all of our examples is about 11,000, we thin each chain by a factor of fifty (from length 500,000 to 10,000). We combine the three thinned chains to obtain our Monte Carlo sample from the posterior distribution.  For all chains, we set the starting value of $\boldsymbol{\beta}$ to the estimated regression coefficients  obtained by regressing $\boldsymbol{y}- \sum_jx_{j}/n$ on a covariate matrix $\boldsymbol{W}_m(\boldsymbol{t}-\Delta^{(0)})$ whose  $j$th row is $\boldsymbol{w}_m^\top(t_j-\Delta^{(0)})$, where $\Delta^{(0)}$ is the initial value of $\Delta$. The initial value of $\boldsymbol{X}(\boldsymbol{t}^\Delta)$ is the combined light curve, that is, $\{\boldsymbol{x}, \boldsymbol{y_{-\Delta^{(0)}}} - \boldsymbol{W}_m^\top(\boldsymbol{t}-\Delta^{(0)})\boldsymbol{\beta}^{(0)}\}$ sorted in time. The starting value of $\mu$ is set to the mean of $\boldsymbol{x}$, that of $\sigma^2$ to $0.01^2$, and that of $\tau$ to 200. We set the initial standard deviations of the proposal distributions to $\psi=10$ days for $\Delta$ and $\phi=3$ for $\log(\tau)$.  (The unit of $\tau$ is days.)
%\footnote{\textcolor{red}{It is more desirable to determine the thinning rate based on the smallest effective sample size among all the parameters. However, we consider the smallest effective sample size of $\Delta$ only because the time delay and most of the other parameters are nearly independent a posteriori as shown in Figure~\ref{fig_corr}.}}.
%We run using  dispersed starting values. 
%The profile likelihood can be used to cross-check the result of the Bayesian analysis. 

We  use  simulated data of doubly- and quadruply-lensed quasars publicly available at the TDC website (http://timedelaychallenge.org) to illustrate our time delay estimation strategy when  prior information for $\Delta$ is not available. We also analyze observed data of quasars \emph{Q0957+561} and \emph{J1029+2623} over the feasible range of $\Delta$ for illustrative purpose, though  prior information is available to limit the range of $\Delta$.
%, referring to the mock LSST catalog of lensed quasar systems \citep{oguri2010gravitationally}. 

We report the CPU time in seconds using a server equipped with two 8-core Intel Xeon E5-2690 at 2.9 GHz and 64 GB of memory.  We report the entire mapping time for $L_{\textrm{prof}}(\Delta)$. %not use a parallel computing.

\subsection{A doubly-lensed quasar simulation}\label{case1} 
\begin{figure}[t!]
\includegraphics[scale = 0.29]{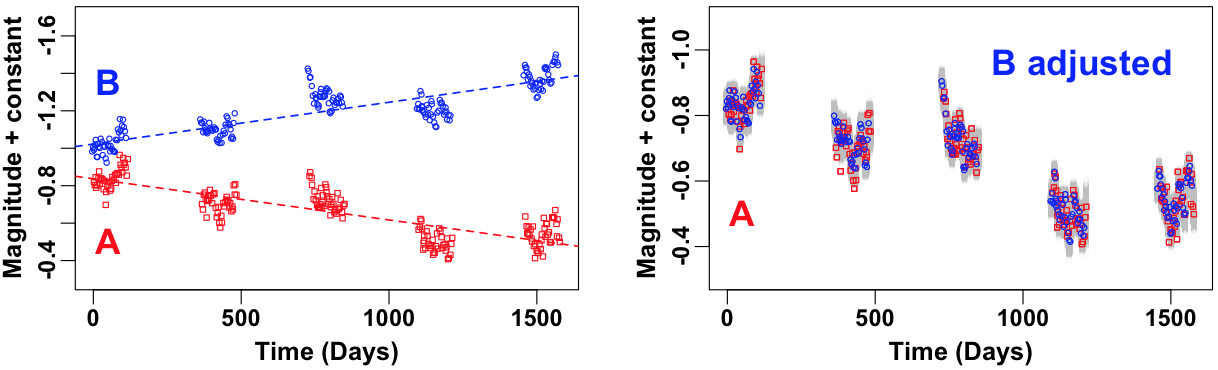}
\caption[]{The first panel shows a TDC data set suffering from microlensing that results in light curves with  different long-term trends. The dashed lines denote fitted linear regression lines. In the second panel, we combine the two light curves by shifting light curve $B$ by E$(\Delta\mid \boldsymbol{x}, \boldsymbol{y})$ in the horizontal axis  and by subtracting the estimated third-order polynomial regression based on E$(\boldsymbol{\beta}\mid \boldsymbol{x}, \boldsymbol{y})$ from light curve $B$.  The microlensing model finds matches between the intrinsic fluctuations of the light curves after removing  the relative microlensing trend from light curve $B$. We plot the posterior sample of $\boldsymbol{X}(\boldsymbol{t}^{\Delta})$ in gray in the right panel to represent the point-wise prediction interval for the combined latent light curve. The gray areas encompass most of the combined observed light curve, indicating that the fitted model predicts the observed data well.}
\label{example1_data}
\end{figure}

The simulated data for a doubly-lensed quasar are plotted in the first panel of Figure~\ref{example1_data}; the median cadence is 3 days, the cadence standard deviation is 1 day,   observations are made  for 4 months in each of 5 years for 200 observations in total, and  measurement errors are heteroskedastic Gaussian. The light curves  suffer from  microlensing  which can be identified from their different long-term linear trends and similar short-term (intrinsic) variability. %No prior information of $\Delta$ is available.
%with light curve B shifted in $x$-axis by the blinded true time delay (5.86 days)

%As a graphical model checking, we plot the latent magnitudes, $\boldsymbol{X}(\boldsymbol{t}^{\Delta})$, denoted by gray dots , on top of the combined light curve. The gray areas encompass most of the combined light curve. Since posterior samples of $\boldsymbol{X}(\boldsymbol{t}^{\Delta})$ represent the range of our predicted values for the combined light curve, we can see how well our model describes the observed data.

\begin{figure}[t!]
\includegraphics[scale = 0.32]{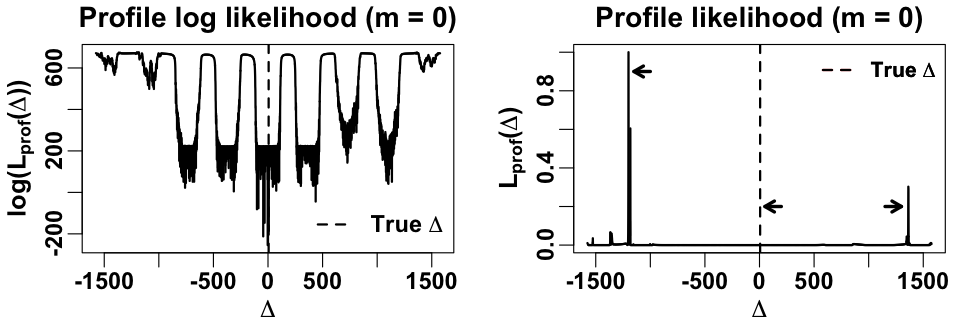}
\caption[]{The profile log likelihood (left) and the profile likelihood (right) of $\Delta$ over its feasible range under the curve-shifted model ($m=0$). We  exponentiate and normalize $\log(L_{\textrm{prof}}(\Delta_i))$ as  $\exp[\log(L_{\textrm{prof}}(\Delta_i))-\max_j(\log(L_{\textrm{prof}}(\Delta_j)))]$ for all $i$.  The vertical  dashed line indicates the  true time delay. The profile likelihood near the true time delay (5.86 days) is overwhelmed by the modes near margins.}
\label{example1_profpost}
\end{figure}

To show the effect of microlensing on the time delay estimation, we fit both the curve-shifted model ($m=0$) in \eqref{curve_shifted}  and the microlensing model with $m=3$ in \eqref{fundamental}. We  plot $\log(L_{\textrm{prof}}(\Delta))$ and $L_{\textrm{prof}}(\Delta)$ based on the curve-shifted model over the   feasible range, $[t_1-t_n,~ t_n-t_1]=[-1575.85,~ 1575.85]$, in the two panels of Figure~\ref{example1_profpost}.  The profile likelihood exhibits large modes near the margins that overwhelm the profile likelihood near the true time delay (5.86 days denoted by the vertical  dashed line). 
%, setting the largest value of $L_{\textrm{prof}}(\Delta)$ to one

\begin{figure}[b!]
\includegraphics[scale = 0.249]{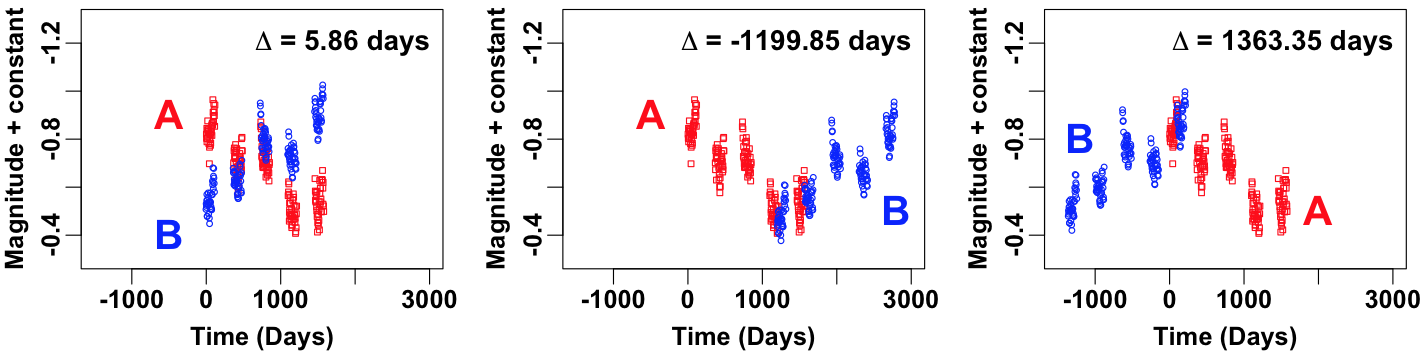}
\caption[]{We shift  light curve B (blue) by the  true time delay (5.86 days) in the first panel, by $-$1,199.85 days  in the second panel, and by 1363.35 days in the third panel. These three time delays correspond to three arrows in the second panel of Figure \ref{example1_profpost}. The shift in magnitude used is the value of $\beta_0$ that maximizes the profile likelihood given each time delay. Without accounting for microlensing, the curve-shifted model fails because the light curves do not match even at the true time delay. The curve-shifted model may produce large modes near the margins because, in the small overlap between the tips of two light curves, spurious matches may be made by chance between similar fluctuation patterns as shown in the second and third panels.}
\label{example1_curveshft}
\end{figure}

In the presence of microlensing, the curve-shifted model cannot identify the time delay because the latent curves are not shifted versions of each other. The modes of $L_{\textrm{prof}}(\Delta)$ near the margins of the range of $\Delta$ occur because, in the small overlap between the tips of two light curves, spurious matches may be made by chance between similar fluctuation patterns. In Figure \ref{example1_curveshft}, for instance,  we shift light curve $B$ in the $x$-axis by the three values of $\Delta$ indicated by three arrows in the second panel of Figure \ref{example1_profpost}. In the first panel of Figure~\ref{example1_curveshft}, the  two light curves shifted by the true time delay do not match for any shift in magnitude. However, given the time delays at around $-1$,200 or 1,360 days, the two light curves look well-connected as shown in the second and third panels. Thus, the profile likelihood near the  true time delay  is overwhelmed by the values of the profile likelihood near $-1$,200 and 1,360 days. %Thus, the profile likelihood can be used as evidence of microlensing. 
%The modes of  near the margins of the range of  occur because a small overlap between the tips of two light curves may exhibit the only similar fluctuation patterns  detectable by shifting one of the light curves.
% in $x$- and $y$-axes due to  the different long-term trends. 
%In this case, the profile likelihood under the curve-shifted model can produce several

%The modes of $L_{\textrm{prof}}(\Delta)$ near the margins of the range of $\Delta$ occur because

\begin{figure}[t!]
\includegraphics[scale = 0.25]{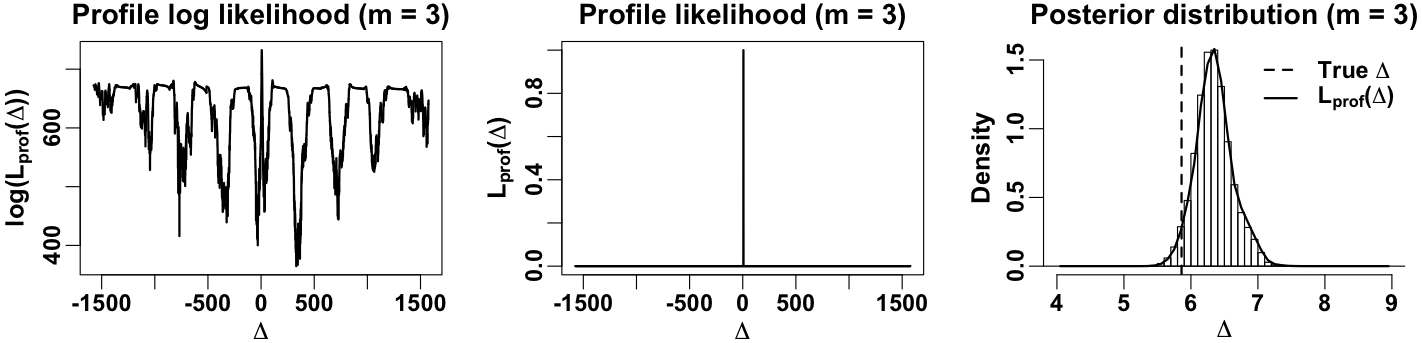}
\caption[]{The profile log likelihood (first panel) and the profile likelihood (second panel) of $\Delta$ over its  feasible  range under the microlensing model ($m=3$). The profile likelihood shows one mode near the true time delay (5.86 days). The third panel shows the marginal posterior distribution of $\Delta$ as a histogram of the MCMC samples with re-normalized $L_{\textrm{prof}}(\Delta)$ superimposed. The vertical  dashed line indicates the  true time delay.}
\label{example1_microlensing_model}
\end{figure}

To correct  this effect, we  fit the microlensing model with a third-order polynomial regression ($m=3$). Both  $\log(L_{\textrm{prof}}(\Delta))$ and $L_{\textrm{prof}}(\Delta)$ are plotted in Figure~\ref{example1_microlensing_model}. One mode clearly dominates $L_{\textrm{prof}}(\Delta)$. Using a uniform prior for $\Delta$ over its feasible range and setting $\sigma^2 \sim \textrm{IG}(1, 2/10^{7})$, we initialize three MCMC chains near  $\hat{\Delta}_\textrm{mean}=6.36$ days.  It took 14,457 seconds to map $L_{\textrm{prof}}(\Delta)$ and 5,115 seconds on average for each MCMC chain. The profile  likelihood and marginal posterior near the dominant mode are almost identical and are consistent with the true value of $\Delta$ as shown in the third panel of Figure~\ref{example1_microlensing_model}. %They are almost identical and are both consistent with the blinded true value of $\Delta$ (marked with a vertical red line). 
%\textcolor{red}{The largest Gelman-Rubin diagnostic statistic among the parameters equals  1.0003, and the  ESS of $\Delta$ in each of three post burn-in chains is 15,211 on average.}

%\begin{figure}[b!]
%\includegraphics[scale = 0.25]{example1_data_micro_adjusted.png}
%\caption[]{}
%\label{example1_data_subtracted}
%\end{figure}

%The value in the parenthesis indicates the asymptotic uncertainty based on the second-derivative of $L_{\textrm{prof}}(\Delta)$ at the MLE.
\begin{table*}[b!]
\caption{Estimates of $\Delta$; the profile likelihood estimates, $\hat{\Delta}_{\textrm{mean}}$ and $\hat{V}^{0.5}$ are given in the E$(\Delta\vert \boldsymbol{x}, \boldsymbol{y})$ and SD$~\equiv SD(\Delta\vert \boldsymbol{x}, \boldsymbol{y})$ columns, where  $Error\equiv\vert \Delta_{\textrm{true}}- E(\Delta\vert \boldsymbol{x}, \boldsymbol{y})\vert$ with $\Delta_{\textrm{true}}$ indicating the true time delay (5.86 days), and $\chi \equiv Error/SD(\Delta\vert \boldsymbol{x}, \boldsymbol{y})$.}
\label{double1res_new}
\begin{tabular}{ccccccc}
\hline
Method & E$(\Delta\vert D_{\textrm{obs}})$  & $\hat{\Delta}_{\textrm{MLE}}$ & SD &  $\Delta_{\textrm{true}}$ & Error & $\chi$\\
\hline
%:
%:
Bayesian & 6.33 & & 0.28  & 5.86 & 0.47 & 1.68\\
Profile likelihood & 6.36 & 6.35 & 0.28  &  5.86 & 0.50 & 1.79\\
\hline
\end{tabular}
\end{table*}
%Profile likelihood (Mode) &  & 0.31 &  45.85 & 0.66 & 4.53\\

In the second panel of Figure \ref{example1_data}, we combine two light curves by shifting light curve $B$ by the posterior mean of $\Delta$  in the horizontal  axis and by subtracting the estimated polynomial regression based on the posterior means of $\boldsymbol{\beta}$ from light curve $B$. The microlensing model finds matches between the intrinsic fluctuations of the light curves after removing  the relative microlensing trend from light curve $B$. We also plot the posterior sample of $\boldsymbol{X}(\boldsymbol{t}^{\Delta})$ in gray in the right panel of Figure~\ref{example1_data}. The gray regions represent the point-wise prediction intervals for the combined latent light curve. The gray areas encompass most of the  combined observed light curve, indicating that the fitted model predicts the observed data well.

%The difference between the adjusted observed light curve ($2n$ observations), e.g., in the second panel of Figure~7 of the previous manuscript,  and this latent light curve can be considered as $2n$ residuals. In the revised manuscript, we have plotted this latent light curve obtained at each iteration within our Gibbs sampler in gray color. Figure \ref{fig1} below is based on the data of Example~1; the posterior samples of latent magnitudes denoted by gray areas encompass most of the observed light curves. Since the gray areas represent our model predictions, we can check how our model describes the observed light curves well.   We also check the residuals defined as observed magnitudes minus corresponding latent magnitudes to summarize the model fit; the proportion of the standardized residuals outside [-2, 2] was 6\%, and that outside [-3, 3] was 1.3\%, which is not perfect but fairly reasonable. We have added these gray areas around the adjusted observed light curve in the second panel of Figure~7, in the third panel of Figure~13, and in the third panel of Figure~14 of the revised manuscript. 

%matches between the intrinsic fluctuations of the light curves after accounting for the additional extrinsic long-term variability via the 3-order polynomial regression. 

We summarize the Bayesian and profile likelihood estimates for $\Delta$ in Table~\ref{double1res_new}. The  true delay is within  two posterior standard deviation of the posterior mean; similar accuracy is obtained with the  profile likelihood approximation. This is anecdotal evidence that our  model works well when  microlensing is properly accounted for; there is no severe multi-modality near edges of the range of $\Delta$ in the second panel of Figure~\ref{example1_microlensing_model}.
% ($\chi=\vert \Delta_{\textrm{true}}- E(\Delta\vert \boldsymbol{x_t}, \boldsymbol{y})\vert/SD(\Delta\vert \boldsymbol{x_t}, \boldsymbol{y})\le 1.76$)
% to the posterior mean (46.26) and standard deviation (0.40) lead to almost the same result.
% additionally with the approximated posterior mode (46.51).

\begin{table*}[t!]
\caption{Coverage estimates calculated from 1,000 simulated data sets; we generate these simulations using \eqref{lik1}, \eqref{lik2}, and \eqref{conditionalOU1} given the posterior median values of $\{\Delta, \boldsymbol{\beta}, \mu, \sigma^2, \tau\}$ as generative  values. After fitting our Bayesian model on each simulation, we check the proportion of interval estimates containing the generative  values.}
\label{double_coverage}
\begin{tabular}{ccccccccc}
\hline
 & $\Delta$  & $\beta_0$ & $\beta_1$ &  $\beta_2$ & $\beta_3$ & $\mu$ & $\sigma^2$ & $\tau$\\
\hline
Coverage estimate & 1.000 & 0.996 & 0.997 & 0.994 & 0.994 & 0.959 & 0.336 & 0.922\\
\hline
\end{tabular}
\end{table*}

\begin{figure}[b!]
\includegraphics[scale = 0.32]{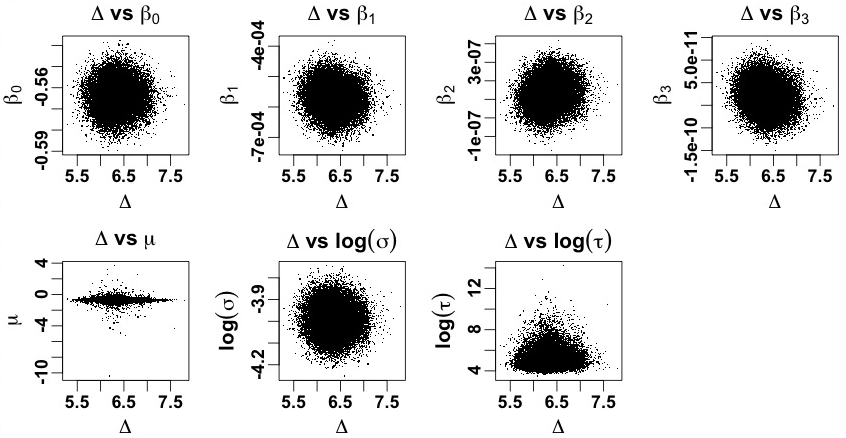}
\caption[]{Scatter plots of the posterior sample of $\Delta$ and each of the other model parameters.}
\label{fig_corr}
\end{figure}

We also conduct a simulation study, generating 1,000 datasets from our final model with an adjustment for microlensing ($m=3$), and report the frequency coverage of the 95\% posterior intervals \citep{tak2016b}. The result is over-coverage for $\Delta$, conservatively meeting the spirit of the frequentist confidence level, reasonable coverage for $\boldsymbol{\beta}$, and under-coverage for both $\sigma^2$ and $\tau$; see Table~\ref{double_coverage}. The severe under-coverage for $\sigma^2$ does not seem to affect the coverage rate of $\Delta$; the scatter plot of $\Delta$ and $\log(\sigma)$ in Figure~\ref{fig_corr}  indicates that the two parameters are almost independent a posteriori. In a numerical sensitivity analysis in Appendix~\ref{sec_sensitivity}, we  show that the posterior mode of $\Delta$ is close to the true time delay even when the posterior mode of $\log(\sigma)$  is substantially different from its true value. (See Figure~\ref{fig_sense_sigma} in Appendix~\ref{sec_sensitivity}.)
%Figure~\ref{fig_sense_sigma} in 

Figure~\ref{fig_corr} displays scatterplots of the posterior sample of $\Delta$ against each of the other model parameters. The time delay $\Delta$ exhibits weak correlations with the regression coefficients and  non-linear relationships with $\mu$ and $\log(\tau)$, though $\beta_0$ and $\log(\sigma)$ appear nearly independent of $\Delta$.
%(three top-right panels) 
%because it can be evidence and justification for the independent prior distribution of $\Delta$ in \eqref{priordeltac}. Figure
%Finally, we check the correlations between posterior sample of $\Delta$ and that of each of the other model parameters in . It shows that $\Delta$ has 

%The profile likelihood can be used as evidence of microlensing. 
%the modes near margins are overwhelmed  Note that the third-order regression in the model makes 

\subsection{A quadruply-lensed quasar simulation}\label{quad1}

\begin{figure}[t!]
\includegraphics[scale = 0.249]{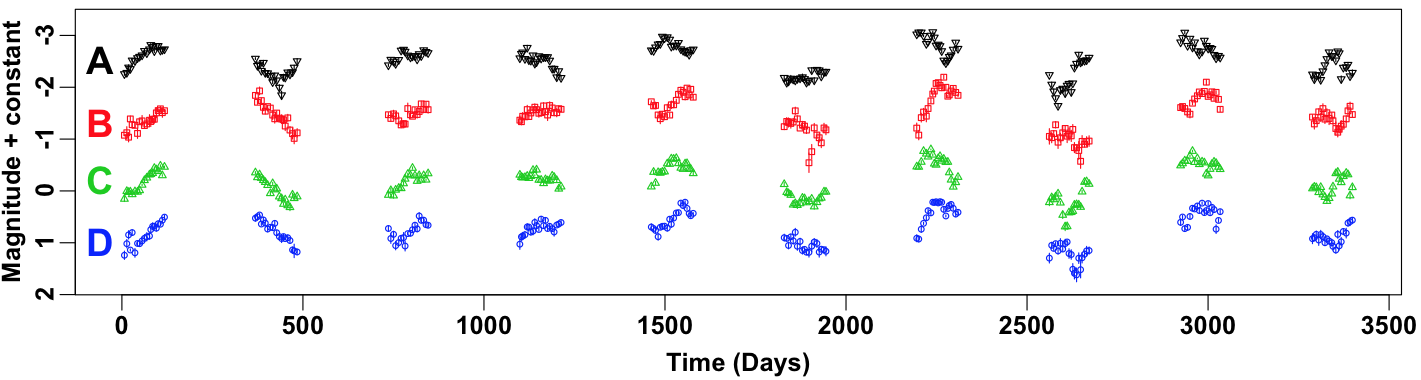}
\caption[]{Simulated quadruply-lensed quasar data used in the TDC.}
\label{fig12}
\end{figure}

The simulated data for a quadruply-lensed quasar are plotted in Figure~\ref{fig12} and are composed of four light curves, $A, B, C$, and $D$; the median cadence is 6 days, the cadence standard deviation is 1 day,  observations are made for 4 months in each of 10 years with 200 observations in total, and  measurement errors are heteroskedastic Gaussian. The feasible range for each $\Delta$ is $[-3391.62,~ 3391.62]$.

%based on the un-thinned whole chains
% For
%a quadruply lensed system, there are four light curves sharing an
%underlying latent process, and thus three delay parameters.  As
%presently formulated, the algorithm presented here can only analyze
%pairs of light curves, independently, so there are six possible
%choices of pairs to analyze.  Since there are only three ``real"
%time delays, the authors will focus on a subset of three pairwise
%delays.

With quadruply lensed data there are three time delay parameters because the four light curves are generated by one underlying process. Our model, however, is designed to analyze pairs of light curves independently, and thus we focus on $\Delta_{\textrm{AB}}$, $\Delta_{\textrm{AC}}$, and $\Delta_{\textrm{AD}}$, where the subscripts index the two light curves being compared, among the six possible pairs. This pair-wise approach  proceeds by applying the method developed for doubly-lensed data in Section~\ref{case1} to  the pair of light curves corresponding to each of $\Delta_{\textrm{AB}}$, $\Delta_{\textrm{AC}}$, and $\Delta_{\textrm{AD}}$ in turn \citep{fassnacht1999}.  %because only time delay estimates and uncertainties are needed in the end for the following astrophysical probes.

By focusing on pairwise comparisons of the four time series, we do not account for the correlations between the time delays.  A coherent model would consider all four light curves in a single  model simultaneously \citep{hojjati2013robust, tewes2013a}; the four light curves are generated from one latent process and the three distinct time delays may have  a posteriori  correlations. It is conceptually straightforward, but properly modeling quadruply lensed data would involve  three time delays, 12 regression coefficients ($m=3$), and three O-U parameters. Extending our model to simultaneously consider  all of the data  is a topic for future research.
%We have not accounted for the correlations between time delays here, leaving it for a future research.
% because  the increased number of parameters and their correlations in a coherent model make its implementation challenging. 

\begin{figure}[t!]
\includegraphics[scale = 0.222]{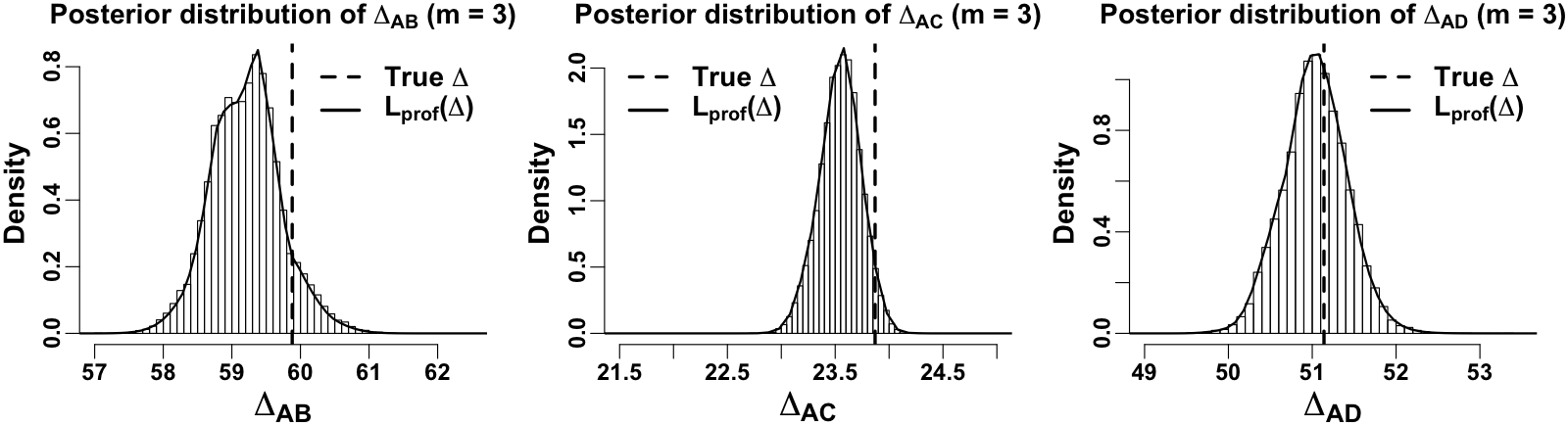}
\caption[]{The marginal posterior distributions  of $\Delta_{\textrm{AB}}$ (first panel), $\Delta_{\textrm{AC}}$ (second panel), and $\Delta_{\textrm{AD}}$ (third panel)  with re-normalized $L_{\textrm{prof}}(\Delta)$ superimposed.  Vertical  dashed lines indicate blinded true time delays.}
\label{fig8}
\end{figure}

%The ESS of $\Delta_{\textrm{AB}}$ is 4,570, that of $\Delta_{\textrm{AC}}$ is 2,044, and that of $\Delta_{\textrm{AD}}$ is 3,917.
% our default strategy to analyze these simulated data
We analyze the quadruply lensed simulated light curves, using the microlensing model ($m=3$). After  confirming a single dominating mode in  the profile likelihood  for each time delay parameter,  we initiate three MCMC chains near this mode. The posterior distributions of $\Delta_{\textrm{AB}}$, $\Delta_{\textrm{AC}}$, and $\Delta_{\textrm{AD}}$ appear in Figure~\ref{fig8} with $L_{\textrm{prof}}(\Delta)$ superimposed. The profile likelihood is almost identical to the posterior distribution of each parameter and both estimate the true time delays well. The average CPU time taken to map $L_{\textrm{prof}}(\Delta)$ is about 73,000 seconds (averaging over the three time delays). The average CPU time taken for each MCMC chain is about 5,500 seconds (averaging over nine chains; three chains for each time delay).  Our estimation results are summarized in Table~\ref{quad1res}. The Bayesian estimates and profile likelihood approximations are quite similar and both produce estimates within two standard deviations of the truth.
% (all $\chi\le1.74$) 
%; \textcolor{red}{the largest Gelman-Rubin diagnostic statistic among  the parameters in three pair-wise cases  is 1.0002 and the average ESS in each post burn-in chain is 18,930 for $\Delta_{\textrm{AB}}$, 15,191 for $\Delta_{\textrm{AC}}$, and 19,419 for $\Delta_{\textrm{AD}}$}. 
%The standard error derived from the second derivative of the profile likelihood at its mode is numerically unstable and sometimes too small.

\begin{table*}[b!]
\caption{Estimates of $\Delta_{\textrm{AB}}$, $\Delta_{\textrm{AC}}$, and $\Delta_{\textrm{AD}}$; the profile likelihood estimates, $\hat{\Delta}_{\textrm{mean}}$ and $\hat{V}^{0.5}$ are given in the E$(\Delta\vert \boldsymbol{x}, \boldsymbol{y})$ and SD$~\equiv SD(\Delta\vert \boldsymbol{x}, \boldsymbol{y})$ columns, where  $Error\equiv\vert \Delta_{\textrm{true}}- E(\Delta\vert \boldsymbol{x}, \boldsymbol{y})\vert$ with $\Delta_{\textrm{true}}$ indicating the true time delay, i.e., $\Delta_{\textrm{AB}}=59.88$, $\Delta_{\textrm{AC}}=23.87$ and $\Delta_{\textrm{AD}}=51.14$, and $\chi \equiv Error/SD(\Delta\vert \boldsymbol{x}, \boldsymbol{y})$.}
\label{quad1res}
\begin{tabular}{cccccccc}
\hline
&Method & E$(\Delta\vert \boldsymbol{x}, \boldsymbol{y})$  & $\hat{\Delta}_{\textrm{MLE}}$ & SD &  $\Delta_{\textrm{true}}$ & Error & $\chi$\\
\hline
\multirow{2}{*}{$\Delta_{\textrm{AB}}$} & Bayesian &  59.21 & & 0.51 & 59.88 &  0.67 & 1.33\\
& Profile likelihood &  59.21 & 59.38 & 0.51 & 59.88 &  0.67 & 1.33\\
\\
\multirow{2}{*}{$\Delta_{\textrm{AC}}$} & Bayesian & 23.55 & & 0.19 & 23.87 & 0.32 & 1.68\\
& Profile likelihood  &23.54 & 23.58 & 0.19 &  23.87 & 0.33 & 1.74\\
\\
\multirow{2}{*}{$\Delta_{\textrm{AD}}$} & Bayesian &  51.04 & & 0.38 & 51.14 & 0.10 & 0.26\\
 & Profile likelihood  &  51.03 & 51.08 & 0.38 &  51.14 & 0.11 & 0.29\\
\hline
\end{tabular}
\end{table*}

 %The value in the parenthesis indicates the asymptotic uncertainty based on the second-derivative of $L_{\textrm{prof}}(\Delta)$ at $\hat{\Delta}_{\textrm{MLE}}$.
%& Profile likelihood (Mode) &  59.36 & 0.02 & 59.88 &  0.52 & 451.00\\
%& Profile likelihood (Mode) &23.58 & 0.07 &  23.87 & 0.29 & 15.26\\
%  & Profile likelihood (Mode) &  51.09 & 0.05 &  51.14 & 0.05 & 1.00\\

\subsection{Quasar \emph{Q0957+561}}\label{sec7}

\begin{figure}[t!]
\includegraphics[scale = 0.248]{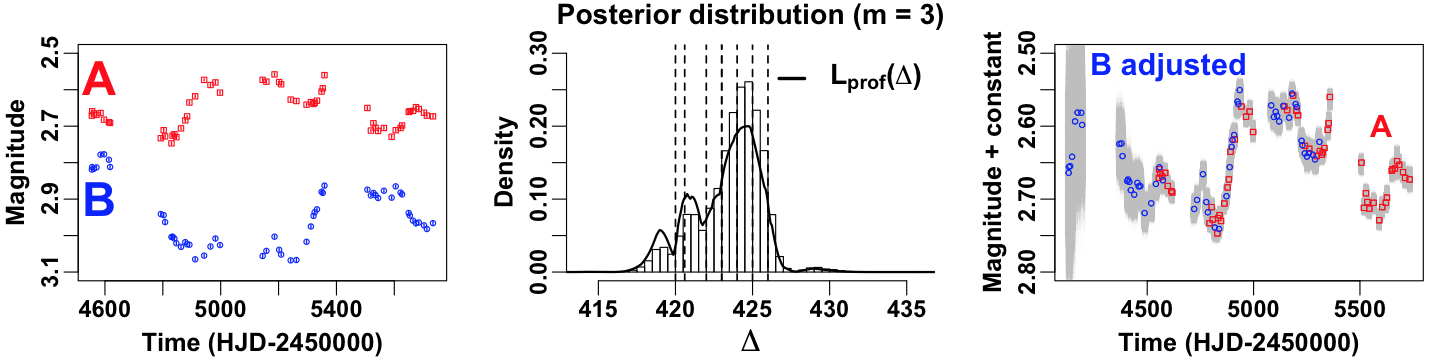}
\caption[]{Observations of Quasar Q0957+561 from \citet{hainline2012new} are plotted in the first panel. The second panel exhibits the marginal posterior distribution of $\Delta$ with re-normalized $L_{\textrm{prof}}(\Delta)$ superimposed.  The vertical dashed lines represent the historical estimates given in Table~\ref{history1} which are concentrated near the highest and the second highest modes.  In the third panel, we combine the two light curves by shifting light curve $B$ by E$(\Delta\mid \boldsymbol{x}, \boldsymbol{y})$ in the  horizontal  axis and by subtracting the estimated third-order polynomial regression based on E$(\boldsymbol{\beta}\mid \boldsymbol{x}, \boldsymbol{y})$. The gray regions represent the point-wise prediction intervals for the combined latent light curve, i.e., the posterior sample of $\boldsymbol{X}(\boldsymbol{t}^{\Delta})$. The gray areas encompass most of the observed light curve, which shows how well the fitted model  predicts the observed data.  HJD indicates the Heliocentric Julian date.}
\label{example3_dataprofpost}
\end{figure}

The first known gravitationally (doubly) lensed quasar \emph{Q0957+561} was discovered by \citet{walsh1979q0957} who suggested that a strong gravitational lensing may have formed the two images. Here we analyze the most recent observations of this quasar. These observations were made by the United States Naval Observatory in 2008--2011 \citep{hainline2012new}. The data were observed on 57 nights and are plotted in the first panel of Figure~\ref{example3_dataprofpost}. The feasible range for $\Delta$ is $[-1178.939, 1178.939]$.
% based on the un-thinned whole chains

Inspection of $L_{\textrm{prof}}(\Delta)$ reveals four modes close to each other near 425 days.  Using a uniform prior distribution  of $\Delta$ over its range and $\sigma^2\sim \textrm{IG}(1, 2/10^7)$, we  ran three MCMC chains near the highest mode. The second panel of Figure~\ref{example3_dataprofpost} shows the marginal posterior distribution of $\Delta$ with $L_{\textrm{prof}}(\Delta)$ superimposed. Here the profile likelihood approximation to the marginal posterior distribution of $\Delta$ is less accurate. This may be because the approximation depends on an asymptotic argument while the data size is small. Nonetheless the profile likelihood identifies each of the dominant modes of the posterior distribution. Mapping  $L_{\textrm{prof}}(\Delta)$ took 1,243 seconds and each MCMC chain took on average 1,955 seconds.
%; \textcolor{red}{the largest Gelman-Rubin diagnostic statistic among  the parameters is 1.0000, and the average ESS of $\Delta$ in each post burn-in chain is 7,707.} 
%In this case the dominant mode of $\Delta$ has more complex structure as detected by both $L_{\textrm{prof}}(\Delta)$ and the MCMC sample; see the second panel of Figure~\ref{example3_dataprofpost}.

 In the third panel of Figure~\ref{example3_dataprofpost}, we shift light curve $B$ by the posterior mean of $\Delta$  in the horizontal  axis and subtract the estimated third-order polynomial regression based on the posterior means of $\boldsymbol{\beta}$. The fitted microlensing model  matches the intrinsic fluctuations of the two light curves well. We also  plot the posterior sample of $\boldsymbol{X}(\boldsymbol{t}^{\Delta})$ in gray  which represents the point-wise prediction intervals for the latent light curves.  The gray areas encompass most of the observed light curve, which shows how well the fitted model predicts the observed data. 
%on top of the combined light curve,

%, and 40 data points observed from February to  with their average measurement error equal to 0.02 magnitudes
% Our work provides the posterior mean (423.69) and standard deviation (2.02) of $\Delta$ and profile likelihood  approximations to them (423.21 $\&$ 2.81). 
\begin{table*}[t!]
\caption{Historical time delay estimates ($\hat{\Delta}$) and standard errors (SE) for Q0957+561 ($r$-band). We compute the posterior mean and standard deviation of $\Delta$ ($423.71\pm2.03$); the profile likelihood approximate the posterior mean and standard deviation as $423.21\pm 2.81$. \cite{pelt1996}, \cite{oscoz1997time, oscoz2001time},  \cite{serra1999bvri}, and \cite{shalyapin20125} adopted various methods to estimate $\Delta$ using different data sets spanning different periods. We report the average measurement  standard deviation (SD)  of their data; two average measurement SDs are reported if their data come from two sources. In all cases except our method, a bootstrapping method was used to calculate the SE.}
\label{history1}
\begin{tabular}{cccccc}
\hline
\multirow{2}{*}{Researchers} & Number of & Observation  & Measurement & \multirow{2}{*}{$\hat{\Delta}$} & \multirow{2}{*}{SE} \\
 & observations &  period & SD (mag.)&&\\
\hline
\cite{pelt1996} & 831 & 1979--1994 & 0.0159&423 & 6\\\\
%\hline
%Kundi$\acute{\textrm{c}}$ et al. (1997) & Dispersion & 417 & 3\\
%g-band
\cite{oscoz1997time} & 86 & 1994--1996  & 0.01, 0.02 &424  & 3\\\\
% &  & \\\\
%Pelt. et al. (1998) & A variation of Dispersion method & 416.3 & 1.7\\
%\hline
%R-band
Serra-Ricart et al. & \multirow{2}{*}{197}  & \multirow{2}{*}{1996--1998} & \multirow{2}{*}{0.023, 0.025}&\multirow{2}{*}{425} & \multirow{2}{*}{4}\\
(1999) &   & & & & \\\\
% & via auto-\& cross-correlation &\\\\
%\hline
%Colley \& Schild (2000) & & 417.4 & N/A\\
%R-band
\multirow{4}{*}{\cite{oscoz2001time}} & \multirow{4}{*}{100}& \multirow{4}{*}{1994--1996} & \multirow{4}{*}{0.009, 0.01} &426 & 5\\
 & & &&423 & 2\\
  & & &&420 & 8\\
  &  & &&422 & 3\\\\
%Shalyapin et al. (2008) & 142  & 2005--2007 &&420.6 & 1.9\\\\

Shalyapin et al. & \multirow{2}{*}{371}  & \multirow{2}{*}{2005--2010} &\multirow{2}{*}{0.012}& \multirow{2}{*}{420.6} & \multirow{2}{*}{1.9}\\
(2012) &   &  && & \\\\

\multirow{2}{*}{This work}  & \multirow{2}{*}{57} &  \multirow{2}{*}{2008--2011} &\multirow{2}{*}{0.004}&423.71 & 2.03\\
&&   && 423.21 & 2.81\\
\hline
\end{tabular}
\end{table*}

% (For reference, with a curve-shifted model, we observed several modes near margins as large as the mode near 423 days, indicating a sign of microlensing, see Appendix \ref{appendix_micro}.) 
%& Profile likelihood (Mode) & 423.05 & 0.01\\
%The posterior distribution and the profile likelihood have similar multimodality. 

% Although we use the smallest number of observations, our estimates are broadly consistent to others possibly because the data we used are the most accurate (the smallest average measurement error). In the revised manuscript, 

%are measured most accurately  in that they 
Estimates based on different observations of Q0957+561 appear in  Table~\ref{history1}.   Though the posterior mean and standard deviation may be difficult to interpret with a multimodal posterior distribution, we include them for comparison. Our estimates are broadly consistent with the others.     We emphasize here that our methods reveal several possible time delays in that there are  four modes in the marginal posterior distribution of $\Delta$, whereas previous analyses  report only a single point estimate for $\Delta$ and its standard error. By investigating the entire posterior distribution, we  learn that  previous estimates, denoted by  vertical dashed lines in the second panel of Figure~\ref{example3_dataprofpost}, are located near the highest and the second highest modes of our marginal posterior distribution of $\Delta$. Thus, our approach is more informative in that it provides a summary of several possible values of $\Delta$ and their relative likelihoods, corresponding to locations and sizes of the several modes of the posterior distribution of~$\Delta$.
%Although our data have the smallest number of observations, our estimates are broadly consistent to the others possibly because our data have the smallest average measurement error.}
%and they are  difficult to compare,

\subsection{Quasar J1029+2623}\label{real2}

\begin{figure}[b!]
\includegraphics[scale = 0.249]{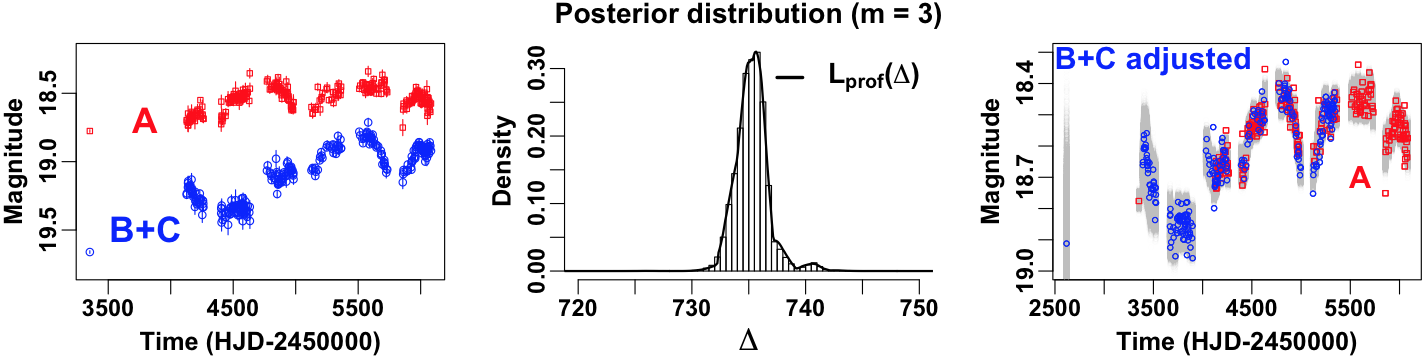}
\caption[]{We plot the observations of Quasar J1029+2623  from \cite{fohlmeister2013two}  in the first panel. The second panel exhibits the marginal posterior distribution of $\Delta$ with re-normalized $L_{\textrm{prof}}(\Delta)$ superimposed. In the last panel, we combine the two light curves by shifting light curve $B+C$ by E$(\Delta\mid\boldsymbol{x}, \boldsymbol{y})$ in the horizontal axis and by subtracting the estimated third-order polynomial regression based on E$(\boldsymbol{\beta}\vert \boldsymbol{x}, \boldsymbol{y})$. The gray areas represent the range of the point-wise prediction intervals for the latent light curve. Comparing the gray regions with the observed data  shows how well the  model fits the observed data.  HJD indicates the Heliocentric Julian date.}
\label{example4_dataprofpost}
\end{figure}

\cite{inada2006sdss} discovered the gravitationally lensed quasar \emph{J1029+2623} whose estimated time delay is the second largest yet observed. Though \emph{J1029+2623} has three images (\emph{A}, \emph{B}, and \emph{C}), \cite{fohlmeister2013two} merged \emph{B} and \emph{C} because they overlap significantly and thus their light curves are difficult to disentangle. They  published its data (\emph{A}, \emph{B+C}) with 279 epochs monitored at the Fred Lawrence Whipple Observatory from January 2007 to June 2012. The time delay estimate obtained by analyzing the combined image can be different from that obtained by analyzing images \emph{B} or \emph{C} separately. Nonetheless, we follow \cite{fohlmeister2013two} in order to provide a fair comparison. The first panel in Figure~\ref{example4_dataprofpost} shows these data. The feasible range of $\Delta$ is $[-2729.759,~ 2729.759]$.

% based on the un-thinned whole chains
%The ESS of $\Delta$ is 4,988. 
%The third panel exhibits the marginal posterior distributionof $\Delta$ with normalized profile likelihood curves superimposed. HJD indicates the Heliocentric Julian date.

%as the Bayesian analysis does}.
We confirm a dominant mode near 735 days  and invisibly small modes near -2,000 and 1,800 days via $L_{\textrm{prof}}(\Delta)$. Since the mode near 735 days  overwhelms the other modes, we focus on the dominant mode. We initiated three MCMC chains near 735 days using  a uniform prior distribution of $\Delta$ over its range and $\sigma^2\sim \textrm{IG}(1, 2\times10^{-7})$.  We display the marginal posterior distribution of $\Delta$ in the second panel of Figure~\ref{example4_dataprofpost} with $L_{\textrm{prof}}(\Delta)$ superimposed. Mapping $L_{\textrm{prof}}(\Delta)$  took 33,683 seconds  and each MCMC chain took an average of 8,555 seconds. The posterior distribution and the profile likelihood are almost identical. In the third panel, we shift light curve $B$ by the posterior mean of $\Delta$  in the horizontal  axis and subtract the estimated third-order polynomial regression based on the posterior mean of $\boldsymbol{\beta}$. Again, the fitted microlensing model is a good match of the two light curves and our graphical model checking shows that the range of our predicted values for the combined latent light curve, denoted by the gray areas,  encompasses the observed light curve well.
%\textcolor{red}{The largest Gelman-Rubin diagnostic statistic among the parameters is 1.0006, and the average ESS of $\Delta$ in each post burn-in chain is 15,151.}
%and thus we ran three MCMC chains near the dominant mode. 
% HJD indicates the Heliocentric Julian date
%predicted values for the combined light curve, i.e., the posterior sample of $\boldsymbol{X}(\boldsymbol{t}^{\Delta})$, which 

%The third panel shows the marginal posterior distribution of $\Delta$ as a histogram of the MCMC samples. We draw  the profile likelihood over the posterior distribution after adjusting the normalizing constant of the profile likelihood to emphasize their similar shapes. The vertical red dashed line indicates the blinded true time delay.

\begin{table*}[t!]
\caption{Historical time delay estimates and 90\% confidence intervals for J1029+2623. Our work provides the posterior mean and 90\% posterior interval of $\Delta$ and profile likelihood  approximations to them. \cite{fohlmeister2013two} did not specify how they produced the sampling distribution of $\Delta$. \cite{kumar2014h0} used a parametric bootstrapping method.  }
\label{history2}
\begin{tabular}{cccc}
\hline
Researchers & Method & Estimate & $90\%$ Interval\\
\hline
%Fohlmeister et al. (2007) & $\chi^2$-minimization & 746 & $\pm6$ (C.I.)\\
% & Dispersion method & 745 & $\pm10$ (C.I.)\\\\
\cite{fohlmeister2013two} & $\chi^2$-minimization (AIC, BIC) & 744 & (734, 754) \\\\
Kumar, Stalin, and & \multirow{2}{*}{Difference-smoothing}  & \multirow{2}{*}{743.5} & \multirow{2}{*}{(734.6, 752.4)}\\
Prabhu (2014) & & & \\\\
\multirow{2}{*}{This work}  & Bayesian  & 735.30 & (733.09, 737.62)\\
& Profile likelihood & 733.11 & (732.94, 738.44)\\
\hline
\end{tabular}
\end{table*}

In Table~\ref{history2} we compare our estimates with historical estimates that are based on the same data. The Bayesian method uses 5\% and 95\% quantiles of the posterior samples of $\Delta$ as the 90\% posterior interval. To obtain the 90\% interval estimate for $\Delta$ via the profile likelihood, we draw a sample of size 50,000 of $\Delta$ using the empirical CDF of the normalized profile likelihood and  report the 5\% and 95\% quantiles.

The shape of the posterior distribution of $\Delta$ is almost identical to that of the profile likelihood in the second panel of Figure~\ref{example4_dataprofpost}. However, the posterior mean of $\Delta$ is larger than the profile approximation, $\hat{\Delta}_{\textrm{mean}}$,  by about two days. This is because of invisibly small modes near -2,000 and 1,800 days. %Since the mode near 735 days  overwhelms} the other modes, it is reasonable to focus on  it as the Bayesian analysis does}. 

%; see the second panel of Figure~\ref{example4_post}.  
%\begin{figure}[t!]
%\includegraphics[scale = 0.251]{example4_post4.png}
%\caption[]{The profile log likelihood (left) and the profile likelihood (right) of $\Delta$ over its feasible range under the microlensing model ($m=3$). Although the profile likelihood shows a dominant mode near 735 days, there are also invisibly} small modes near -2,000 and 1,800 days. }
%\label{example4_post}
%\end{figure}

Overall, our point estimates are smaller than the historical estimates by about ten days and our 90\% posterior intervals are much shorter than the  historical 90\%  confidence intervals in Table~\ref{history2}. We suspect that the discrepancy between the estimates might arise from the overly simple microlensing models used in the historical analyses. \cite{fohlmeister2013two}, for example, combine the output obtained from fitting two different models, one with a linear model for the microlensing polynomial (which was optimal with respect to AIC) and the other with no adjustment for  microlensing  (which was optimal with respect to BIC). Unfortunately, they do not describe \emph{how} they combine the fits. Since they also apply high-order splines for each season in addition to the microlensing polynomial, neither of their models are directly comparable to ours. \cite{kumar2014h0}, on the other hand, account for the microlensing by using a Gaussian kernel smoothing technique. 

% model.  factors: (i) Different choice of point estimates and (ii) different methods for accounting for microlensing. The $\chi^2$ minimization of \cite{fohlmeister2013two} involves high-dimensional optimization to compute the point estimate. They also combined the output obtained from fitting two different models, one with a linear model for the microlensing polynomial (which was optimal with respect to AIC) and the other with no adjustment for  microlensing  (which was optimal with respect to BIC). Unfortunately, they do not describe \emph{how} they combine the fits. Since they also applied high-order splines for each season in addition to the microlensing polynomial, neither of their models are directly comparable to our model with $m=1$ or $m=0$. \cite{kumar2014h0}'s method also involves high-dimensional optimization in the computation of their point estimate. They account for the microlensing by using a smoothing technique based on a Gaussian kernel. }
% The reason that the estimates of \cite{kumar2014h0} and  \cite{fohlmeister2013two} are similar may be  that \cite{kumar2014h0} initialized their optimization method near the  estimate of \cite{fohlmeister2013two}; \cite{kumar2014h0} do not specify whether they try multiple starting values for their optimization method.

\begin{figure}[t!]
\includegraphics[scale = 0.395]{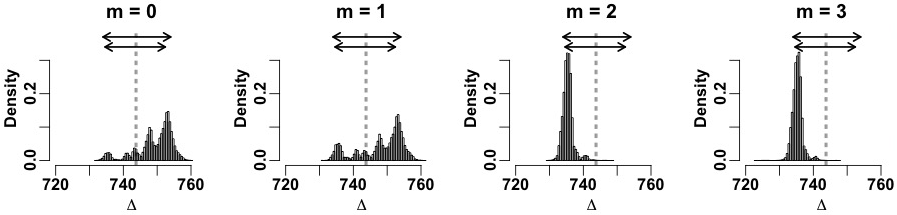}
\caption[]{The marginal posterior distributions of $\Delta$ with $m=0, 1, 2$, and $3$. The 90\% intervals of  \cite{fohlmeister2013two} and \cite{kumar2014h0} are denoted by the lengths of the arrows at the top of each panel and cover all the modes of the posterior distributions of $\Delta$ shown in the first ($m=0$) and second ($m=1$) panels. This implies that their microlensing models might have produced  results similar to  our microlensing model with either $m=0$ or ${m=1}$. Their point estimates are denoted by vertical dashed lines. As we increase the order of the polynomial regression for microlensing, the severe multi-modality dissipates and one mode becomes dominant.  This implies why our point and interval estimates are quite different from theirs.}
\label{example4_micro}
\end{figure}

% have worked similarly to
% and are near the third highest mode when $m=0$ and the smallest mode when $m=1$.
Figure~\ref{example4_micro}  shows our marginal posterior distribution of $\Delta$ with $m=0, 1, 2$, and $3$, respectively. The 90\% intervals of  \cite{fohlmeister2013two} and \cite{kumar2014h0} denoted by the lengths of the arrows at the top of each panel cover all the modes of the posterior distributions of $\Delta$ shown in the first ($m=0$) and second ($m=1$) panels.  This implies that their microlensing models might have produced  results similar to our microlensing model with either $m=0$ or ${m=1}$. As we increase the order of the polynomial regression for microlensing, the severe multi-modality dissipates and one mode becomes dominant.  Thus, the discrepancy between their estimates and ours with $m=3$  might be due to their use of an overly simple microlensing model.

\section{Concluding remarks}\label{disc} Accurately estimating time delays among  gravitationally lensed quasar images is a key to making fundamental measurements of the current expansion rate of the Universe and  dark energy \citep{ref1964, Linder2011}.  The Large Synoptic Survey Telescope \citep{lsst2009lsst} will produce extensive time series data on thousands of multiply lensed quasars starting in 2022.  Anticipating this era of the LSST, we have improved the fully Bayesian model of Harva and Raychaudhury (2006) by leveraging recent advances in astrophysical and statistical modeling. We have added an Ornstein-Uhlenbeck process to model the fluctuations in quasar light curves, a polynomial regression to account for microlensing, and a profile-likelihood-guided Bayesian strategy.
%Accurately estimating time delays among the gravitationally lensed quasar images is a key goal for astronomical observations aimed at making fundamental measurements of the current expansion rate of the Universe and the dark energy \citep{ref1964, Linder2011}. The upcoming large-scale astronomical survey via the Large Synoptic Survey Telescope \citep[LSST,][]{lsst2009lsst}, the top-ranked  ground-based telescope project in the 2010 Astrophysics Decadal Survey, will start producing extensive time series data on thousands of multiply lensed quasars in 2022.  To improve current existing time delay estimation methods in preparation for the era of the LSST, \cite{dobler2013} organized a blind competition called the Time Delay Challenge (TDC). The TDC motivated us to improve the  fully Bayesian model of \citet[H\&R,][]{harva2006} because their approach unifies parameter estimation and uncertainty quantification into a single coherent analysis based on the posterior distribution of the time delay unlike other existing methods. 

We proposed our original model in the context of the Time Delay Challenge \citep[TDC, ][]{dobler2013, kai2014}. This original model worked well for data without severe microlensing, leading to the best precision and targeted average bias level among the methods submitted to the TDC. Our original model, however, did not properly account for microlensing and thus produced poor time delay estimates in some cases resulting in a mediocre performance in one of the evaluation criteria. This motivated us to develop our current microlensing model.

The upcoming second TDC, called the TDC2, aims to further improve estimation methods  under a more realistic setting. Since the LSST will produce multi-band optical data observed in filters centered at six different wavelengths for each lens system, proper analysis will require jointly modeling a vector of light curves to estimate the common time delay in each system.  Modeling microlensing will be more challenging in TDC2, because its effect depends on the wavelength of the quasar light. 

There are several opportunities to build upon our work in preparation for the TDC2 and eventually for the LSST. It is desirable to implement more sophisticated methods of model selection such as information criteria to choose the complexity of the microlensing trend. Though astrophysicists have used a cubic polynomial trend for microlensing models for some quasars so far, it would be better to have a fast and principled mechanism to determine the order given any data of gravitationally lensed quasars.  Another avenue for further improvement is to constrain the range of the time delay by incorporating additional astrophysical information such as spatial positions of the images relative to the lensing galaxy, and an astrophysical model for the mass distribution of the lens. For quadruply-lensed quasar systems, constructing a Bayesian model to simultaneously analyze the four light curves, would allow us to coherently estimate the relative time delays without loss of information. Further improvements to the computational efficiency of our profile likelihood and MCMC strategies for analyzing extensive vector time series will enhance their effectiveness in the era of the LSST.

%The upcoming large-scale astronomical survey via the , the top-ranked  ground-based telescope project in the 2010 Astrophysics Decadal Survey, will start producing extensive time series data on thousands of multiply lensed quasars in 2022.

% of the LSST.

% This is because the effect of microlensing arises differently according to quasars. Our model does not account for 

\section*{Acknowledgements}
This work was conducted under the auspices of the CHASC International Astrostatistics Center. CHASC is supported by NSF grants DMS 1208791 and DMS 1209232. In addition, HT and XLM acknowledge support from the Harvard Statistics Department, and HT also acknowledges partial support from the NSF under Grant DMS 1127914 to the Statistical and Applied Mathematical Sciences Institute. KM is supported at Harvard by NSF grants AST-1211196 and AST-156854, VLK and AS from a NASA contract to the Chandra X-Ray Center NAS8-03060, and DvD from a Wolfson Research Merit Award (WM110023) provided by the British Royal Society, a Marie-Curie Career Integration Grant (FP7-PEOPLE-2012-CIG-321865) provided by the European Commission, and a Marie-Sklodowska-Curie RISE Grant (H2020-MSCA-RISE-2015-691164) provided by the European Commission. In addition, we thank CHASC members for many helpful discussions and the editor, associate editor, and reviewer for their careful reading and insightful suggestions.

\begin{supplement} 
\label{sup}
\sname{Supplement}
\stitle{R codes and data}
%\slink[doi]{COMPLETED BY THE TYPESETTER}
\sdatatype{Rcode\_data.zip}
\sdescription{This zip file contains all the computer code (Rcode.R) and data (Data.zip) used in this article. An R package, \texttt{timedelay}, that implements the Bayesian and profile likelihood methods is publicly available at CRAN (https://cran.r-project.org/package=timedelay).}
\end{supplement}

\bibliography{bibliography}

\begin{thebibliography}{54}
% BibTex style file: imsart-nameyear.bst, 2013-01-28
% Default style options (sort=1,type=nameyear).
% Used options (sort=1,type=nameyear).

\bibitem[\protect\citeauthoryear{Berger, Bernardo and
  Sun}{2015}]{berger2015overall}
\begin{barticle}[author]
\bauthor{\bsnm{Berger},~\bfnm{J.~O.}\binits{J.~O.}},
  \bauthor{\bsnm{Bernardo},~\bfnm{J.~M.}\binits{J.~M.}} \AND
  \bauthor{\bsnm{Sun},~\bfnm{D.}\binits{D.}}
(\byear{2015}).
\btitle{{Overall Objective Priors}}.
\bjournal{Bayesian Analysis}
\bvolume{10}
\bpages{189--221}.
\end{barticle}
\endbibitem

\bibitem[\protect\citeauthoryear{Berger, Liseo and
  Wolpert}{1999}]{berger1999integrated}
\begin{barticle}[author]
\bauthor{\bsnm{Berger},~\bfnm{J.~O.}\binits{J.~O.}},
  \bauthor{\bsnm{Liseo},~\bfnm{B.}\binits{B.}} \AND
  \bauthor{\bsnm{Wolpert},~\bfnm{R.~L.}\binits{R.~L.}}
(\byear{1999}).
\btitle{Integrated Likelihood Methods for Eliminating Nuisance Parameters}.
\bjournal{Statistical Science}
\bvolume{14}
\bpages{1--28}.
\end{barticle}
\endbibitem

\bibitem[\protect\citeauthoryear{Berk et~al.}{2004}]{berk2004ensemble}
\begin{barticle}[author]
\bauthor{\bsnm{Berk},~\bfnm{D.~E.~V.}\binits{D.~E.~V.}},
  \bauthor{\bsnm{Wilhite},~\bfnm{Brian~C}\binits{B.~C.}},
  \bauthor{\bsnm{Kron},~\bfnm{Richard~G}\binits{R.~G.}},
  \bauthor{\bsnm{Anderson},~\bfnm{Scott~F}\binits{S.~F.}},
  \bauthor{\bsnm{Brunner},~\bfnm{Robert~J}\binits{R.~J.}},
  \bauthor{\bsnm{Hall},~\bfnm{Patrick~B}\binits{P.~B.}},
  \bauthor{\bsnm{Ivezi{\'c}},~\bfnm{{\v{Z}}eljko}\binits{{\v{Z}}.}},
  \bauthor{\bsnm{Richards},~\bfnm{Gordon~T}\binits{G.~T.}},
  \bauthor{\bsnm{Schneider},~\bfnm{Donald~P}\binits{D.~P.}},
  \bauthor{\bsnm{York},~\bfnm{Donald~G}\binits{D.~G.}},
  \bauthor{\bsnm{Brinkmann},~\bfnm{J.~V.}\binits{J.~V.}},
  \bauthor{\bsnm{Lamb},~\bfnm{D.~Q.}\binits{D.~Q.}},
  \bauthor{\bsnm{Nichol},~\bfnm{R.~C.}\binits{R.~C.}} \AND
  \bauthor{\bsnm{Schlegel},~\bfnm{D.~J.}\binits{D.~J.}}
(\byear{2004}).
\btitle{The Ensemble Photometric Variability of $\sim$25,000 Quasars in the
  Sloan Digital Sky Survey}.
\bjournal{The Astrophysical Journal}
\bvolume{601}
\bpages{692}.
\end{barticle}
\endbibitem

\bibitem[\protect\citeauthoryear{Blandford and Narayan}{1992}]{blandford1992}
\begin{barticle}[author]
\bauthor{\bsnm{Blandford},~\bfnm{RD}\binits{R.}} \AND
  \bauthor{\bsnm{Narayan},~\bfnm{R}\binits{R.}}
(\byear{1992}).
\btitle{Cosmological Applications of Gravitational Lensing}.
\bjournal{Annual Review of Astronomy \& Astrophysics}
\bvolume{30}
\bpages{311--358}.
\end{barticle}
\endbibitem

\bibitem[\protect\citeauthoryear{Brooks et~al.}{1997}]{brooks1997finite}
\begin{barticle}[author]
\bauthor{\bsnm{Brooks},~\bfnm{Stephen~P}\binits{S.~P.}},
  \bauthor{\bsnm{Morgan},~\bfnm{Byron~JT}\binits{B.~J.}},
  \bauthor{\bsnm{Ridout},~\bfnm{Martin~S}\binits{M.~S.}} \AND
  \bauthor{\bsnm{Pack},~\bfnm{SE}\binits{S.}}
(\byear{1997}).
\btitle{{Finite Mixture Models for Proportions}}.
\bjournal{Biometrics}
\bvolume{53}
\bpages{1097--1115}.
\end{barticle}
\endbibitem

\bibitem[\protect\citeauthoryear{Brooks et~al.}{2011}]{brooks2011handbook}
\begin{bbook}[author]
\bauthor{\bsnm{Brooks},~\bfnm{Steve}\binits{S.}},
  \bauthor{\bsnm{Gelman},~\bfnm{Andrew}\binits{A.}},
  \bauthor{\bsnm{Jones},~\bfnm{Galin}\binits{G.}} \AND
  \bauthor{\bsnm{Meng},~\bfnm{Xiao-Li}\binits{X.-L.}}
(\byear{2011}).
\btitle{Handbook of Markov Chain Monte Carlo}.
\bpublisher{CRC press}.
\end{bbook}
\endbibitem

\bibitem[\protect\citeauthoryear{Chang and Refsdal}{1979}]{chang1979flux}
\begin{barticle}[author]
\bauthor{\bsnm{Chang},~\bfnm{K}\binits{K.}} \AND
  \bauthor{\bsnm{Refsdal},~\bfnm{S}\binits{S.}}
(\byear{1979}).
\btitle{Flux Variations of QSO 0957+561 A, B and Image Splitting by Stars Near
  the Light Path}.
\bjournal{Nature}
\bvolume{282}
\bpages{561--564}.
\end{barticle}
\endbibitem

\bibitem[\protect\citeauthoryear{{LSST Science
  Collaboration}}{2009}]{lsst2009lsst}
\begin{barticle}[author]
\bauthor{\bsnm{{LSST Science Collaboration}}}
(\byear{2009}).
\btitle{LSST Science Book, Version 2.0}.
\bjournal{arXiv:0912.0201}.
\end{barticle}
\endbibitem

\bibitem[\protect\citeauthoryear{Courbin et~al.}{2013}]{courbin2011}
\begin{barticle}[author]
\bauthor{\bsnm{Courbin},~\bfnm{F}\binits{F.}},
  \bauthor{\bsnm{Chantry},~\bfnm{Virginie}\binits{V.}},
  \bauthor{\bsnm{Revaz},~\bfnm{Y}\binits{Y.}},
  \bauthor{\bsnm{Sluse},~\bfnm{D}\binits{D.}},
  \bauthor{\bsnm{Faure},~\bfnm{C}\binits{C.}},
  \bauthor{\bsnm{Tewes},~\bfnm{M}\binits{M.}},
  \bauthor{\bsnm{Eulaers},~\bfnm{Eva}\binits{E.}},
  \bauthor{\bsnm{Koleva},~\bfnm{M}\binits{M.}},
  \bauthor{\bsnm{Asfandiyarov},~\bfnm{I}\binits{I.}},
  \bauthor{\bsnm{Dye},~\bfnm{S}\binits{S.}},
  \bauthor{\bsnm{Magain},~\bfnm{P.}\binits{P.}}, \bauthor{\bparticle{van}
  \bsnm{Winckel},~\bfnm{H.}\binits{H.}},
  \bauthor{\bsnm{Coles},~\bfnm{J.}\binits{J.}},
  \bauthor{\bsnm{Saha},~\bfnm{P.}\binits{P.}},
  \bauthor{\bsnm{Ibrahimov},~\bfnm{M.}\binits{M.}} \AND
  \bauthor{\bsnm{Meylan},~\bfnm{G.}\binits{G.}}
(\byear{2013}).
\btitle{COSMOGRAIL: the COSmological MOnitoring of GRAvItational Lenses IX.
  Time Delays, Lens Dynamics and Baryonic Fraction in HE 0435-1223}.
\bjournal{Astronomy \& Astrophysics}
\bvolume{536}
\bpages{A53}.
\end{barticle}
\endbibitem

\bibitem[\protect\citeauthoryear{Davison}{2003}]{davison2003statistical}
\begin{bbook}[author]
\bauthor{\bsnm{Davison},~\bfnm{Anthony~Christopher}\binits{A.~C.}}
(\byear{2003}).
\btitle{Statistical Models}.
\bpublisher{Cambridge University Press}.
\end{bbook}
\endbibitem

\bibitem[\protect\citeauthoryear{Dobler et~al.}{2015}]{dobler2013}
\begin{barticle}[author]
\bauthor{\bsnm{Dobler},~\bfnm{Gregory}\binits{G.}},
  \bauthor{\bsnm{Fassnacht},~\bfnm{Christopher}\binits{C.}},
  \bauthor{\bsnm{Treu},~\bfnm{Tommaso}\binits{T.}},
  \bauthor{\bsnm{Marshall},~\bfnm{Phillip~J}\binits{P.~J.}},
  \bauthor{\bsnm{Liao},~\bfnm{Kai}\binits{K.}},
  \bauthor{\bsnm{Hojjati},~\bfnm{Alireza}\binits{A.}},
  \bauthor{\bsnm{Linder},~\bfnm{Eric}\binits{E.}} \AND
  \bauthor{\bsnm{Rumbaugh},~\bfnm{Nicholas}\binits{N.}}
(\byear{2015}).
\btitle{Strong Lens Time Delay Challenge. I. Experimental Design}.
\bjournal{The Astrophysical Journal}
\bvolume{799}
\bpages{168}.
\end{barticle}
\endbibitem

\bibitem[\protect\citeauthoryear{Fassnacht et~al.}{1999}]{fassnacht1999}
\begin{barticle}[author]
\bauthor{\bsnm{Fassnacht},~\bfnm{CD}\binits{C.}},
  \bauthor{\bsnm{Pearson},~\bfnm{TJ}\binits{T.}},
  \bauthor{\bsnm{Readhead},~\bfnm{ACS}\binits{A.}},
  \bauthor{\bsnm{Browne},~\bfnm{IWA}\binits{I.}},
  \bauthor{\bsnm{Koopmans},~\bfnm{LVE}\binits{L.}},
  \bauthor{\bsnm{Myers},~\bfnm{ST}\binits{S.}} \AND
  \bauthor{\bsnm{Wilkinson},~\bfnm{PN}\binits{P.}}
(\byear{1999}).
\btitle{A Determination of H$_o$ with the CLASS Gravitational Lens B1608+ 656.
  I. Time Delay Measurements with the VLA}.
\bjournal{The Astrophysical Journal}
\bvolume{527}
\bpages{498}.
\end{barticle}
\endbibitem

\bibitem[\protect\citeauthoryear{Fischer et~al.}{1997}]{fischer1997}
\begin{barticle}[author]
\bauthor{\bsnm{Fischer},~\bfnm{P.}\binits{P.}},
  \bauthor{\bsnm{Bernstein},~\bfnm{G.}\binits{G.}},
  \bauthor{\bsnm{Rhee},~\bfnm{G.}\binits{G.}} \AND
  \bauthor{\bsnm{Tyson},~\bfnm{J.~A.}\binits{J.~A.}}
(\byear{1997}).
\btitle{The Mass Distribution of the Cluster Q0957+561 from Gravitational
  Lensing}.
\bjournal{The Astronomical Journal}
\bvolume{113}
\bpages{521}.
\end{barticle}
\endbibitem

\bibitem[\protect\citeauthoryear{Fohlmeister et~al.}{2013}]{fohlmeister2013two}
\begin{barticle}[author]
\bauthor{\bsnm{Fohlmeister},~\bfnm{Janine}\binits{J.}},
  \bauthor{\bsnm{Kochanek},~\bfnm{Christopher~S}\binits{C.~S.}},
  \bauthor{\bsnm{Falco},~\bfnm{Emilio~E}\binits{E.~E.}},
  \bauthor{\bsnm{Wambsganss},~\bfnm{Joachim}\binits{J.}},
  \bauthor{\bsnm{Oguri},~\bfnm{Masamune}\binits{M.}} \AND
  \bauthor{\bsnm{Dai},~\bfnm{Xinyu}\binits{X.}}
(\byear{2013}).
\btitle{A Two-year Time Delay for the Lensed Quasar SDSS J1029+ 2623}.
\bjournal{The Astrophysical Journal}
\bvolume{764}
\bpages{186}.
\end{barticle}
\endbibitem

\bibitem[\protect\citeauthoryear{Gelman and Rubin}{1992}]{gelman1992inference}
\begin{barticle}[author]
\bauthor{\bsnm{Gelman},~\bfnm{Andrew}\binits{A.}} \AND
  \bauthor{\bsnm{Rubin},~\bfnm{Donald~B}\binits{D.~B.}}
(\byear{1992}).
\btitle{Inference from Iterative Simulation Using Multiple Sequences}.
\bjournal{Statistical Science}
\bvolume{7}
\bpages{457--472}.
\end{barticle}
\endbibitem

\bibitem[\protect\citeauthoryear{Gelman et~al.}{2013}]{gelman2013bayesian}
\begin{bbook}[author]
\bauthor{\bsnm{Gelman},~\bfnm{Andrew}\binits{A.}},
  \bauthor{\bsnm{Carlin},~\bfnm{John~B}\binits{J.~B.}},
  \bauthor{\bsnm{Stern},~\bfnm{Hal~S}\binits{H.~S.}},
  \bauthor{\bsnm{Dunson},~\bfnm{David~B}\binits{D.~B.}},
  \bauthor{\bsnm{Vehtari},~\bfnm{Aki}\binits{A.}} \AND
  \bauthor{\bsnm{Rubin},~\bfnm{Donald~B}\binits{D.~B.}}
(\byear{2013}).
\btitle{Bayesian Data Analysis}.
\bpublisher{CRC press}.
\end{bbook}
\endbibitem

\bibitem[\protect\citeauthoryear{Hainline et~al.}{2012}]{hainline2012new}
\begin{barticle}[author]
\bauthor{\bsnm{Hainline},~\bfnm{Laura~J}\binits{L.~J.}},
  \bauthor{\bsnm{Morgan},~\bfnm{Christopher~W}\binits{C.~W.}},
  \bauthor{\bsnm{Beach},~\bfnm{Joseph~N}\binits{J.~N.}},
  \bauthor{\bsnm{Kochanek},~\bfnm{CS}\binits{C.}},
  \bauthor{\bsnm{Harris},~\bfnm{Hugh~C}\binits{H.~C.}},
  \bauthor{\bsnm{Tilleman},~\bfnm{Trudy}\binits{T.}},
  \bauthor{\bsnm{Fadely},~\bfnm{Ross}\binits{R.}},
  \bauthor{\bsnm{Falco},~\bfnm{Emilio~E}\binits{E.~E.}} \AND
  \bauthor{\bsnm{Le},~\bfnm{Truong~X}\binits{T.~X.}}
(\byear{2012}).
\btitle{A New Microlensing Event in the Doubly Imaged Quasar Q 0957+ 561}.
\bjournal{The Astrophysical Journal}
\bvolume{744}
\bpages{104}.
\end{barticle}
\endbibitem

\bibitem[\protect\citeauthoryear{Harva and Raychaudhury}{2006}]{harva2006}
\begin{bbook}[author]
\bauthor{\bsnm{Harva},~\bfnm{Markus}\binits{M.}} \AND
  \bauthor{\bsnm{Raychaudhury},~\bfnm{Somak}\binits{S.}}
(\byear{2006}).
\btitle{Bayesian Estimation of Time Delays Between Unevenly Sampled Signals}.
\bpublisher{IEEE}.
\end{bbook}
\endbibitem

\bibitem[\protect\citeauthoryear{Hojjati, Kim and
  Linder}{2013}]{hojjati2013robust}
\begin{barticle}[author]
\bauthor{\bsnm{Hojjati},~\bfnm{Alireza}\binits{A.}},
  \bauthor{\bsnm{Kim},~\bfnm{Alex~G}\binits{A.~G.}} \AND
  \bauthor{\bsnm{Linder},~\bfnm{Eric~V}\binits{E.~V.}}
(\byear{2013}).
\btitle{Robust Strong Lensing Time Delay Estimation}.
\bjournal{Physical Review D}
\bvolume{87}
\bpages{123512}.
\end{barticle}
\endbibitem

\bibitem[\protect\citeauthoryear{Inada et~al.}{2006}]{inada2006sdss}
\begin{barticle}[author]
\bauthor{\bsnm{Inada},~\bfnm{Naohisa}\binits{N.}},
  \bauthor{\bsnm{Oguri},~\bfnm{Masamune}\binits{M.}},
  \bauthor{\bsnm{Morokuma},~\bfnm{Tomoki}\binits{T.}},
  \bauthor{\bsnm{Doi},~\bfnm{Mamoru}\binits{M.}},
  \bauthor{\bsnm{Yasuda},~\bfnm{Naoki}\binits{N.}},
  \bauthor{\bsnm{Becker},~\bfnm{Robert~H}\binits{R.~H.}},
  \bauthor{\bsnm{Richards},~\bfnm{Gordon~T}\binits{G.~T.}},
  \bauthor{\bsnm{Kochanek},~\bfnm{Christopher~S}\binits{C.~S.}},
  \bauthor{\bsnm{Kayo},~\bfnm{Issha}\binits{I.}},
  \bauthor{\bsnm{Konishi},~\bfnm{Kohki}\binits{K.}} \betal{et~al.}
(\byear{2006}).
\btitle{SDSS J1029+ 2623: A Gravitationally Lensed Quasar with an Image
  Separation of 225}.
\bjournal{The Astrophysical Journal Letters}
\bvolume{653}
\bpages{L97}.
\end{barticle}
\endbibitem

\bibitem[\protect\citeauthoryear{Kelly, Bechtold and
  Siemiginowska}{2009}]{kelly2009variations}
\begin{barticle}[author]
\bauthor{\bsnm{Kelly},~\bfnm{Brandon~C}\binits{B.~C.}},
  \bauthor{\bsnm{Bechtold},~\bfnm{Jill}\binits{J.}} \AND
  \bauthor{\bsnm{Siemiginowska},~\bfnm{Aneta}\binits{A.}}
(\byear{2009}).
\btitle{Are the Variations in Quasar Optical Flux Driven by Thermal
  Fluctuations?}
\bjournal{The Astrophysical Journal}
\bvolume{698}
\bpages{895}.
\end{barticle}
\endbibitem

\bibitem[\protect\citeauthoryear{Kochanek et~al.}{2006}]{kochanek2006time}
\begin{barticle}[author]
\bauthor{\bsnm{Kochanek},~\bfnm{CS}\binits{C.}},
  \bauthor{\bsnm{Morgan},~\bfnm{ND}\binits{N.}},
  \bauthor{\bsnm{Falco},~\bfnm{EE}\binits{E.}},
  \bauthor{\bsnm{McLeod},~\bfnm{BA}\binits{B.}},
  \bauthor{\bsnm{Winn},~\bfnm{JN}\binits{J.}},
  \bauthor{\bsnm{Dembicky},~\bfnm{J}\binits{J.}} \AND
  \bauthor{\bsnm{Ketzeback},~\bfnm{B}\binits{B.}}
(\byear{2006}).
\btitle{The Time Delays of Gravitational Lens HE 0435--1223: An Early-Type
  Galaxy with a Rising Rotation Curve}.
\bjournal{The Astrophysical Journal}
\bvolume{640}
\bpages{47}.
\end{barticle}
\endbibitem

\bibitem[\protect\citeauthoryear{Koz{\l}owski and
  Kochanek}{2009}]{kozlowski2009discovery}
\begin{barticle}[author]
\bauthor{\bsnm{Koz{\l}owski},~\bfnm{Szymon}\binits{S.}} \AND
  \bauthor{\bsnm{Kochanek},~\bfnm{Christopher~S}\binits{C.~S.}}
(\byear{2009}).
\btitle{Discovery of 5000 Active Galactic Nuclei Behind the Magellanic Clouds}.
\bjournal{The Astrophysical Journal}
\bvolume{701}
\bpages{508}.
\end{barticle}
\endbibitem

\bibitem[\protect\citeauthoryear{Koz{\l}owski
  et~al.}{2010}]{kozlowski2010quantifying}
\begin{barticle}[author]
\bauthor{\bsnm{Koz{\l}owski},~\bfnm{Szymon}\binits{S.}},
  \bauthor{\bsnm{Kochanek},~\bfnm{Christopher~S}\binits{C.~S.}},
  \bauthor{\bsnm{Udalski},~\bfnm{A}\binits{A.}},
  \bauthor{\bsnm{Soszy{\'n}ski},~\bfnm{I}\binits{I.}},
  \bauthor{\bsnm{Szyma{\'n}ski},~\bfnm{MK}\binits{M.}},
  \bauthor{\bsnm{Kubiak},~\bfnm{M}\binits{M.}},
  \bauthor{\bsnm{Pietrzy{\'n}ski},~\bfnm{G}\binits{G.}},
  \bauthor{\bsnm{Szewczyk},~\bfnm{O}\binits{O.}},
  \bauthor{\bsnm{Ulaczyk},~\bfnm{K}\binits{K.}} \AND
  \bauthor{\bsnm{Poleski},~\bfnm{R}\binits{R.}}
(\byear{2010}).
\btitle{Quantifying Quasar Variability as Part of a General Approach to
  Classifying Continuously Varying Sources}.
\bjournal{The Astrophysical Journal}
\bvolume{708}
\bpages{927}.
\end{barticle}
\endbibitem

\bibitem[\protect\citeauthoryear{Kumar, Stalin and Prabhu}{2014}]{kumar2014h0}
\begin{barticle}[author]
\bauthor{\bsnm{Kumar},~\bfnm{S~Rathna}\binits{S.~R.}},
  \bauthor{\bsnm{Stalin},~\bfnm{CS}\binits{C.}} \AND
  \bauthor{\bsnm{Prabhu},~\bfnm{TP}\binits{T.}}
(\byear{2014}).
\btitle{H$_0$ from 11 Well Measured Time-delay Lenses}.
\bjournal{Astronomy \& Astrophysics}
\bvolume{580}
\bpages{A38}.
\end{barticle}
\endbibitem

\bibitem[\protect\citeauthoryear{{Liao} et~al.}{2015}]{kai2014}
\begin{barticle}[author]
\bauthor{\bsnm{{Liao}},~\bfnm{K.}\binits{K.}},
  \bauthor{\bsnm{{Treu}},~\bfnm{T.}\binits{T.}},
  \bauthor{\bsnm{{Marshall}},~\bfnm{P.}\binits{P.}},
  \bauthor{\bsnm{{Fassnacht}},~\bfnm{C.~D.}\binits{C.~D.}},
  \bauthor{\bsnm{{Rumbaugh}},~\bfnm{N.}\binits{N.}},
  \bauthor{\bsnm{{Dobler}},~\bfnm{G.}\binits{G.}},
  \bauthor{\bsnm{{Aghamousa}},~\bfnm{A.}\binits{A.}},
  \bauthor{\bsnm{{Bonvin}},~\bfnm{V.}\binits{V.}},
  \bauthor{\bsnm{{Courbin}},~\bfnm{F.}\binits{F.}},
  \bauthor{\bsnm{{Hojjati}},~\bfnm{A.}\binits{A.}},
  \bauthor{\bsnm{{Jackson}},~\bfnm{N.}\binits{N.}},
  \bauthor{\bsnm{{Kashyap}},~\bfnm{V.}\binits{V.}}, \bauthor{\bsnm{{Rathna
  Kumar}},~\bfnm{S.}\binits{S.}},
  \bauthor{\bsnm{{Linder}},~\bfnm{E.}\binits{E.}},
  \bauthor{\bsnm{{Mandel}},~\bfnm{K.}\binits{K.}},
  \bauthor{\bsnm{{Meng}},~\bfnm{X.~L.}\binits{X.~L.}},
  \bauthor{\bsnm{{Meylan}},~\bfnm{G.}\binits{G.}},
  \bauthor{\bsnm{{Moustakas}},~\bfnm{L.~A.}\binits{L.~A.}},
  \bauthor{\bsnm{{Prabhu}},~\bfnm{T.~P.}\binits{T.~P.}},
  \bauthor{\bsnm{{Romero-Wolf}},~\bfnm{A.}\binits{A.}},
  \bauthor{\bsnm{{Shafieloo}},~\bfnm{A.}\binits{A.}},
  \bauthor{\bsnm{{Siemiginowska}},~\bfnm{A.}\binits{A.}},
  \bauthor{\bsnm{{Stalin}},~\bfnm{C.~S.}\binits{C.~S.}},
  \bauthor{\bsnm{{Tak}},~\bfnm{H.}\binits{H.}},
  \bauthor{\bsnm{{Tewes}},~\bfnm{M.}\binits{M.}} \AND \bauthor{\bsnm{{van
  Dyk}},~\bfnm{D.}\binits{D.}}
(\byear{2015}).
\btitle{{Strong Lens Time Delay Challenge: II. Results of TDC1}}.
\bjournal{The Astrophysical Journal}
\bvolume{800}
\bpages{11}.
\end{barticle}
\endbibitem

\bibitem[\protect\citeauthoryear{Linder}{2011}]{Linder2011}
\begin{barticle}[author]
\bauthor{\bsnm{Linder},~\bfnm{Eric~V.}\binits{E.~V.}}
(\byear{2011}).
\btitle{Lensing Time Delays and Cosmological Complementarity}.
\bjournal{Phys. Rev. D}
\bvolume{84}
\bpages{123529}.
\bdoi{10.1103/PhysRevD.84.123529}
\end{barticle}
\endbibitem

\bibitem[\protect\citeauthoryear{Liu}{2008}]{liu2008monte}
\begin{bbook}[author]
\bauthor{\bsnm{Liu},~\bfnm{Jun~S}\binits{J.~S.}}
(\byear{2008}).
\btitle{Monte Carlo Strategies in Scientific Computing}.
\bpublisher{Springer Science \& Business Media}.
\end{bbook}
\endbibitem

\bibitem[\protect\citeauthoryear{MacLeod et~al.}{2010}]{macleod2010modeling}
\begin{barticle}[author]
\bauthor{\bsnm{MacLeod},~\bfnm{CL}\binits{C.}},
  \bauthor{\bsnm{Ivezi{\'c}},~\bfnm{{\v{Z}}}\binits{{\v{Z}}.}},
  \bauthor{\bsnm{Kochanek},~\bfnm{CS}\binits{C.}},
  \bauthor{\bsnm{Koz{\l}owski},~\bfnm{S}\binits{S.}},
  \bauthor{\bsnm{Kelly},~\bfnm{B}\binits{B.}},
  \bauthor{\bsnm{Bullock},~\bfnm{E}\binits{E.}},
  \bauthor{\bsnm{Kimball},~\bfnm{A}\binits{A.}},
  \bauthor{\bsnm{Sesar},~\bfnm{B}\binits{B.}},
  \bauthor{\bsnm{Westman},~\bfnm{D}\binits{D.}},
  \bauthor{\bsnm{Brooks},~\bfnm{K}\binits{K.}},
  \bauthor{\bsnm{Gibson},~\bfnm{R.}\binits{R.}},
  \bauthor{\bsnm{Becker},~\bfnm{A.~C.}\binits{A.~C.}} \AND
  \bauthor{\bparticle{de} \bsnm{Vries},~\bfnm{W.~H.}\binits{W.~H.}}
(\byear{2010}).
\btitle{Modeling the Time Variability of SDSS Stripe 82 Quasars as a Damped
  Random Walk}.
\bjournal{The Astrophysical Journal}
\bvolume{721}
\bpages{1014}.
\end{barticle}
\endbibitem

\bibitem[\protect\citeauthoryear{Morgan et~al.}{2012}]{morgan2012further}
\begin{barticle}[author]
\bauthor{\bsnm{Morgan},~\bfnm{Christopher~W}\binits{C.~W.}},
  \bauthor{\bsnm{Hainline},~\bfnm{Laura~J}\binits{L.~J.}},
  \bauthor{\bsnm{Chen},~\bfnm{Bin}\binits{B.}},
  \bauthor{\bsnm{Tewes},~\bfnm{Malte}\binits{M.}},
  \bauthor{\bsnm{Kochanek},~\bfnm{Christopher~S}\binits{C.~S.}},
  \bauthor{\bsnm{Dai},~\bfnm{Xinyu}\binits{X.}},
  \bauthor{\bsnm{Kozlowski},~\bfnm{Szymon}\binits{S.}},
  \bauthor{\bsnm{Blackburne},~\bfnm{Jeffrey~A}\binits{J.~A.}},
  \bauthor{\bsnm{Mosquera},~\bfnm{Ana~M}\binits{A.~M.}},
  \bauthor{\bsnm{Chartas},~\bfnm{George}\binits{G.}},
  \bauthor{\bsnm{Courbin},~\bfnm{F.}\binits{F.}} \AND
  \bauthor{\bsnm{Meylan},~\bfnm{G.}\binits{G.}}
(\byear{2012}).
\btitle{Further Evidence that Quasar X-ray Emitting Regions are Compact: X-ray
  and Optical Microlensing in the Lensed Quasar Q J0158-4325}.
\bjournal{The Astrophysical Journal}
\bvolume{756}
\bpages{52}.
\end{barticle}
\endbibitem

\bibitem[\protect\citeauthoryear{Mosquera and
  Kochanek}{2011}]{mosquera2011microlensing}
\begin{barticle}[author]
\bauthor{\bsnm{Mosquera},~\bfnm{Ana~M}\binits{A.~M.}} \AND
  \bauthor{\bsnm{Kochanek},~\bfnm{Christopher~S}\binits{C.~S.}}
(\byear{2011}).
\btitle{The Microlensing Properties of a Sample of 87 Lensed Quasars}.
\bjournal{The Astrophysical Journal}
\bvolume{738}
\bpages{96}.
\end{barticle}
\endbibitem

\bibitem[\protect\citeauthoryear{Munoz et~al.}{1998}]{munoz1998castles}
\begin{barticle}[author]
\bauthor{\bsnm{Munoz},~\bfnm{JA}\binits{J.}},
  \bauthor{\bsnm{Falco},~\bfnm{EE}\binits{E.}},
  \bauthor{\bsnm{Kochanek},~\bfnm{CS}\binits{C.}},
  \bauthor{\bsnm{Leh{\'a}r},~\bfnm{J}\binits{J.}},
  \bauthor{\bsnm{McLeod},~\bfnm{BA}\binits{B.}},
  \bauthor{\bsnm{Impey},~\bfnm{CD}\binits{C.}},
  \bauthor{\bsnm{Rix},~\bfnm{H-W}\binits{H.-W.}} \AND
  \bauthor{\bsnm{Peng},~\bfnm{CY}\binits{C.}}
(\byear{1998}).
\btitle{The CASTLES project}.
\bjournal{Astrophysics and Space Science}
\bvolume{263}
\bpages{51--54}.
\end{barticle}
\endbibitem

\bibitem[\protect\citeauthoryear{Oguri and
  Marshall}{2010}]{oguri2010gravitationally}
\begin{barticle}[author]
\bauthor{\bsnm{Oguri},~\bfnm{Masamune}\binits{M.}} \AND
  \bauthor{\bsnm{Marshall},~\bfnm{Philip~J}\binits{P.~J.}}
(\byear{2010}).
\btitle{Gravitationally Lensed Quasars and Supernovae in Future Wide-field
  Optical Imaging Surveys}.
\bjournal{Monthly Notices of the Royal Astronomical Society}
\bvolume{405}
\bpages{2579--2593}.
\end{barticle}
\endbibitem

\bibitem[\protect\citeauthoryear{Oscoz et~al.}{1997}]{oscoz1997time}
\begin{barticle}[author]
\bauthor{\bsnm{Oscoz},~\bfnm{Alejandro}\binits{A.}},
  \bauthor{\bsnm{Mediavilla},~\bfnm{Evencio}\binits{E.}},
  \bauthor{\bsnm{Goicoechea},~\bfnm{Luis~Julian}\binits{L.~J.}},
  \bauthor{\bsnm{Serra-Ricart},~\bfnm{Miquel}\binits{M.}} \AND
  \bauthor{\bsnm{Buitrago},~\bfnm{Jesus}\binits{J.}}
(\byear{1997}).
\btitle{Time Delay of QSO 0957+ 561 and Cosmological Implications}.
\bjournal{The Astrophysical Journal Letters}
\bvolume{479}
\bpages{L89}.
\end{barticle}
\endbibitem

\bibitem[\protect\citeauthoryear{Oscoz et~al.}{2001}]{oscoz2001time}
\begin{barticle}[author]
\bauthor{\bsnm{Oscoz},~\bfnm{A}\binits{A.}},
  \bauthor{\bsnm{Alcalde},~\bfnm{D}\binits{D.}},
  \bauthor{\bsnm{Serra-Ricart},~\bfnm{M}\binits{M.}},
  \bauthor{\bsnm{Mediavilla},~\bfnm{E}\binits{E.}},
  \bauthor{\bsnm{Abajas},~\bfnm{C}\binits{C.}},
  \bauthor{\bsnm{Barrena},~\bfnm{R}\binits{R.}},
  \bauthor{\bsnm{Licandro},~\bfnm{J}\binits{J.}},
  \bauthor{\bsnm{Motta},~\bfnm{V}\binits{V.}} \AND
  \bauthor{\bsnm{Munoz},~\bfnm{JA}\binits{J.}}
(\byear{2001}).
\btitle{Time Delay in QSO 0957+ 561 from 1984-1999 Optical Data}.
\bjournal{The Astrophysical Journal}
\bvolume{552}
\bpages{81}.
\end{barticle}
\endbibitem

\bibitem[\protect\citeauthoryear{Pelt et~al.}{1994}]{pelt1994}
\begin{barticle}[author]
\bauthor{\bsnm{Pelt},~\bfnm{J}\binits{J.}},
  \bauthor{\bsnm{Hoff},~\bfnm{W}\binits{W.}},
  \bauthor{\bsnm{Kayser},~\bfnm{R}\binits{R.}},
  \bauthor{\bsnm{Refsdal},~\bfnm{S}\binits{S.}} \AND
  \bauthor{\bsnm{Schramm},~\bfnm{T}\binits{T.}}
(\byear{1994}).
\btitle{Time Delay Controversy on QSO 0957+ 561 not yet Decided}.
\bjournal{Astronomy \& Astrophysics}
\bvolume{286}
\bpages{775--785}.
\end{barticle}
\endbibitem

\bibitem[\protect\citeauthoryear{Pelt et~al.}{1996}]{pelt1996}
\begin{barticle}[author]
\bauthor{\bsnm{Pelt},~\bfnm{Jaan}\binits{J.}},
  \bauthor{\bsnm{Kayser},~\bfnm{Rainer}\binits{R.}},
  \bauthor{\bsnm{Refsdal},~\bfnm{Sjur}\binits{S.}} \AND
  \bauthor{\bsnm{Schramm},~\bfnm{Thomas}\binits{T.}}
(\byear{1996}).
\btitle{The Light Curve and the Time Delay of QSO 0957+ 561}.
\bjournal{Astronomy \& Astrophysics}
\bvolume{305}
\bpages{97--106}.
\end{barticle}
\endbibitem

\bibitem[\protect\citeauthoryear{Refsdal}{1964}]{ref1964}
\begin{barticle}[author]
\bauthor{\bsnm{Refsdal},~\bfnm{Sjur}\binits{S.}}
(\byear{1964}).
\btitle{The Gravitational Lens Effect}.
\bjournal{Monthly Notices of the Royal Astronomical Society}
\bvolume{128}
\bpages{295--306}.
\end{barticle}
\endbibitem

\bibitem[\protect\citeauthoryear{Roberts and
  Rosenthal}{2007}]{roberts2007coupling}
\begin{barticle}[author]
\bauthor{\bsnm{Roberts},~\bfnm{Gareth~O}\binits{G.~O.}} \AND
  \bauthor{\bsnm{Rosenthal},~\bfnm{Jeffrey~S}\binits{J.~S.}}
(\byear{2007}).
\btitle{Coupling and Ergodicity of Adaptive Markov Chain Monte Carlo
  Algorithms}.
\bjournal{Journal of Applied Probability}
\bvolume{44}
\bpages{458--475}.
\end{barticle}
\endbibitem

\bibitem[\protect\citeauthoryear{Schneider, Ehlers and
  Falco}{1992}]{schneider1992}
\begin{bbook}[author]
\bauthor{\bsnm{Schneider},~\bfnm{P}\binits{P.}},
  \bauthor{\bsnm{Ehlers},~\bfnm{J}\binits{J.}} \AND
  \bauthor{\bsnm{Falco},~\bfnm{EE}\binits{E.}}
(\byear{1992}).
\btitle{Gravitational Lenses}.
\bpublisher{Springer}.
\end{bbook}
\endbibitem

\bibitem[\protect\citeauthoryear{Schneider, Wambsganss and
  Kochanek}{2006}]{schneider2006}
\begin{bbook}[author]
\bauthor{\bsnm{Schneider},~\bfnm{Peter}\binits{P.}},
  \bauthor{\bsnm{Wambsganss},~\bfnm{Joachim}\binits{J.}} \AND
  \bauthor{\bsnm{Kochanek},~\bfnm{Christopher~S}\binits{C.~S.}}
(\byear{2006}).
\btitle{Gravitational Lensing: Strong, Weak and Micro}.
\bpublisher{Springer}.
\end{bbook}
\endbibitem

\bibitem[\protect\citeauthoryear{Serra-Ricart et~al.}{1999}]{serra1999bvri}
\begin{barticle}[author]
\bauthor{\bsnm{Serra-Ricart},~\bfnm{Miquel}\binits{M.}},
  \bauthor{\bsnm{Oscoz},~\bfnm{Alejandro}\binits{A.}},
  \bauthor{\bsnm{Sanch{\'\i}s},~\bfnm{Teresa}\binits{T.}},
  \bauthor{\bsnm{Mediavilla},~\bfnm{Evencio}\binits{E.}},
  \bauthor{\bsnm{Goicoechea},~\bfnm{Luis~Juli{\'a}n}\binits{L.~J.}},
  \bauthor{\bsnm{Licandro},~\bfnm{Javier}\binits{J.}},
  \bauthor{\bsnm{Alcalde},~\bfnm{David}\binits{D.}} \AND
  \bauthor{\bsnm{Gil-Merino},~\bfnm{Rodrigo}\binits{R.}}
(\byear{1999}).
\btitle{BVRI Photometry of QSO 0957+ 561A, B: Observations, New Reduction
  Method, and Time Delay}.
\bjournal{The Astrophysical Journal}
\bvolume{526}
\bpages{40}.
\end{barticle}
\endbibitem

\bibitem[\protect\citeauthoryear{Shalyapin, Goicoechea and
  Gil-Merino}{2014}]{shalyapin20125}
\begin{barticle}[author]
\bauthor{\bsnm{Shalyapin},~\bfnm{VN}\binits{V.}},
  \bauthor{\bsnm{Goicoechea},~\bfnm{LJ}\binits{L.}} \AND
  \bauthor{\bsnm{Gil-Merino},~\bfnm{R}\binits{R.}}
(\byear{2014}).
\btitle{A 5.5-year Robotic Optical Monitoring of Q0957+ 561: Substructure in a
  Non-local cD Galaxy}.
\bjournal{Astronomy \& Astrophysics}
\bvolume{540}
\bpages{A132}.
\end{barticle}
\endbibitem

\bibitem[\protect\citeauthoryear{Suyu et~al.}{2013}]{suyu2013two}
\begin{barticle}[author]
\bauthor{\bsnm{Suyu},~\bfnm{SH}\binits{S.}},
  \bauthor{\bsnm{Auger},~\bfnm{MW}\binits{M.}},
  \bauthor{\bsnm{Hilbert},~\bfnm{S}\binits{S.}},
  \bauthor{\bsnm{Marshall},~\bfnm{PJ}\binits{P.}},
  \bauthor{\bsnm{Tewes},~\bfnm{M}\binits{M.}},
  \bauthor{\bsnm{Treu},~\bfnm{T}\binits{T.}},
  \bauthor{\bsnm{Fassnacht},~\bfnm{CD}\binits{C.}},
  \bauthor{\bsnm{Koopmans},~\bfnm{LVE}\binits{L.}},
  \bauthor{\bsnm{Sluse},~\bfnm{D}\binits{D.}},
  \bauthor{\bsnm{Blandford},~\bfnm{RD}\binits{R.}},
  \bauthor{\bsnm{Courbin},~\bfnm{F.}\binits{F.}} \AND
  \bauthor{\bsnm{Meylan},~\bfnm{G.}\binits{G.}}
(\byear{2013}).
\btitle{Two Accurate Time-delay Distances from Strong Lensing: Implications for
  Cosmology}.
\bjournal{The Astrophysical Journal}
\bvolume{766}
\bpages{70}.
\end{barticle}
\endbibitem

\bibitem[\protect\citeauthoryear{Tak, Kelly and Morris}{2017+}]{tak2016b}
\begin{barticle}[author]
\bauthor{\bsnm{Tak},~\bfnm{H.}\binits{H.}},
  \bauthor{\bsnm{Kelly},~\bfnm{J.}\binits{J.}} \AND
  \bauthor{\bsnm{Morris},~\bfnm{C.~N.}\binits{C.~N.}}
(\byear{2017}+).
\btitle{Rgbp: An R Package for Gaussian, Poisson, and Binomial Random Effects
  Models with Frequency Coverage Evaluations}.
\bjournal{Journal of Statistical Software}
\bpages{arXiv:1612.01595}.
\end{barticle}
\endbibitem

\bibitem[\protect\citeauthoryear{{R Core Team}}{2016}]{r2014}
\begin{bmanual}[author]
\bauthor{\bsnm{{R Core Team}}}
(\byear{2016}).
\btitle{{R: A Language and Environment for Statistical Computing}}
\bpublisher{R Foundation for Statistical Computing},
\baddress{Vienna, Austria}.
\end{bmanual}
\endbibitem

\bibitem[\protect\citeauthoryear{Tewes, Courbin and Meylan}{2013}]{tewes2013a}
\begin{barticle}[author]
\bauthor{\bsnm{Tewes},~\bfnm{M}\binits{M.}},
  \bauthor{\bsnm{Courbin},~\bfnm{F}\binits{F.}} \AND
  \bauthor{\bsnm{Meylan},~\bfnm{G}\binits{G.}}
(\byear{2013}).
\btitle{COSMOGRAIL: the COSmological MOnitoring of GRAvItational Lenses XI.
  Techniques for time delay measurement in presence of microlensing}.
\bjournal{The Astrophysical Journal}
\bvolume{605}
\bpages{58}.
\end{barticle}
\endbibitem

\bibitem[\protect\citeauthoryear{Tierney}{1994}]{tierney1994markov}
\begin{barticle}[author]
\bauthor{\bsnm{Tierney},~\bfnm{Luke}\binits{L.}}
(\byear{1994}).
\btitle{Markov Chains for Exploring Posterior Distributions}.
\bjournal{The Annals of Statistics}
\bvolume{22}
\bpages{1701--1728}.
\end{barticle}
\endbibitem

\bibitem[\protect\citeauthoryear{Treu and Marshall}{2016}]{treu2016time}
\begin{barticle}[author]
\bauthor{\bsnm{Treu},~\bfnm{T.}\binits{T.}} \AND
  \bauthor{\bsnm{Marshall},~\bfnm{P.~J.}\binits{P.~J.}}
(\byear{2016}).
\btitle{{Time Delay Cosmography}}.
\bjournal{The Astronomy and Astrophysics Review}
\bvolume{24}
\bpages{11}.
\end{barticle}
\endbibitem

\bibitem[\protect\citeauthoryear{Uhlenbeck and
  Ornstein}{1930}]{uhlenbeck1930theory}
\begin{barticle}[author]
\bauthor{\bsnm{Uhlenbeck},~\bfnm{George~E}\binits{G.~E.}} \AND
  \bauthor{\bsnm{Ornstein},~\bfnm{Leonard~Salomon}\binits{L.~S.}}
(\byear{1930}).
\btitle{On the Theory of the Brownian Motion}.
\bjournal{Physical review}
\bvolume{36}
\bpages{823}.
\end{barticle}
\endbibitem

\bibitem[\protect\citeauthoryear{van Dyk and Meng}{2001}]{van2001art}
\begin{barticle}[author]
\bauthor{\bparticle{van} \bsnm{Dyk},~\bfnm{David~A}\binits{D.~A.}} \AND
  \bauthor{\bsnm{Meng},~\bfnm{Xiao-Li}\binits{X.-L.}}
(\byear{2001}).
\btitle{The Art of Data Augmentation}.
\bjournal{Journal of Computational and Graphical Statistics}
\bvolume{10}
\bpages{1--50}.
\end{barticle}
\endbibitem

\bibitem[\protect\citeauthoryear{Walsh, Carswell and
  Weymann}{1979}]{walsh1979q0957}
\begin{barticle}[author]
\bauthor{\bsnm{Walsh},~\bfnm{D}\binits{D.}},
  \bauthor{\bsnm{Carswell},~\bfnm{RF}\binits{R.}} \AND
  \bauthor{\bsnm{Weymann},~\bfnm{RJ}\binits{R.}}
(\byear{1979}).
\btitle{0957+ 561 A, B- Twin Quasistellar Objects or Gravitational Lens}.
\bjournal{Nature}
\bvolume{279}
\bpages{381--384}.
\end{barticle}
\endbibitem

\bibitem[\protect\citeauthoryear{Yu and Meng}{2011}]{yu2011}
\begin{barticle}[author]
\bauthor{\bsnm{Yu},~\bfnm{Yaming}\binits{Y.}} \AND
  \bauthor{\bsnm{Meng},~\bfnm{Xiao-Li}\binits{X.-L.}}
(\byear{2011}).
\btitle{To Center or not to Center: That is not the Question? An
  Ancillarity--Sufficiency Interweaving Strategy (ASIS) for Boosting MCMC
  Efficiency}.
\bjournal{Journal of Computational and Graphical Statistics}
\bvolume{20}
\bpages{531--570}.
\end{barticle}
\endbibitem

\bibitem[\protect\citeauthoryear{Zu et~al.}{2013}]{zu2013quasar}
\begin{barticle}[author]
\bauthor{\bsnm{Zu},~\bfnm{Ying}\binits{Y.}},
  \bauthor{\bsnm{Kochanek},~\bfnm{CS}\binits{C.}},
  \bauthor{\bsnm{Koz{\l}owski},~\bfnm{Szymon}\binits{S.}} \AND
  \bauthor{\bsnm{Udalski},~\bfnm{Andrzej}\binits{A.}}
(\byear{2013}).
\btitle{Is Quasar Optical Variability a Damped Random Walk?}
\bjournal{The Astrophysical Journal}
\bvolume{765}
\bpages{106}.
\end{barticle}
\endbibitem

\end{thebibliography}
\bibliographystyle{imsart-nameyear}

\appendix

\section{The  likelihood function}\label{appcond}
We define a combined light curve $\boldsymbol{z}=(z_1, z_2, \ldots, z_{2n})^\top$ as follows. The observed magnitude at time $t^\Delta_i$ ($i=1, 2, \ldots, 2n$) is denoted by $z_i$, which is either $x_j$ or $y_j-\boldsymbol{w}_m^\top(t_j-\Delta) \boldsymbol{\beta}$ for some $j$ ($j=1, 2, \ldots, n$) depending on whether $t^\Delta_i$ is one of the elements of $\boldsymbol{t}$ or of $\boldsymbol{t}-\Delta$. The  measurement  standard deviation  is denoted by $\xi_{i}$, which is either $\delta_{j}$ for $x_{j}$ or $\eta_{j}$ for $y_{j}-\boldsymbol{w}_m^\top(t_j-\Delta) \boldsymbol{\beta}$ for some $j$.  We also define $z'_{i}$ as the centered observed magnitude at time $t^\Delta_i$, which is either $x_{j}-\mu$ or $y_{j}-\boldsymbol{w}_m^\top(t_j-\Delta) \boldsymbol{\beta}-\mu$ for some $j$.  Let $D_i=\{z'_{1}, z'_{2}, \ldots, z'_{i}\}$ and derive the likelihood
\begin{equation}\label{marginal_likelihood}
L(\Delta, \boldsymbol{\beta}, \mu, \sigma^2, \tau) \propto p(z'_1) \times \prod_{i=2}^{2n}p(z'_i\mid D_{i-1})
\end{equation}
with $\boldsymbol{X}(\boldsymbol{t}^\Delta)$ integrated out. The sampling distribution of $D_{2n}$ given $\Delta, \boldsymbol{\beta}, \mu, \sigma^2$, and $\tau$ factors as
\begin{align}
z'_{1}  &\sim \textrm{N}\left[0,~ \xi_{1}^2+\tau\sigma^2/2\right],\label{latent_obs_density1}\\
z'_{i}\mid D_{i-1}  &\sim \textrm{N}\left[a_i\mu_{i-1},~ \xi_{i}^2+a_i^2\Omega_{i-1}+\tau\sigma^2(1-a_i^2)/2\right],\label{latent_obs_density2}
\end{align}
\!\!\!where $\mu_1=(1-B_1)z'_{1}$, $\mu_i=(1-B_i)z'_{i} + B_ia_i\mu_{i-1}$, $\Omega_i=(1-B_i)\xi_{i}^2$, $B_1=\xi_{1}^2 / [\xi_{1}^2 +\tau\sigma^2/2]$,  $B_i=\xi_{i}^2 / [\xi_{i}^2 +a^2_i\Omega_{i-1}+\tau\sigma^2(1-a^2_i)/2]$. Thus, the likelihood function  of ($\Delta, \boldsymbol{\beta}, \mu, \sigma^2, \tau$) in \eqref{marginal_likelihood} is the product of the Gaussian densities.

%.  For $i=1, 2, \ldots, 2n$, the posterior predictive distributions  of $z'_{i}$ with $\boldsymbol{X}(\boldsymbol{t}^\Delta)$ integrated out are
%The posterior density is 
%\begin{equation}
%p(\Delta, \boldsymbol{\beta}, \mu, \sigma^2, \tau\mid \boldsymbol{x_t}, \boldsymbol{y})\propto L(\Delta, \boldsymbol{\beta}, \mu, \sigma^2, \tau)p(\Delta, \boldsymbol{\beta})p(\mu, \sigma^2, \tau),
%\end{equation}
%where the prior density for  $(\Delta, \boldsymbol{\beta})$ is given in \eqref{priordeltac} and and that for $(\mu, \sigma^2, \tau)$ is given in \eqref{hyp}.
%where the  distributions of the density functions in the product are given in  \eqref{latent_obs_density1} and \eqref{latent_obs_density2} .

By multiplying \eqref{marginal_likelihood} by the prior density functions for $\Delta$ and $\boldsymbol{\beta}$ in \eqref{priordeltac} and those for  $\mu, \sigma^2$, and $\tau$ in \eqref{hyp}, we can obtain their  joint posterior density function  with the latent  magnitudes $\boldsymbol{X}(\boldsymbol{t}^\Delta)$ marginalized out. Given the values of $(\boldsymbol{\beta}, \mu, \sigma^2, \tau)$ and a Uniform$[u_1, u_2]$ prior distribution for $\Delta$, $L(\Delta, \boldsymbol{\beta}, \mu, \sigma^2, \tau)$ is proportional to the marginalized conditional posterior density $p(\Delta\mid \boldsymbol{\beta}, \mu, \sigma^2, \tau, \boldsymbol{x}, \boldsymbol{y})$ used in \eqref{eq:step1} and \eqref{MH}.
% for $\Delta\in [u_1, u_2]$ and is zero elsewhere
 %for $\Delta$ defined in \eqref{priordeltac} and zero outside $[u_1, u_2]$.

\section{Conditional posterior distributions of the latent magnitudes}\label{app1}

We use the same notation for the observed data as is defined in Appendix~\ref{appcond}, i.e., $z'_{i}$ and $\xi_{i}$.  We introduce the centered latent magnitudes $\boldsymbol{X}'(\boldsymbol{t}^\Delta)=\boldsymbol{X}(\boldsymbol{t}^\Delta) - \mu$ for notational simplicity. Also, let ``$< t^\Delta_i$'' denote the set $\{t^\Delta_j:~j=1, 2, \ldots, i-1\}$ and ``$> t^\Delta_i$'' denote $\{t^\Delta_j:~j=i+1, i+2, \ldots, 2n\}$. To sample $p(\boldsymbol{X}'(\boldsymbol{t}^\Delta)\mid \Delta, \boldsymbol{\beta}, \mu, \sigma^2, \tau, \boldsymbol{x}, \boldsymbol{y})$ used in \eqref{eq:step1}, we sample the following conditional posterior distributions of each latent magnitude. (We suppress conditioning on $\Delta, \boldsymbol{\beta}, \mu, \sigma^2, \tau, \boldsymbol{x}, \boldsymbol{y}$.)
 \begin{equation}
X'(t^\Delta_1) \mid \boldsymbol{X'}(>t^\Delta_1)\sim \textrm{N}\left[(1-B_1)z'_{1}+B_1a_2X'(t^\Delta_2), ~(1-B_1)\xi^2_{1}\right],
\end{equation}
where $B_1=\xi^2_{1}~/~[\xi^2_{1}+\tau\sigma^2(1-a_2^2)/2]$. For $i=2, 3, \ldots, 2n-1$,
\begin{equation}
X'(t^\Delta_i) \mid \boldsymbol{X'}(<t^\Delta_i), \boldsymbol{X'}(>t^\Delta_i)~~~~~~~~~~~~~~~~~~~~~~~~~~~~~~~~~~~~~~~~~~~~~~~~~~~~~~~~~~~~
\end{equation}
\begin{equation}
\sim\! \textrm{N}\!\left[(1-B_i)z'_{i}+B_i\left((1-B_i^\ast)\frac{X'(t^\Delta_{i+1})}{a_{i+1}}+B_i^\ast a_iX'(t^\Delta_{i-1})\right),~ (1-B_i)\xi^2_{i}\right]\!,\nonumber
\end{equation}
where $B_i=\xi^2_{i}\bigg/\left[\xi^2_{i}+\frac{\tau\sigma^2}{2}\frac{(1-a^2_{i})(1-a^2_{i+1})}{1-a^2_{i}a^2_{i+1}}\right]$ and $B_i^\ast=\frac{1-a^2_{i+1}}{1-a^2_i a^2_{i+1}}$. Lastly,
\begin{equation}
X'(t^\Delta_{2n}) \mid \boldsymbol{X'}(<t^\Delta_{2n})\sim \textrm{N}\left[(1-B_{2n})z'_{2n} + B_{2n}a_{2n}X'(t^\Delta_{2n-1}),~ (1-B_{2n})\xi^2_{2n}\right],
\end{equation}
where $B_{2n}=\xi^2_{2n}/[\xi^2_{2n}+\tau\sigma^2(1-a_{2n}^2)/2]$ and $a_i=\exp(-(t_i^\Delta - t_{i-1}^\Delta)/\tau)$.  Having sampled $\boldsymbol{X}'(\boldsymbol{t}^\Delta)$, we set $\boldsymbol{X}(\boldsymbol{t}^\Delta)=\boldsymbol{X}'(\boldsymbol{t}^\Delta)+\mu$. 
%return $\boldsymbol{X}'(\boldsymbol{t}^\Delta)$  to the non-centered latent magnitudes, i.e., $\boldsymbol{X}(\boldsymbol{t}^\Delta)=\boldsymbol{X}'(\boldsymbol{t}^\Delta) + \mu$ at the end of sampling.  To save space, we suppress conditioning on $\Delta, \boldsymbol{\beta}, \mu, \sigma^2, \tau, \boldsymbol{x}, \boldsymbol{y}$. The complete conditional distributions are given by
 %Conditioning on all the model parameters, observed data, and the $n$th sample of $\boldsymbol{X'(t^\Delta)}$ represented by the superscript $(n)$, we can sample the $(n+1)$st sample of $\boldsymbol{X'(t^\Delta)}$ by

%For notational simplicity, let's define $D_{t_j}\equiv(x(t_1), x(t_2), \ldots, x_{j})$. Then,
%\begin{equation}
%X(t_1)\mid \mu, \sigma^2, \tau, D_{t_1} \sim \textrm{N}[\mu^\Delta_1, \Sigma_1],
%\end{equation}

%where $\mu^\Delta_1\equiv (1-B_1)x(t_1)+B_1\mu$,  $\Sigma_1\equiv (1-B_1)\delta^2(t_1)$, and $B_1\equiv \frac{\delta^2(t_1)}{\delta^2(t_1)+\tau\sigma^2/2}$. In general, for $j=2, 3, \ldots, n$, once we observe $x_{j}$, we update the posterior predictive distribution of $X(t_j)$;
%\begin{equation}
%X(t_j)\mid \mu, \sigma^2, \tau, D_{t_j} \sim \textrm{N}[\mu^\Delta_j, \Sigma_j],
%\end{equation}

%where $\mu^\Delta_j\equiv (1-B_j)x_{j}+B_j\big[(1-a_j)\mu+a_j \mu^\Delta_{j-1} \big]$,  $\Sigma_j\equiv (1-B_j)\delta^2_{t_j}$, and $B_j\equiv \frac{\delta^2_{t_j}}{\delta^2_{t_j}+a^2_j\Sigma_{j-1}+\tau\sigma^2(1-a^2_j)/2}$.

\section{Conditional posterior distributions of $\boldsymbol{\beta}$, $\mu$, $\sigma^2$, and $\tau$ used in Algorithm~1}\label{detail_metro}

%The second part of the factorization, i.e., $p(\Delta \mid c, \mu, \sigma^2, \tau, D_{\textrm{obs}})$, is the marginalized conditional posterior distribution of $\Delta$  with $\boldsymbol{X}(\boldsymbol{t}^\Delta)$ integrated out, see Appendix \ref{appcond}. 
%Note that  we must jointly update $\boldsymbol{X}(\boldsymbol{t}^\Delta)$ and $\Delta$  because the sorted observation times $\boldsymbol{t}^\Delta$ depend on $\Delta$; $p(\Delta\mid \boldsymbol{X}(\boldsymbol{t}^\Delta), \cdots)$ is not a valid conditional distribution because $\boldsymbol{X}(\boldsymbol{t}^\Delta)$ changes according to $\Delta$.We suppress $\Delta, c, \mu, \sigma^2, \tau, \boldsymbol{x_t}, \boldsymbol{y}$ in the condition to save space.
We specify the conditional posterior distributions of $\boldsymbol{\beta}$, $\mu$, $\sigma^2$, and $\tau$ used in Steps 2--5 of the MHwG+ASIS sampler in Algorithm~\ref{algorithm_sampler1}. We suppress explicitly conditioning on $ \boldsymbol{x}$ and  $\boldsymbol{y}$  in the condition.

Step~2 of Algorithm~\ref{algorithm_sampler1} first samples $\boldsymbol{\beta}$ from the following  Gaussian conditional posterior distribution (Step 2$_a$ in \eqref{eq:asis_step1}); with an $n$ by $n$ diagonal matrix $V$ whose diagonal elements are $\boldsymbol{\eta}^2$, 
\begin{equation}\label{ancil_c}
\boldsymbol{\beta}\mid \mu, \sigma, \tau, \boldsymbol{X}(\boldsymbol{t}^\Delta), \Delta\sim \textrm{N}_{m+1}\!\left[J^{-1}\boldsymbol{W}_m(\boldsymbol{t}-\Delta)^{\top}V^{-1}\boldsymbol{u},~J^{-1}\right]\!,
\end{equation}
where $J\equiv\boldsymbol{W}_m^{\top}(\boldsymbol{t}-\Delta)V^{-1}\boldsymbol{W}_m(\boldsymbol{t}-\Delta)+10^{-5}I_{m+1}$ and $\boldsymbol{u}\equiv \boldsymbol{y}-\boldsymbol{X}(\boldsymbol{t}-\Delta)$. To implement ASIS, we need the conditional posterior distribution for $\boldsymbol{\beta}$ given $\boldsymbol{K}(\boldsymbol{t}^\Delta)$ used in \eqref{eq:asis_step3}. Let $\boldsymbol{K}'(\boldsymbol{t}^\Delta)\equiv \boldsymbol{K}(\boldsymbol{t}^\Delta)-\mu$, $B$ be a $2n$ by $(m+1)$ matrix  whose $j$th row is $(\boldsymbol{w}_m(t^\Delta_j)-a_j\times \boldsymbol{w}_m(t^\Delta_{j-1}))^\top$  with $a_1 = 0$, $L$ be a $2n$ by $2n$ diagonal matrix whose $j$th diagonal element is $\tau\sigma^2(1-a_j^2)/2$,  $\boldsymbol{b}$ be a $2n$ by 1 vector whose $j$ th element is $K'(t^\Delta_j) - a_jK'(t^\Delta_{j-1})$, and finally $A\equiv B^\top L^{-1}B+10^{-5}I_{m+1}$. Then, 
\begin{equation}\label{suff_c}
\boldsymbol{\beta}\mid \mu, \sigma^2, \tau, \boldsymbol{K}(\boldsymbol{t}^\Delta), \Delta, \boldsymbol{x}, \boldsymbol{y}\sim 
 \textrm{N}_{m+1}\left[ A^{-1}B^\top L^{-1}\boldsymbol{b},~A^{-1}\right].
\end{equation}

In Step~3 of Algorithm~\ref{algorithm_sampler1}, we sample $\mu$ from a truncated Gaussian conditional posterior distribution whose support is $[-30, 30]$; 
\begin{align}
&\mu \mid \sigma^2, \tau, \boldsymbol{X}(\boldsymbol{t}^\Delta), \Delta, \boldsymbol{\beta} \sim\\
&~~~~~~~~~~~~~~~~~~\textrm{N}\!\left[\frac{X(t_1^\Delta) + \sum_{i=2}^{2n} \frac{X(t_i^\Delta) - a_i X(t_{i-1}^\Delta)}{1 + a_i}}{1 + \sum_{i=2}^{2n} \frac{1-a_i}{1 + a_i}},~\frac{\tau\sigma^2/2}{1 + \sum_{i=2}^n \frac{1-a_i}{1 + a_i}}\right]\!.\nonumber
\end{align}

%; with its prior distribution $p(\sigma^2)\propto \exp(-b_\sigma/\sigma^2)/(\sigma^2)^2\times I_{\{\sigma^2>0\}}$, 
In Step~4 of Algorithm~\ref{algorithm_sampler1}, the parameter $\sigma^2$ has an inverse-Gamma conditional posterior distribution,  i.e.,
\begin{equation}\label{sigma_conditional_post}
\sigma^2 \mid \mu,  \tau, \boldsymbol{X}(\boldsymbol{t}^\Delta), \Delta, \boldsymbol{\beta} \sim ~~~~~~~~~~~~~~~~~~~~~~~~~~~~~~~~~~~~~~~~~~~~~~~~~~~~~~~
\end{equation}
\begin{displaymath}
\textrm{IG}\!\left(n + 1,~ b_\sigma+\frac{(X(t_1^\Delta)-\mu)^2}{\tau}+\sum_{i=2}^{2n} \frac{\big[(X(t_i^\Delta) - \mu) - a_i(X(t_{i-1}^\Delta)-\mu)\big]^2}{\tau(1-a_i^2)}\right)\!.
\end{displaymath}

%; with its prior distribution $p(\tau)\propto \exp(-1/\tau)/\tau^{2}\times I_{\{\tau>0\}}$, 
The conditional posterior density function of $\tau$ used in Step~5 of Algorithm~\ref{algorithm_sampler1} is known up to a normalizing constant,  i.e., 
\begin{equation}\label{cond_tau2}
p(\tau \mid \mu, \sigma^2,  \boldsymbol{X}(\boldsymbol{t}^\Delta), \Delta, \boldsymbol{\beta})\propto ~~~~~~~~~~~~~~~~~~~~~~~~~~~~~~~~~~~~~~~~~~~~~~
\end{equation}
\vskip-20pt
\begin{displaymath}
~~~~~~~\frac{\exp\left(-\frac{1}{\tau}-\frac{(X(t_1^\Delta)-\mu)^2}{\tau\sigma^2}-\sum_{i=2}^{2n}\frac{\big[(X(t_i^\Delta) - \mu) - a_i(X(t_{i-1}^\Delta)-\mu)\big]^2}{\tau\sigma^2(1-a^2_i)}\right)}{\tau^{n+2}\prod_{i=2}^{2n}(1-a_i^2)^{1/2}}\times I_{\{\tau>0\}}.
\end{displaymath}

To sample $\tau$ from \eqref{cond_tau2}, we use an M-H step with a Gaussian proposal density N$[\log(\tau), \phi^2]$ on a logarithmic scale where $\phi$ is a proposal scale tuned to produce a  reasonable acceptance rate.

%\subsection{Metropolis-Hastings within Gibbs  in \ref{double1}}\label{appC1}
%From the first row,  trace plots, autocorrelation plots, and histograms are presented. Each column represents $\Delta$, $c$, $\mu$, $\sigma$, and $\tau$ from the left. Gelman-Rubin statistics for ($\Delta$, $c$, $\mu$, $\sigma$, and $\tau$) are (1.0025, 1.0088, 1.0013, 1.0011, 1.0020).\\
%\includegraphics[scale = 0.28]{appC1.png}

%\frac{ \sum_{j=1}^{2n}\frac{(K'(t_j^\Delta)-a_j K'(t_{j-1}^\Delta))(I_y(t_j^\Delta)-a_j I_y(t_{j-1}^\Delta))}{1-a_j^2} }{\sum_{j=1}^{2n}\frac{(I_y(t_j^\Delta)-a_j I_y(t_{j-1}^\Delta))^2}{1-a_j^2}},~ \frac{ \tau\sigma^2/2 }{\sum_{j=1}^{2n}\frac{(I_y(t_j^\Delta)-a_j I_y(t_{j-1}^\Delta))^2}{1-a_j^2}}
% where $a_j\equiv \exp(-(t_j^\Delta - t_{j-1}^\Delta)/\tau)$ and $a_0=0$.

\section{Profile likelihood approximation to the marginal posterior distribution of $\Delta$}\label{proof_prof_likelihood}
We show that $L_{\textrm{prof}}(\Delta)$  with a uniform prior distribution on $\Delta$  is approximately proportional to  $p(\Delta\mid \boldsymbol{x}, \boldsymbol{y})$. Let $\boldsymbol{\nu}\equiv (\boldsymbol{\beta}^\top, \boldsymbol{\theta}^\top)^\top$. Then,
\begin{equation}\label{berger}
p(\Delta\mid \boldsymbol{x}, \boldsymbol{y}) = \int L(\Delta, \boldsymbol{\nu}) p(\Delta, \boldsymbol{\nu}) d\boldsymbol{\nu} ~= k\int L(\Delta, \boldsymbol{\nu}) p(\boldsymbol{\nu}\mid \Delta) d\boldsymbol{\nu},
\end{equation}
where $k$ is a  normalizing constant of the uniform prior distribution for $\Delta$ and the likelihood function is the marginal likelihood function defined in \eqref{marginal_data_augmentation} or  \eqref{marginal_likelihood}. We specify a Jeffreys' prior on $\boldsymbol{\nu}$ given $\Delta$, i.e., $p(\boldsymbol{\nu}\mid\Delta)\propto\vert I_\Delta(\boldsymbol{\nu})\vert^{0.5}d\boldsymbol{\nu}$, where $I_\Delta(\boldsymbol{\nu})$ is the Fisher information defined as $-E\left[\partial^2 \log(L(\Delta, \boldsymbol{\nu}))/\partial \boldsymbol{\nu} \boldsymbol{\nu}^\top\right]$. The resulting $p(\Delta\mid \boldsymbol{x}, \boldsymbol{y})$ is a  Jeffreys-integrated marginal likelihood under a uniform prior \citep{berger1999integrated}. (With a uniform prior on $\boldsymbol{\nu}$ given $\Delta$, i.e., $p(\boldsymbol{\nu}\mid\Delta)\propto1$, the  likelihood is a uniform-integrated marginal likelihood which can be approximated by the Laplace method\footnote{The Laplace approximation based on the uniform prior requires the Hessian but most optimizers numerically evaluate the Hessian  (or its approximation) automatically. However, the closed-form of the Hessian matrix is not available and a numerical approximation to the Hessian matrix is unstable in the small data example in Section \ref{sec7}. The profile likelihood approximation based on the Jeffreys' prior does not require calculating the Hessian, which is a  computational advantage especially since we must evaluate the profile likelihood at every point on a grid of values of $\Delta$. We note, however, that the  Jeffreys' prior can be inappropriate for a reference prior because it sometimes becomes too informative in high dimensions \citep[e.g.,][]{berger2015overall}.}.) If  we can approximate  $l(\Delta, \boldsymbol{\nu})\equiv\log(L(\Delta, \boldsymbol{\nu}))$ with respect to $\boldsymbol{\nu}$  by a second-order Taylor's series, e.g., under standard asymptotic arguments, then
\begin{equation}
l(\Delta, \boldsymbol{\nu})\approx l(\Delta, \hat{\boldsymbol{\nu}}_{\Delta})-(\boldsymbol{\nu}-\hat{\boldsymbol{\nu}}_{\Delta})^\top[-l''(\Delta, \hat{\boldsymbol{\nu}}_{\Delta})](\boldsymbol{\nu}-\hat{\boldsymbol{\nu}}_{\Delta})/2,
\end{equation}
where $\hat{\boldsymbol{\nu}}_\Delta=\argmax_{\boldsymbol{\nu}} l(\Delta, \boldsymbol{\nu})$, and $l''(\Delta, \hat{\boldsymbol{\nu}}_\Delta)\equiv\partial^2 l(\Delta, \boldsymbol{\nu})/\partial \boldsymbol{\nu} \boldsymbol{\nu}^\top\vert_{\boldsymbol{\nu}=\hat{\boldsymbol{\nu}}_\Delta}$, which results in
% Conditioning on $\Delta$, we use a Taylor approximation to $L(\Delta, \boldsymbol{\nu})$, the part of integrand in \eqref{berger}, with respect to $\boldsymbol{\nu}$ as follows.
\begin{equation}\label{taylor}
L(\Delta, \boldsymbol{\nu})\approx \exp\left(l(\Delta, \hat{\boldsymbol{\nu}}_{\Delta})-(\boldsymbol{\nu}-\hat{\boldsymbol{\nu}}_{\Delta})^\top[-l''(\Delta, \hat{\boldsymbol{\nu}}_{\Delta})](\boldsymbol{\nu}-\hat{\boldsymbol{\nu}}_{\Delta})/2\right).
\end{equation}
Using this, we approximate the  marginal posterior density function of $\Delta$ by 
\begin{align}\label{laplace}
p(\Delta\mid &\boldsymbol{x}, \boldsymbol{y})\approx ~k\times L(\Delta, \hat{\boldsymbol{\nu}}_{\Delta})\\
&\times \int \exp\bigg(-(\boldsymbol{\nu}-\hat{\boldsymbol{\nu}}_{\Delta})^\top[-l''(\Delta, \hat{\boldsymbol{\nu}}_{\Delta})](\boldsymbol{\nu}-\hat{\boldsymbol{\nu}}_{\Delta})/2\bigg) \vert I_\Delta(\boldsymbol{\nu})\vert^{0.5}d\boldsymbol{\nu}.\nonumber
\end{align}
If we replace the Fisher information in \eqref{laplace}, i.e., $I_\Delta(\boldsymbol{\nu})$, with the observed information, $-l''_\Delta(\hat{\boldsymbol{\nu}}_\Delta)$, under standard asymptotic arguments,  the integral in  \eqref{laplace} converges to $(2\pi)^2$ because  the integrand converges to a multivariate Gaussian  density  up to $(2\pi)^{-2}$. Finally,
\begin{equation}
p(\Delta\mid \boldsymbol{x}, \boldsymbol{y})\approx k\times(2\pi)^2\times L(\Delta, \hat{\boldsymbol{\nu}}_{\Delta})=k\times(2\pi)^2\times L_{\textrm{prof}}(\Delta)\propto L_{\textrm{prof}}(\Delta).
\end{equation}

%from the first example of Section~\ref{sec6}
%\section{Histograms and ACFs of posterior samples of model parameters }\label{hist_acf}
%We plot histograms and the auto-correlation functions (ACF) of the posterior samples of all  the model parameters obtained from Section~\ref{case1} in Figure~\ref{example1_allpara}.  The convergence rates of Markov chains are satisfactory, considering fast-decaying ACFs and GRDs close to one.
%The Gelman-Rubin diagnostic statistic  (GRD) appears on the upper-right corner of each ACF. 

%\begin{figure}[h!]
%\includegraphics[scale = 0.18]{example1_allpara.png}
%\caption[]{The histograms (1st row) and auto-correlation functions (ACFs) (2nd row) of posterior samples of $\Delta_{\textrm{AB}}, c_{BA}, \mu, \log(\sigma)$, and $\log(\tau)$ from the left. The Gelman-Rubin diagnostic statistics (GRD) appear on the upper right corner of ACF plots. The  vertical red line in the histogram of $\Delta_{\textrm{AB}}$ indicates the blinded true time delay.}
%\label{example1_allpara}
%\end{figure}

\section{Sensitivity analyses}\label{sec_sensitivity} To assess the influence of prior distributions of $\tau$ and $\sigma^2$ on the posterior distribution of $\Delta$, we conduct  sensitivity analyses, varying the scale and shape parameters of their IG prior distributions. 

%It turns out that the posterior distribution of the time delay is robust to the hyper-prior distributions as long as the shape parameter of the IG of $\tau$ is moderately small and the scale parameter of the IG of $\sigma^2$ is small enough to set a lower bound.  The hyper-prior distributions of $\tau$ and $\sigma^2$ in Section~\ref{hyper_prior_description} work well in this specific example in that the model recovers the generative true parameters near the modes of their posterior distributions.

As an example, we generate 80 observations with $(\Delta, \beta_0, \mu, \sigma^2, \tau)=($50, 2, 0, 0.03$^2$, 100). The median observation cadence is 3 days and the measurement standard deviations  are set to 0.005 magnitude. When fitting the Bayesian model, we assume for simplicity  that $\Delta\sim$ Uniform[0, 100]  a priori. We run three Markov chains, each for 10,000 iterations after 10,000 burn-in iterations.

%When fitting} the Bayesian model, we assume for simplicity  that $\Delta\sim$ Uniform[0, 100]  a priori. We run three Markov chains, each for 10,000 iterations after 10,000 burn-in iterations. The initial values of $\Delta$ for  three chains are 20, 50, and 80, and those of ($\mu, \sigma, \tau$) are (0, 0.01, 200) for every chain. 
%The initial value of $\beta_0$ is the average of $\boldsymbol{y}$ minus that of $\boldsymbol{x}$.  The initial proposal scales, $\psi$ and $\phi$, are 10 and 3 days, respectively. 

%We conduct  sensitivity analyses for the shape parameter of the inverse-Gamma (hereafter IG) distribution of $\tau$ and the scale parameter of the IG distribution of $\sigma^2$ using a simulated data set. 

\subsection{Sensitivity analysis of the prior distribution of $\tau$}\label{app.e1} We investigate the sensitivity of the posterior distribution of $\Delta$ to the shape parameter of the IG prior distribution of $\tau$. We denote the shape parameter by $a_\tau$ and fix the scale parameter at one day. A reasonably small value of the scale parameter does not make any differences in the resultant posterior distributions of $\tau$ or $\Delta$ because $a_j=\exp(-(t^\Delta_j-t^\Delta_{j-1})/\tau)$  dominates the scale parameter in the conditional posterior density of $\tau$ in  \eqref{cond_tau2}.  We fix  the IG($1, b_\sigma$) prior distribution for $\sigma^2$, where $b_\sigma=2\times10^{-7}$ mag$^2/$day as  described in Section~\ref{hyper_prior_description}.
%found that the posterior distribution of $\Delta$ was robust to the IG distribution of $\tau$ unless the shape parameter was extremely large. The effect of its scale parameter is negligible because 

\begin{figure}[b!]
\includegraphics[scale = 0.25]{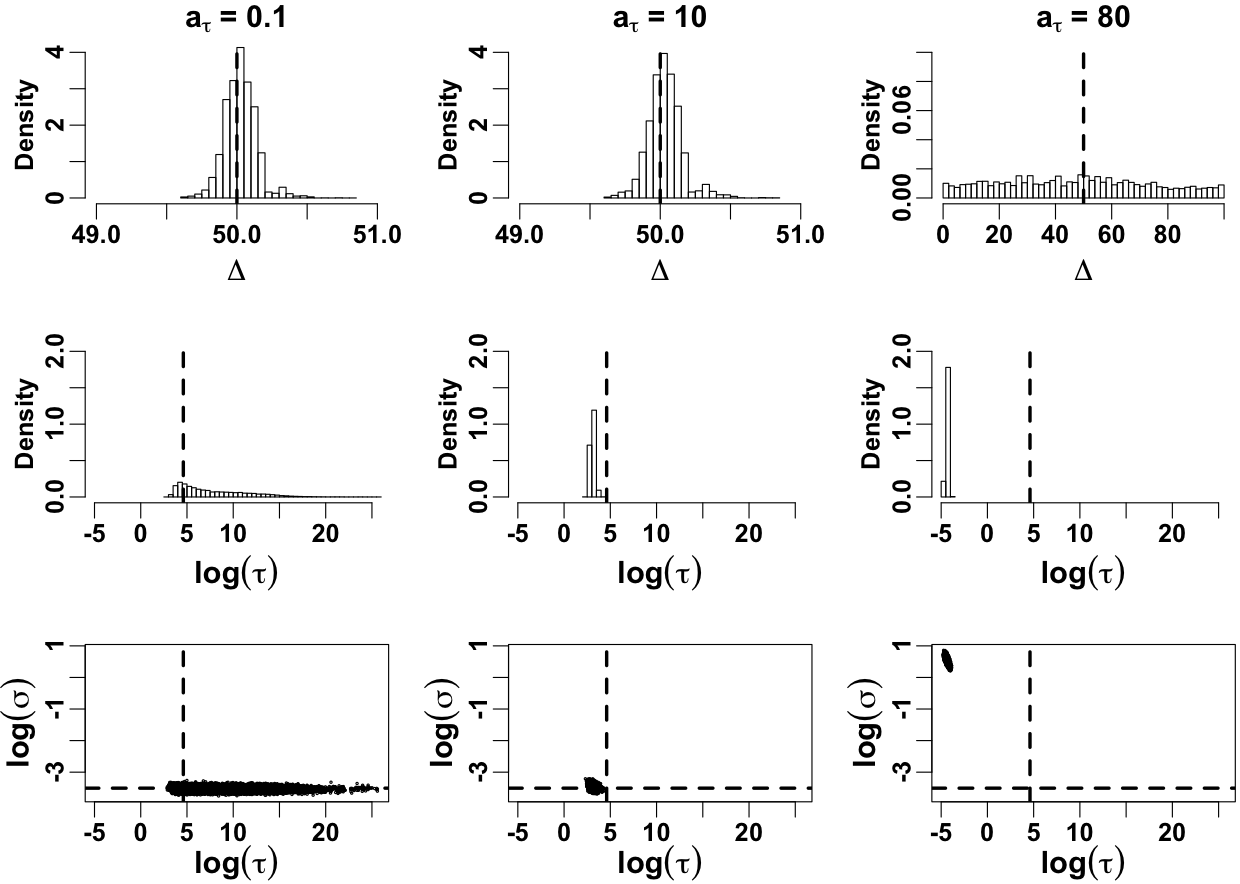}
\caption[]{Each column shows posterior distribution of $\Delta$ (first row), that of $\log(\tau)$ (second row), and a scatter plot of $\log(\sigma)$ over $\log(\tau)$ (third row) obtained under three  values of $a_\tau$ (columns, $a_\tau$=0.1, 10, and 80). The generative values of $(\Delta, \log(\sigma), \log(\tau))$ are (50, -3.5, 4.6) and represented by the dashed  lines on each plot. The posterior distribution of the time delay is robust to the shape parameter ($a_\tau$) as long as it is reasonably small. The ESS of $\Delta$ is 3235, 3068, 3080, 478, and 3625 from the left.}
\label{fig_sense_tau}
\end{figure}

%, the half of the degrees of freedom in the corresponding inverse-$\chi^2$ distribution;  from the first column.
Figure~\ref{fig_sense_tau} shows the result of sensitivity analysis with three  values of the shape parameter, $a_\tau= 0.1, 10$, and $80$ (columns). Each column shows the posterior distribution of $\Delta$ (first row), that of $\log(\tau)$ (second row), and a scatter plot of posterior samples of $\log(\sigma)$ and $\log(\tau)$ (third row) obtained under each shape parameter. The dashed  lines indicate the generative true values.  

% we assume too much information in 
%the prior distribution by setting $a_\tau$ to 80 ($=n$), 
The modes of the first two posterior distributions of $\Delta$ are near the generative value of $\Delta$. However, with the informative choice of $a_{\tau}=80$, the posterior distribution of $\Delta$ is flat.  A large value  of $a_\tau$ concentrates the prior density on the O-U processes on mean-reversion timescales $\tau$ much shorter than the observational cadence. Moreover, a large value results in a prior mode, $1/(1 + a_\tau)$, that is close to zero, and a large value of the degrees of freedom ($2 \times a_\tau$) for the prior distribution strongly influences the posterior of $\tau$. Hence, the latent light curves governed by these O-U processes with small $\tau$ will effectively appear as white noise time series.  The result is a model that is ineffective at constraining the time delay because it is unable to  match serially correlated fluctuation patterns in the light curves. The second row in Figure~\ref{fig_sense_tau} shows that as $a_\tau$ increases, the mode of the posterior distribution of $\log(\tau)$ becomes smaller with a shorter right tail and thus moves away from the generative value of $\log(\tau)=4.6$. When the mode of $\log(\tau)$ reaches $-5$ ($\tau=\exp(-5)=0.007<<$ 3-day observation cadence), the posterior distribution of $\Delta$  becomes flat.

% As a result, .  Although the posterior distributions of $\tau$ and $\sigma$ are not robust to $a_\tau$ (partly due to their negative correlation), the unit shape parameter ($a_\tau=1$) in the second column seems to help the model recover the true values of $\tau$ and $\sigma$ near the modes, effectively reducing the extreme right-skewness of the posterior distribution of $\tau$.

%The posterior distribution of $\Delta$ is robust to $a_\tau$ as long as it is moderately small.  
%The widely flat posterior distribution of $\Delta$ in the last column reflects on this point.
%As a default, we used $a_\tau=1$ in the TDC because it was the smallest integer preventing severe right-skewness in its posterior distribution when the data size was small; the conditional posterior distribution of $\tau$ behaves like 

\begin{figure}[t!]
\includegraphics[scale = 0.25]{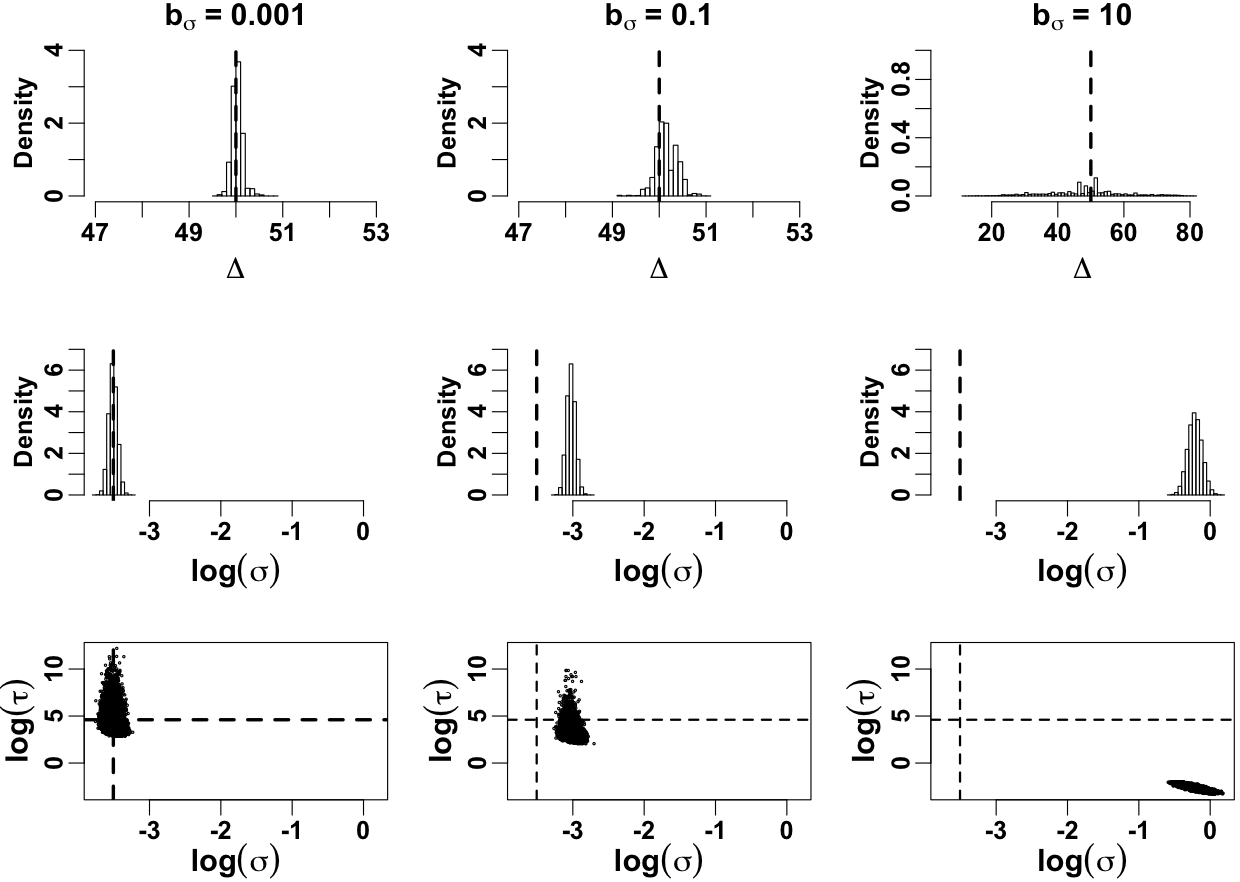}
\caption[]{Each column shows posterior distribution of $\Delta$ (first row), that of $\log(\sigma)$ (second row), and a scatter plot of $\log(\tau)$ over $\log(\sigma)$ (third row) obtained under three values of $b_\sigma$ (columns, $b_\sigma$ = 0.001, 0.1, and 10). The generative values of $(\Delta, \log(\sigma), \log(\tau))$ are (50, -3.5, 4.6) and represented by the dashed  lines on each plot. The modes of the posterior distributions of parameters are near the generative values as the scale parameter (soft lower bound) decreases. The ESS of $\Delta$ is 2957, 3082, 4148, 2459, and 23 from the left.}
\label{fig_sense_sigma}
\end{figure}

\subsection{Sensitivity analysis of the prior distribution of $\sigma^2$}\label{sub_sensitivity_sigma}
We check the sensitivity of the posterior distribution of $\Delta$ to the scale parameter $b_\sigma$ of the IG(1, $b_\sigma$) prior distribution for $\sigma^2$. The effect of the unit shape parameter is negligible because the resultant shape parameter of the IG conditional posterior distribution of $\sigma^2$ in  \eqref{sigma_conditional_post} is $n+1$ so that $n$ plays a dominant role in controlling the right tail behavior. We fix the IG(1, $b_\tau$) prior distribution for $\tau$, where $b_\tau$ is fixed at one day, as described in Section~\ref{hyper_prior_description}.

%It turns out that the posterior distribution of $\Delta$ is robust to the IG distribution of $\sigma^2$ as long as its scale parameter (a soft lower bound) is as small as we suggested (with a unit shape parameter). The effect of the shape parameter, denoted by $a_\sigma$, on the posterior distribution of $\sigma^2$ is negligible unless it is dogmatically large. This is because the shape parameter of its conditional posterior distribution in Appendix \ref{cond_sigma} is $n+a_\sigma$. As a result, the motivation to control the right-tail behavior is weaker than the case of $\tau$ because the right-tail is sufficiently controlled by $n$. Thus,  while 

%the mode of the posterior distribution of $\log(\tau)$ becomes} smaller with a shorter right tail and thus moves away} from the generative value of $\log(\tau)=4.6$.

We display the result of the sensitivity analysis  in Figure~\ref{fig_sense_sigma}, where the values of $b_\sigma$ are increasing from $0.001$ to $10$ from the first column. As the soft lower bound ($=b_\sigma/2$) increases from the left, the posterior distribution of the time delay becomes flatter. This is because the generative value of $\sigma^2~(=0.03^2)$ is less than the soft lower bound for large values of $b_\sigma$. For example, when $b_\sigma=10$ in the right most column, the IG(1, 10) prior distribution of $\sigma^2$ exponentially cuts off the probability density in the region to left of the mode, 5 mag$^2/$day, which excludes the generative value of $\sigma^2$ $(=0.03^2)$. Because the generative $\sigma^2$ is much smaller than the soft lower bound, the posterior distribution of $\sigma^2$ has negligible mass near the generative value of $\sigma^2$. Also, because the posterior samples of $\sigma$ and $\tau$ are negatively correlated a posteriori as shown in the scatter plots, posterior distributions that favor large values of $\sigma^2$ also favor small values of $\tau$. As discussed in Appendix~\ref{app.e1},  when the posterior distribution of $\tau$ is concentrated on values smaller than the observational cadence, the posterior latent light curve $\boldsymbol{X}(\boldsymbol{t}^\Delta)$ effectively becomes a white noise sequence.  In this case, it is difficult to constrain $\Delta$.
% by overcoming the exponential cut-off as the sample size is small. 
% larger posterior samples of $\sigma^2$ than the true $\sigma^2$ lead to relatively smaller posterior samples of $\tau$ than the true $\tau$. 

The second row of Figure~\ref{fig_sense_sigma} shows that as the soft lower bound decreases from the right, the posterior distribution of $\log(\sigma)$ moves towards the generative value of $\log(\sigma)=-3.5$. Though not shown here,  posterior distributions  obtained under a value of $b_\sigma$ smaller than 0.001 do not noticeably differ from that obtained with $b_\sigma=0.001$. With the small soft lower bound,  the modes of the posterior distributions of the other parameters tend to be near their generative values.  We also found that the choice of $b_\sigma$ is less important for large data sets, e.g., with $n>400$.

%For reference, although it may depend on data, the results became insensitive to the scale $b_\sigma$ as the number of observations was more than 400.

%
%\section{The profile likelihood with a curve-shifted model} \label{appendix_micro}
%The curve-shifted model assumes all the regression coefficients except $\beta_0$ are zero, allowing only the shifts in $x$- and $y$-axes. This model produces several modes near margins for quasars $Q0957+561$ and $J1029+2623$, which is evidence of the microlensing variability. The first panel of Figure~\ref{fig_appendix_micro} displays the profile likelihood with a curve-shifted model for the data of $Q0957+561$ and the second panel plots that for the data of $J1029+2623$.

%\begin{figure}[h!]
%\includegraphics[scale = 0.3]{appendix_micro.png}
%\caption[]{The profile likelihood with a curve-shifted model for the data of $Q0957+561$ (left) and that for the data of $J1029+2623$ (right).}
%\label{fig_appendix_micro}
%:
%\end{figure}

\end{document}